\documentclass[journal]{IEEEtran}
\usepackage{cite}
\usepackage{float}
\usepackage[cmex10]{amsmath}
\usepackage{amssymb}
\usepackage{amsfonts}
\usepackage{psfrag}
\usepackage{subfigure}
\usepackage{epsfig}
\usepackage{algorithm}
\usepackage{algorithmic}

\usepackage{fancyhdr}
\usepackage[normalem]{ulem}
\usepackage[usenames,dvipsnames]{color}
\newcommand{\rd}{\textcolor{Red}}

\begin{document}

\title{Full-Wave Iterative Image Reconstruction
in Photoacoustic Tomography with Acoustically
Inhomogeneous Media}

\author{Chao~Huang, Kun~Wang, Liming~Nie, Lihong~V.~Wang,~\IEEEmembership{Fellow,~IEEE,}
 and~Mark~A.~Anastasio,~\IEEEmembership{Senior Member,~IEEE,}
\thanks{
Department of Biomedical Engineering, Washington University in St. Louis, 
St. Louis, MO 63130
}}
\maketitle

\begin{abstract}
Existing approaches to image reconstruction
in photoacoustic computed tomography (PACT) with
acoustically heterogeneous media are limited to weakly varying media, are
computationally burdensome, and/or 
cannot effectively mitigate the effects of measurement data incompleteness and noise.
In this work, we develop and investigate a
discrete imaging model
for PACT  that is based on the exact photoacoustic (PA) wave equation
 and facilitates
the circumvention of these limitations.
A key contribution of the work is the establishment of a
procedure to implement a matched  forward and backprojection
operator pair associated with the discrete imaging model,
which permits application of a wide-range of modern image
reconstruction algorithms that can mitigate 
the effects of data incompleteness and noise.
The forward and backprojection operators are based on the k-space pseudospectral method
for computing numerical solutions to
the PA wave equation in the time domain.
The developed reconstruction methodology is investigated by use of both
computer-simulated and experimental PACT measurement data.
\end{abstract}



\thispagestyle{fancy}

\footnote{Copyright (c) 2010 IEEE. Personal use of this material is permitted. However, permission to use this material for any other purposes must be obtained from the IEEE by sending a request to pubs-permissions@ieee.org.}

\section{Introduction}
\label{sect:intro}

Photoacoustic computed tomography (PACT), also known as optoacoustic
or thermoacoustic tomography, is a rapidly emerging
 hybrid imaging modality that combines optical image contrast
 with ultrasound detection. \cite{WanglihongBook2009,OraevskyBook2003,XuminghuaRSI2006,XuzhunJBO2010}
 In PACT, the to-be-imaged object is illuminated
 with a pulsed optical wavefield.  Under conditions
of thermal confinement \cite{GusevBook1993, OraevskyBook2003}, the absorption
of the optical energy results  in the generation of
 acoustic wavefields via the thermoacoustic effect.
These wavefields propagate out of the object and
are measured by use of wide-band ultrasonic transducers.
 From these measurements,
a tomographic reconstruction algorithm
 is employed to obtain an image that depicts the spatially variant
 absorbed optical energy density distribution within the object,
which will be denoted by the function $A(\mathbf r)$.
 Because the optical absorption properties of tissue
 are highly related to its hemoglobin concentration and molecular constitution,
PACT holds great potential for a wide-range of anatomical, functional,
and molecular imaging tasks in preclinical and clinical medicine
\cite{XuminghuaRSI2006,KrugerMP1999,HaltmeierIP2004,CoxOSA2006,XuzhunJBO2011}.

The majority of currently available PACT reconstruction algorithms 
are based on idealized imaging models that assume a lossless and acoustically
homogeneous medium.   However, in many applications of PACT these
assumptions are violated and the induced photoacoustic (PA) wavefields
are scattered and absorbed as they propagate to the receiving transducers.
In small animal imaging applications of PACT, for example,
the presence of bone and/or gas pockets can strongly perturb
the photoacoustic wavefield.
Another example is
 transcranial PACT brain imaging of primates \cite{HuangchaoJBObrain},
in which the PA wavefields
can be strongly aberrated and attenuated \cite{FryJASA1978,XingjinMP2008,HuangchaoJBOatten} by the skull.
In these and other biomedical applications of PACT, the reconstructed images can
contain significant distortions and artifacts if the inhomogeneous
acoustic properties of the object are not accounted for in the
reconstruction algorithm.

Several image reconstruction methods 
have been proposed to compensate for weak variations
in a medium's speed-of-sound (SOS) distribution
 \cite{XuyuanIEEE2003,DimpleJBO2010,JoseOE2011}.
These methods are based on geometrical acoustic
approximations to the PA wave equation, 
which stipulate
that the PA wavefields propagate along well-defined rays.
For these ray-based propagation models
to be valid, variations in the SOS distribution must
occur on length scales that are large compared to
the effective acoustic wavelength.
These assumptions can be violated
in preclinical and clinical applications of PACT.
To compensate for strong SOS variations, a statistical approach has been 
proposed \cite{DeanBenAPL2011} to mitigate the artifacts in the reconstructed images
caused by the wavefront distortions by use of \emph{a priori} information
regarding the acoustic heterogeneities. However, 
this method neglected variations in the medium's mass density 
and the effects of acoustic attenuation.


A few works have reported the development of
full-wave PACT reconstruction algorithms that are based on
solutions to the exact PA wave equation
\cite{Yuanzhen2007MP,Yaolei2011AO,HristovaIP2008,TreebyIP2010,StefanovIP2009,QianiJIS2011}.
 While these methods are grounded in accurate models of the imaging physics
and therefore have a broader domain of applicability than ray-based
methods, they also possess certain practical limitations.
Finite element methods (FEMs) have been applied for inverting
the PA wave equation in both the time and temporal frequency
domains \cite{Yuanzhen2007MP,Yaolei2011AO}.  However, a very large computational burden
accompanies these methods, which is especially problematic
for three-dimensional (3D) applications of PACT.
Image reconstruction methods based on
time-reversal (TR)  are mathematically exact
in their continuous forms
in homogeneous media for the 3D case \cite{HristovaIP2008}.
While these methods possess significantly lower
 computational burdens then FEM-based approaches,
they possess other limitations for use with practical PACT
applications.
For example, TR methods
are predicated upon the assumption that the measured PA signals
are densely sampled on a measurement surface that
encloses the object,  
which is seldom  achievable in biomedical applications of PACT.
More recently, a Neumann series-based reconstruction method has been reported
\cite{StefanovIP2009,QianiJIS2011}
for media containing SOS variations 
that is based on a discretization of a  mathematically exact
inversion formula.    
The robustness of the method to practical sparse sampling of
PA signals, however, has not been established.

In this work, we develop and investigate a full-wave approach to iterative
image reconstruction in PACT with media possessing
inhomogeneous SOS and mass density distributions as well as 
 acoustic attenuation described by a frequency power law.
The primary contributions of the work are the establishment of a
discrete imaging model that is based on the exact PA wave equation 
and a procedure to implement an associated 
matched discrete forward and backprojection
operator pair.
The availability of efficient numerical procedures to implement these 
 operators
permits a variety of modern
iterative reconstruction methods to be employed that can 
effectively mitigate image artifacts due to  data incompleteness,
noise, finite sampling , and modeling errors.
Specifically, the k-space pseudospectral method is adopted
 \cite{TreebyIP2010} for implementing the forward operator and
a numerical procedure for implementing 
the exact adjoint of this operator is provided.
The k-space pseudospectral method possesses significant
computational advantages over real space finite-difference 
and finite-element methods,
as it allows fewer mesh points per wavelength and allows larger time steps
without reducing accuracy or introducing instability \cite{CoxJASA2007}.
An iterative image reconstruction algorithm that seeks to minimize
a total variation (TV)-regularized penalized least
 squares (PLS) cost function is implemented by use of the developed
projection operators and investigated in computer-simulation and
experimental studies of PACT in inhomogeneous acoustic media.
Also, the performance of this algorithm is compared to that of an existing
TR  method.



The paper is organized as follows. In Section \ref{sect:background}, the salient
imaging physics and image reconstruction principles are briefly reviewed. 
The explicit formulation of the discrete imaging model
is described in Section \ref{sec:model}. Section \ref{sect:studies} gives
a description of the numerical and experimental studies, which includes
the implementation of the forward and backprojection operators, and the  
iterative reconstruction algorithm. The numerical and experimental results 
are given in Section \ref{sect:results}. The paper concludes with 
a summary and discussion in Section \ref{sect:summary}.

\section{Background}
\label{sect:background}

Below we review descriptions of photoacoustic wavefield generation
and propagation in their continuous and discrete forms.
The discrete description is based on the k-space pseudospectral
method \cite{MastIEEE2001,CoxJASA2007,TreebyIP2010}.
We present the pseudospectral k-space method by
 use of matrix notation,
which facilitates the establishment of a discrete PACT imaging model in Section \ref{sec:model}.
  We also  summarize a discrete formulation
of the image reconstruction problem for PACT in acoustically
inhomogeneous media.
Unless otherwise indicated, 
 lowercase and uppercase symbols in bold font will 
denote vectors and matrices, respectively.

\subsection{Photoacoustic wavefield propagation: Continuous formulation}

Let $p(\mathbf r, t)$ denote the thermoacoustically-induced pressure
wavefield at location $\mathbf r \in \mathbb{R}^3$
and time $t\ge 0$.
Additionally, let
 $A(\mathbf r)$ denote the absorbed optical energy
density within the object, { $\Gamma(\mathbf r)$ } denote the
dimensionless Grueneisen parameter,
$\mathbf{u} (\mathbf{r},t)\equiv \left(u^1 (\mathbf{r},t), u^2 (\mathbf{r},t), u^3 (\mathbf{r},t)\right)$ denote the vector-valued acoustic particle
velocity, $c_0(\mathbf{r})$ denote the medium's SOS distribution,
and $\rho(\mathbf{r},t)$ and $\rho_0(\mathbf{r})$
denote  the distributions of the
medium's acoustic and ambient densities, respectively.
The object function $A(\mathbf r)$ and all quantities
that describe properties of the medium
are assumed to be represented by bounded functions  possessing
compact supports.

In many applications, acoustic absorption is not negligible
\cite{TreebyIP2010,PatrickOL2006,BurgholzerSPIE2007,DimpleSPIE2009,BenPMB2011}.
For a wide variety of lossy materials, including biological tissues,
the acoustic attenuation coefficient $\alpha$
can be described by
a frequency power law of the form \cite{SzaboJASA1994}
\begin{equation}
\label{eq:powerlaw}
\alpha(\mathbf r, f)=\alpha_0(\mathbf r) f^y,
\end{equation}
where $f$ is the  temporal frequency in MHz,
 $\alpha_0$ is the frequency-independent attenuation coefficient
in dB MHz$^{-y}$ cm$^{-1}$,
and $y$ is the power law exponent
which is typically in the range of 0.9-2.0 
in tissues \cite{SzaboBook2004}.

In a heterogeneous lossy fluid medium 
in which the acoustic absorption  is described by
the frequency power law, the  propagation of $p(\mathbf r,t)$
can be modeled by the following three coupled equations
\cite{MorseBook1987,TreebyIP2010}
\begin{equation}
\label{eq:motion}
\frac{\partial}{\partial t} \mathbf{u} (\mathbf{r},t) = - \frac{1}{\rho_0(\mathbf{r})} \nabla p(\mathbf{r},t),
\end{equation}
\begin{equation}
\label{eq:cont}
\frac{\partial}{\partial t} \rho (\mathbf{r},t) = - \rho_0(\mathbf{r}) \nabla \cdot \mathbf{u} (\mathbf{r},t),
\end{equation}
\begin{equation}
\label{eq:lossy}
\begin{split}
p(\mathbf{r},t) = c_0(\mathbf{r})^2\big\{  1 & - \mu(\mathbf{r}) \frac{\partial}{\partial t} (-\nabla^2)^{y/2-1}  \\
                                             & - \eta(\mathbf{r}) (-\nabla^2)^{(y-1)/2} \big\} \rho(\mathbf{r},t),
\end{split}
\end{equation}
subject to the initial conditions:
\begin{equation}
\label{eq:initialc}
 p_0(\mathbf{r})\equiv p(\mathbf{r},t) |_{t=0} = { \Gamma(\mathbf r) } A(\mathbf r), \qquad \mathbf{u} (\mathbf{r},t)  |_{t=0} = 0. 
\end{equation}
where the quantities $\mu(\mathbf{r})$ and $\eta(\mathbf{r})$ describe
the acoustic absorption and dispersion proportionality coefficients that are defined as
\begin{equation}
\label{eq:mu_eta}
\begin{split}
\mu(\mathbf{r}) = -2 \alpha_0 c_0(\mathbf{r})^{y-1}, \qquad \eta(\mathbf{r}) = 2 \alpha_0 c_0(\mathbf{r})^y \text{tan}(\pi y/2).
\end{split}
\end{equation}
Note that acoustic absorption and dispersion are modeled by
the second and third terms in the bracket,
which employ two lossy derivative operators
based on the fractional Laplacian
to separately account for the acoustic absorption and dispersion
in a way that is consistent  with Eqn. (\ref{eq:powerlaw}).
When acoustic attenuation can be neglected, 
$\mu(\mathbf{r})=0$ and $\eta(\mathbf{r})=0$,
and Eqn. (\ref{eq:lossy}) reduces to
\begin{equation}
\label{eq:lossless}
p(\mathbf{r},t) = c_0(\mathbf{r})^2 \rho(\mathbf{r},t).
\end{equation}

\subsection{Photoacoustic wavefield propagation: Discrete formulation}
\label{sec:df}

The k-space pseudospectral method 
can be employed to propagate a photoacoustic wavefield
forward in space and time by computing numerical solutions
to the coupled equations described by Eqn. \eqref{eq:motion},
 \eqref{eq:cont}, \eqref{eq:lossy}, and
\eqref{eq:initialc}.
This method can be significantly more computationally efficient
 than real space finite-element and finite-difference methods
because it employs the fast Fourier transform (FFT) algorithm to compute the
spatial partial derivatives and possesses less restrictive spatial and temporal sampling
requirements.  Applications of the  k-space pseudospectral method
in studies of PACT can be found in references
\cite{CoxJASA2007,TreebyIP2010,HuangchaoJBOatten,HuangchaoJBObrain}.

The salient features of the k-space pseudospectral method
that will underlie the discrete PACT imaging model 
are described below. Additional details regarding the application
of this method to PACT have been published by Treeby and Cox
in references \cite{CoxJASA2007,TreebyIP2010}.
Let $\mathbf{r}_1,\cdots,\mathbf{r}_N \in \mathbb{R}^3$ 
specify the locations of the $N=N_1 N_2 N_3$ vertices 
of a 3D Cartesian grid,  {where $N_i$ denotes 
the number of vertices along the $i$-th dimension.}
Additionally, let
  $m \Delta t$, $m\in \mathbb Z^*$, $\Delta t\in \mathbb{R}^+$, denote
discretized values of the temporal coordinate $t$, where $ Z^*$
and $ \mathbb{R}^+$ denote the sets of non-negative integers
and positive real numbers.
The sampled values of 
$p(\mathbf{r}, t=m\Delta t)$ and ${u}^i (\mathbf{r},t=m\Delta t)$, $i=1,2$ or 3,
corresponding to spatial locations on the 3D Cartesian grid
will be described by
the 3D matrices  $\mathbf{P}_m$ and $\mathbf{U}^i_m$, respectively,
where the subscript $m$ indicates that these quantities depend on the temporal sample index.
Unless otherwise indicated, the dimensions of all 3D matrices will be
$N_1\times N_2\times N_3$.
Lexicographically ordered vector representations of these matrices
will be denoted as
\begin{equation}
\label{eq:u}
\mathbf{u}_m^i \equiv (u^i(\mathbf{r}_1,m\Delta t),\cdots, u^i(\mathbf{r}_N,m\Delta t))^{\rm T}, 
\end{equation}
and 
\begin{equation}
\label{eq:p}
\mathbf{p}_m \equiv (p(\mathbf{r}_1,m\Delta t),\cdots, p(\mathbf{r}_N,m\Delta t))^{\rm T}.
\end{equation}
 The sampled values of the ambient density $\rho_0(\mathbf r)$
 and squared SOS distribution $c_0^2(\mathbf r)$ will be represented
as
\begin{equation}
\label{eq:rho0}
\mathbf{Q} \equiv \text{diag}(\rho_0(\mathbf{r}_1), \cdots, \rho_0(\mathbf{r}_N)),
\end{equation}
and
\begin{equation}
\mathbf{C} \equiv
\text{diag}(c_0^2(\mathbf{r}_1), \cdots, c_0^2(\mathbf{r}_N)),
\end{equation}
where diag($a_1,...,a_N$)
 defines a diagonal 2D matrix whose diagonal entries starting in the upper left corner
 are $a_1,...,a_N$.

In the k-space pseudospectral method,
the 1D discrete spatial derivatives of the sampled fields
with respect to the $i$-th dimension ($i=1,2,$ or $3$)
are computed in the Fourier domain as
\begin{equation}
\label{eq:Dp}
\boldsymbol\nabla_{i}^{\text{Mat}} \mathbf{P}_m \equiv
\mathbf{F}^{-1}\{ j \mathbf{K}^i \circ \boldsymbol\kappa \circ \mathbf{F} \{ \mathbf{P}_m \}\},
\end{equation}
and
\begin{equation}
\label{eq:Du}
\boldsymbol\nabla_{i}^{\text{Mat}}
\mathbf{U}_{m}^i \equiv
\mathbf{F}^{-1} \{ j \mathbf{K}^i \circ \boldsymbol\kappa \circ \mathbf{F} 
\{ \mathbf{U}_{m}^i \} \},
\end{equation}
where $j\equiv\sqrt{-1}$, {the superscript `Mat' indicates
that the 1D discrete derivative operator $\boldsymbol\nabla_{i}^{\text{Mat}}$ acts on a 3D matrix},  
 $\mathbf{F}$ and $\mathbf{F}^{-1}$ denote the 3D
forward and inverse discrete Fourier transforms (DFTs),
and $\circ$ denotes Hadamard product.
The elements of the 3D matrix
$\mathbf{K}^i$ ($i=1,2,3$) are given by
\begin{equation}
\begin{split}
& \mathbf{K}^1_{n_1n_2n_3}=2 \pi \frac{n_1-1}{L_1}, \\
& \mathbf{K}^2_{n_1n_2n_3}=2 \pi \frac{n_2-1}{L_2}, \\
& \mathbf{K}^3_{n_1n_2n_3}=2 \pi \frac{n_3-1}{L_3},
\end{split}
\end{equation}
where $n_i = 1,\cdots,N_i$ ($i=1,2,3$), 
and $L_i$ denotes the length of 
the spatial grid in the $i$-th dimension.

The 3D matrix
$\boldsymbol\kappa=\text{sinc}(\frac{1}{2}\Delta t c_{\text{min}} \mathbf{K})$
is the k-space operator,
where $\text{sinc}(x)=\frac{\sin(x)}{x}$, 
$c_{\text{min}}$ is the minimum of $c_0(\mathbf{r})$,
$\mathbf{K}$ is a 3D matrix defined as
\begin{equation}
\mathbf{K}\equiv \sqrt{\sum_{i=1}^3 \mathbf{K}^i \circ \mathbf{K}^i},
\end{equation}
and the sinc function and square root function
are both element-wise operations.

{
Consider the operators
$\boldsymbol\Phi_{i}^\text{Mat}$ and $\boldsymbol\Psi_{i}^{\text{Mat}}$
that
 are defined as
\begin{equation}
\label{eq:Phi}
\boldsymbol\Phi_i^{\text{Mat}} \mathbf{P}_m \equiv
- {\Delta t}\,\mathbf{Q}^{-1}
\boldsymbol\nabla_{i}^{\text{Mat}} \mathbf{P}_m,
\end{equation}
and
\begin{equation}
\label{eq:Psi}
\boldsymbol\Psi_i^{\text{Mat}} \mathbf{U}_m \equiv
- {\Delta t}\,\mathbf{Q}
\boldsymbol\nabla_{i}^{\text{Mat}} \mathbf{U}_m^i.
\end{equation}

It will prove convenient to introduce the $N \times N$ matrices
$\boldsymbol\Phi_{i}$ and $\boldsymbol\Psi_{i}$
that act on the vector representations of the matrices
 $\mathbf{P}_m$ and $\mathbf{U}^i_m$, respectively.
Specifically, $\boldsymbol\Phi_{i}$ and $\boldsymbol\Psi_{i}$
are defined such that
 $\boldsymbol\Phi_{i} \mathbf{p}_m $ and 
$\boldsymbol\Psi_{i} \mathbf{u}_{m}^i$ are lexicographically ordered
vector representations of the matrices $\boldsymbol\Phi_i^{\text{Mat}} \mathbf{P}_m $
and $\boldsymbol\Psi_i^{\text{Mat}} \mathbf{U}_{m}^i$, respectively.
In terms of these quantities, the discretized forms of Eqn. \eqref{eq:motion}, \eqref{eq:cont},  
and \eqref{eq:lossy} can be expressed as}
\begin{equation}
\label{eq:d_motion}
\mathbf{u}_{m+1}^i = 
\mathbf{u}_m^i + \boldsymbol\Phi_i \mathbf{p}_m,
\end{equation}
\begin{equation}
\label{eq:d_cont}
\boldsymbol\rho_{m+1}^i = 
\boldsymbol\rho_m^i + \boldsymbol\Psi_i \mathbf{u}_{m+1}^i,
\end{equation}
where $\boldsymbol\rho_{m}^i$ is an
$N\times 1$ vector whose elements are  defined to be zero for $m=0$, and
\begin{equation}
\label{eq:d_loss}
\mathbf{p}_{m+1}
= \mathbf{C} \sum_{i=1}^3 
\{
\boldsymbol\rho_{m+1}^i 
+ \mathbf{A} \mathbf{u}_{m+1}^i
+ \mathbf{B} \boldsymbol\rho_{m+1}^i
\}.
\end{equation}
{The quantities $\mathbf{A} \mathbf{u}_{m+1}^i$ and 
$\mathbf{B} \mathbf{\rho}_{m+1}^i$ in Eqn.\ (\ref{eq:d_loss}) 
represent the absorption and dispersion terms in the equation
of state.  They are defined as
 lexicographically ordered vector representations of 
$\mathbf{A}^{\text{Mat}} \mathbf{U}_{m+1}^i$ and $\mathbf{B}^{\text{Mat}} \mathbf{N}_{m+1}^i$,
which are defined in analogy to Eqn.\ (\ref{eq:lossy}) as
} %
\begin{equation}
\label{eq:abs}
\mathbf{A}^{\text{Mat}} \mathbf{U}_{m+1}^i \equiv 
\boldsymbol\mu \mathbf{F}^{-1} \left\{
\mathbf{K}^{y-2} \mathbf{F} \Big \{
\mathbf{Q} \sum_{i=1}^3
\boldsymbol\nabla_i^{\text{Mat}} \mathbf{U}_{m+1}^i
\Big \}
\right \},
\end{equation}
\begin{equation}
\label{eq:disp}
\mathbf{B}^{\text{Mat}} \mathbf{N}_{m+1}^i \equiv 
\boldsymbol\eta \mathbf{F}^{-1} \left \{
\mathbf{K}^{y-1} \mathbf{F} \Big \{
\sum_{i=1}^3 \mathbf{N}_{m+1}^i
\Big \}
\right \},
\end{equation}
where $\mathbf{N}_{m+1}^i$ 
is the 3D matrix form of $\boldsymbol\rho_{m}^i$,
and $\boldsymbol\mu$ and $\boldsymbol\eta$ are defined as
\begin{equation}
\boldsymbol\mu \equiv
\text{diag}(\mu_0(\mathbf{r}_1), \cdots, \mu_0(\mathbf{r}_N)),
\end{equation}
\begin{equation}
\boldsymbol\eta \equiv
\text{diag}(\eta_0(\mathbf{r}_1), \cdots, \eta_0(\mathbf{r}_N)),
\end{equation}
and $\mathbf{K}^{y-2}$ and $\mathbf{K}^{y-1}$ are powers
of  $\mathbf{K}$ that are computed on an element-wise basis.

\subsection{The image reconstruction problem}
Here, for simplicity, we neglect the acousto-electrical
impulse response (EIR) of the ultrasonic transducers and assume
each transducer is point-like. However, a description of how to
incorporate the transducer responses in the developed 
imaging model is provided in Appendix-A. 
With these assumptions, we can define 
$\mathbf{\hat{p}}_m \equiv ({p}(\mathbf{r}_1^d,m \Delta t), 
\cdots, {p}(\mathbf{r}_L^d,m \Delta t))^{\rm T}$ 
as the measured pressure wavefield data 
at time $t=m \Delta t$ ($m=0,\cdots,M-1$), where 
$M$ is the total number of time steps and $\mathbf{r}_l^d \in \mathbb{R}^3$
($l=1,\cdots,L$) denotes the positions of the $L$ ultrasonic transducers 
that  reside outside the support of the object.
The PACT image reconstruction problem we address is to obtain an estimate of
$ p_0(\mathbf{r})$ or, equivalently,  $A(\mathbf r)$,
from knowledge of $\mathbf{\hat{p}}_m$, $m=0,\cdots,M-1$,
$c_0(\mathbf{r})$, $\rho_0(\mathbf{r})$, $\alpha_0(\mathbf{r})$, and $y$.
The development of image reconstruction methods for addressing this
problem is an active area of research
\cite{StefanovIP2009,HristovaIP2008,CoxJASA2007,HuangSPIE2010,HuangchaoJBObrain}.
Note that the acoustic parameters of the medium can be estimated by use of adjunct
ultrasound tomography image data 
\cite{AubryJASA2003, XingjinPMB, JoseMP2012} and are assumed to be known.
The effects of errors in these quantities on the accuracy of the reconstructed
PACT image will be investigated in Section \ref{sect:studies}.

The discrete form of the imaging model for PACT can be expressed generally as 
\begin{equation}
\hat{\mathbf{p}}=\mathbf H \mathbf{{p}_0},
\label{eqn:imagingmodel}
\end{equation}
where the $LM \times 1$ vector 
\begin{equation}
\label{eq:phat}
\hat{\mathbf{p}}
\equiv
\begin{bmatrix}
\hat{\mathbf{p}}_0 \\
\hat{\mathbf{p}}_1 \\
\vdots \\
\hat{\mathbf{p}}_{M-1}
\end{bmatrix},
\end{equation}
represents the measured pressure data corresponding to
all transducer locations and temporal samples, and the $N\times 1$ vector $\mathbf p_0$ is the discrete
representation of the sought
after initial pressure distribution within the object (i.e., Eqn.\ (\ref{eq:p}) with $m=0$).
The $LM \times N$ matrix $\mathbf H$ represents the discrete imaging operator, also referred
to as the system matrix.

The image reconstruction task is to determine an estimate of
$\mathbf p_0$ from knowledge of the measured data $\hat{\mathbf{p}}$.
This can be accomplished by computing an appropriately regularized
inversion of Eqn.\ (\ref{eqn:imagingmodel}).
When iterative methods are employed to achieve this by minimizing
a penalized least squares cost function \cite{Fessler94}, 
the action of the operators $\mathbf H$
and its adjoint $\mathbf H^{\dagger }$  must be computed.
Methods for implementing these operators are described below.

\section{Explicit formulation of discrete imaging model}
\label{sec:model}

The k-space pseudospectral method for numerically solving
the photoacoustic wave equation described in Section \ref{sec:df}
will be employed to implement the action of the system
matrix $\mathbf H$.  In this section,
we provide an explicit matrix representation of 
 $\mathbf H$  that will subsequently be employed to 
determine $\mathbf H^{\dagger}$.

Equations \eqref{eq:d_motion} - \eqref{eq:d_loss} 
can be described by a single matrix equation 
to determine the updated wavefield variables after 
a time step $\Delta t$ as
\begin{equation}
\label{eq:update}
\mathbf{v}_{m+1} = \mathbf{W}\mathbf{v}_{m},
\end{equation}
where $\mathbf{v}_{m} = (\mathbf{u}_m^1,
\mathbf{u}_m^2, \mathbf{u}_m^3,
\boldsymbol\rho_m^1,
\boldsymbol\rho_m^2,
\boldsymbol\rho_m^3,
\mathbf{p}_m)^{\rm T}$
is a $7N \times 1$ vector containing 
all the wavefield variables at the time step $m\Delta t$.
The $7N \times 7N$ propagator matrix $\mathbf{W}$ is defined as
\begin{equation}
\label{eq:W}
\begin{split}
& \mathbf{W} \equiv \\
& \begin{bmatrix}
\mathbf{I}_{N \times N} & \mathbf{0}_{N \times N} &  \mathbf{0}_{N \times N} & \mathbf{0}_{N \times N} & \mathbf{0}_{N \times N} &  \mathbf{0}_{N \times N} & \boldsymbol\Phi_1 \\
\mathbf{0}_{N \times N} & \mathbf{I}_{N \times N} &  \mathbf{0}_{N \times N} & \mathbf{0}_{N \times N} & \mathbf{0}_{N \times N} &  \mathbf{0}_{N \times N} & \boldsymbol\Phi_2 \\
\mathbf{0}_{N \times N} & \mathbf{0}_{N \times N} &  \mathbf{I}_{N \times N} & \mathbf{0}_{N \times N} & \mathbf{0}_{N \times N} &  \mathbf{0}_{N \times N} & \boldsymbol\Phi_3 \\
\boldsymbol\Psi_1 &  \mathbf{0}_{N \times N} & \mathbf{0}_{N \times N} & \mathbf{I}_{N \times N} &  \mathbf{0}_{N \times N} & \mathbf{0}_{N \times N} & \boldsymbol\Psi_1 \boldsymbol\Phi_1 \\
\mathbf{0}_{N \times N} & \boldsymbol\Psi_2 & \mathbf{0}_{N \times N} & \mathbf{0}_{N \times N} & \mathbf{I}_{N \times N} & \mathbf{0}_{N \times N} & \boldsymbol\Psi_2 \boldsymbol\Phi_2 \\
\mathbf{0}_{N \times N} & \mathbf{0}_{N \times N} & \boldsymbol\Psi_3 & \mathbf{0}_{N \times N} & \mathbf{0}_{N \times N} & \mathbf{I}_{N \times N} & \boldsymbol\Psi_3 \boldsymbol\Phi_3 \\
\mathbf{D}_1 & \mathbf{D}_2 & \mathbf{D}_3 & \mathbf{E} & \mathbf{E} & \mathbf{E} & \mathbf{G}
\end{bmatrix},
\end{split}
\end{equation}
where $\mathbf{D}_i \equiv
\mathbf{C}(\mathbf{A}+\boldsymbol\Psi_i+\mathbf{B}\boldsymbol\Psi_i)$ ($i=1,2,3$), 
$\mathbf{E} \equiv  \mathbf{C}+\mathbf{C}\mathbf{B}$,
$\mathbf{G} \equiv \mathbf{C} 
\sum\limits_{i=1}^3 \mathbf{A} \boldsymbol\Phi_i 
+ (\mathbf{I}+\mathbf{B})\boldsymbol\Psi_i \boldsymbol\Phi_i$, 
$\mathbf{I}_{N \times N}$ is the  $N \times N$ identity matrix, 
and $\mathbf{0}_{N \times N}$ is the $N \times N$ zero matrix.
Recall that $\mathbf \Psi_i$ was defined below Eqn.\ (\ref{eq:Psi}).

The wavefield quantities can be propagated forward in time
from $t = 0$ to $t = (M-1) \Delta t$ as 
\begin{equation}
\label{eq:cascade}
\begin{bmatrix}
\mathbf{v}_0 \\
\mathbf{v}_1 \\
\vdots \\
\mathbf{v}_{M-1}
\end{bmatrix}
=
\mathbf{T}_{M-1} \cdots \mathbf{T}_1
\begin{bmatrix}
\mathbf{v}_0 \\
\mathbf{0}_{7N \times 1} \\
\vdots \\
\mathbf{0}_{7N \times 1}
\end{bmatrix},
\end{equation}
where the $7NM \times 7NM$ matrices 
$\mathbf{T}_m$ ($m=1,\cdots,M-1$) are defined in terms of $\mathbf W$ as
\begin{equation}
\label{eq:T}
\begin{split}
& \mathbf{T}_m \equiv \\
& \left[
\begin{array} {c c}
\begin{array} {c c c}
\mathbf{I}_{7N \times 7N} & \cdots & \mathbf{0}_{7N \times 7N} \\
\vdots & \ddots &\vdots \\
\mathbf{0}_{7N \times 7N} & \cdots & \mathbf{I}_{7N \times 7N} \\
\mathbf{0}_{7N \times 7N} & \cdots & \mathbf{W}
\end{array}
& \mathbf{0}_{(m+1) \cdot 7N \times (M-m) \cdot 7N} \\
\mathbf{0}_{(M-m-1) \cdot 7N \times m \cdot 7N}
& \mathbf{0}_{(M-m-1) \cdot 7N \times (M-m) \cdot 7N} \\
\end{array}
\right],
\end{split}
\end{equation}
with $\mathbf{W}$ residing between the 
$(7N(m-1)+1)$-th to $7Nm$-th rows and
the $(7Nm+1)$-th to $7N(m+1)$-th columns of $\mathbf{T}_m$.

From the equation of state in Eqn. \eqref{eq:lossless} and 
initial conditions Eqn.\ \eqref{eq:initialc}, 
the vector $(\mathbf{v}_0, \mathbf{0}, \cdots , \mathbf{0})^{\rm T}$
can be computed from the initial pressure distribution $\mathbf{p}_0$ as 
\begin{equation}
\label{eq:v0}
\begin{bmatrix}
\mathbf{v}_0 \\
\mathbf{0}_{7N \times 1} \\
\vdots \\
\mathbf{0}_{7N \times 1}
\end{bmatrix}
= \mathbf{T}_0 \mathbf{p}_0,
\end{equation}
where
\begin{equation}
\label{eq:t0}
\mathbf{T}_0 \equiv 
(\boldsymbol\tau, \mathbf{0}_{7N \times N},
\cdots, \mathbf{0}_{7N \times N})^{\rm T},
\end{equation} 
\begin{equation}
\label{eq:tau}
\boldsymbol\tau \equiv 
(\mathbf{0}_{N \times N},\mathbf{0}_{N \times N}, \mathbf{0}_{N \times N}, 
\frac{1}{3}\mathbf{C}^{-1}, \frac{1}{3}\mathbf{C}^{-1}, \frac{1}{3}\mathbf{C}^{-1}, 
\mathbf{I}_{N \times N})^{\rm T},
\end{equation}
and
$\mathbf p_0$ is the initial pressure distribution as defined
by Eqn.\ (\ref{eq:p}) with $m=0$.

In general, the transducer locations $\mathbf r_l^d$ at which
the PA data $\hat{\mathbf p}$ are recorded
 will not coincide with the vertices of the Cartesian grid at which the values
of the propagated field quantities are computed.
The measured PA data $\hat{\mathbf{p}}$ 
can  be related to the computed field quantities
via an interpolation operation as
\begin{equation}
\label{eq:p_hat}
\mathbf{\hat{p}}
=
\mathbf{S}
\begin{bmatrix}
\mathbf{v}_0 \\
\mathbf{v}_1 \\
\vdots \\
\mathbf{v}_{M-1}
\end{bmatrix},
\end{equation}
where
\begin{equation}
\label{eq:S}
\mathbf{S} \equiv
\begin{bmatrix}
\boldsymbol\Theta & \mathbf{0}_{L \times 7N} & \cdots & \mathbf{0}_{L \times 7N}\\
\mathbf{0}_{L \times 7N} & \boldsymbol\Theta & \cdots & \mathbf{0}_{L \times 7N}\\
\vdots & \vdots & \ddots & \vdots \\
\mathbf{0}_{L \times 7N} & \mathbf{0}_{L \times 7N} & \cdots & \boldsymbol\Theta
\end{bmatrix}.
\end{equation}
Here,
$\boldsymbol\Theta \equiv [\mathbf{s}_1, \cdots, \mathbf{s}_L]^{\rm T}$,
where $\mathbf{s}_l$ ($l=1,\cdots, L$) is a $1 \times 7N$
row vector in which all elements are zeros except the 4 
corresponding to acoustic pressure at 4 grid nodes
$\mathbf{r}_{l,1},\mathbf{r}_{l,2},\mathbf{r}_{l,3}, \mathbf{r}_{l,4}$
that are nearest to the transducer location $\mathbf{r}_l^d$.
In other words, these 4 entries are interpolation coefficients
to compute the acoustic pressure at the $l$-th transducer,
and their values are given by the barycentric coordinates
of $\mathbf{r}_l^d$ with respect to
$\mathbf{r}_{l,1},\mathbf{r}_{l,2},\mathbf{r}_{l,3}, \mathbf{r}_{l,4}$,
which are determined by Delaunay triangulation \cite{Lee1980IJPP}.

By use of Eqns.\ \eqref{eq:cascade}, \eqref{eq:v0}, and \eqref{eq:p_hat},
one obtains
\begin{equation}
\label{eq:H}
\hat{\mathbf{p}} = 
\mathbf{S} \mathbf{T}_{M-1} \cdots \mathbf{T}_1 \mathbf{T}_0 \mathbf{p}_0.
\end{equation}
Finally, upon comparison of this result to Eqn.\ (\ref{eqn:imagingmodel}), the sought-after
explicit form of the system matrix is identified as
\begin{equation}
\label{eqn:fr}
\mathbf H\equiv \mathbf{S} \mathbf{T}_{M-1} \cdots \mathbf{T}_1 \mathbf{T}_0.
\end{equation}

Commonly employed iterative image reconstruction methods involve
use of a backprojection matrix $\mathbf{H}^{\dagger}$ 
that corresponds to the adjoint of the system matrix. 
Since $\mathbf H$ contains real-valued elements in our case, 
$\mathbf{H}^{\dagger}$ is equivalent to the transpose $\mathbf{H}^{\rm T}$.
According to Eqn.\ (\ref{eqn:fr}), the explicit form of 
$\mathbf{H}^{\rm T}$ is given by
\begin{equation}
\label{eq:HT}
\mathbf{H}^{\rm T} =
\mathbf{T}_0^{\rm T}  \mathbf{T}_1^{\rm T} \cdots  \mathbf{T}_{M-1}^{\rm T} \mathbf{S}^{\rm T}.
\end{equation}
The implementations of $\mathbf H$ and $\mathbf{H}^{\rm T}$
are described in Section \ref{sec:implement}.
Note that, although the  descriptions  of
$\mathbf H$ and $\mathbf{H}^{\rm T}$ above
are based on the 3D PA wave equation,
the two-dimensional formulation is contained as a special case.

\section{Descriptions of numerical and experimental studies}
\label{sect:studies}

Numerical studies were
conducted to demonstrate the effectiveness
and robustness
of the proposed discrete imaging model 
in studies of iterative image reconstruction from
incomplete data sets in 2D and 3D PACT.
Specifically, the system matrix and its adjoint,
as formulated in Section \ref{sec:model}, were
employed with an iterative image reconstruction
algorithm that was designed to minimize a
PLS cost function that contained a total variation (TV)
penalty term.
The performance of the reconstruction algorithm was
compared to an existing TR-based reconstruction algorithm.

\subsection{Implementation of the forward and backprojection operators}
\label{sec:implement}
The k-space pseudospectral method for 
numerically solving the photoacoustic
wave equation has been implemented 
in the MATLAB k-Wave toolbox
\cite{treeby2010k}. This toolbox was employed 
to compute the action of $\mathbf{H}$.  
To prevent acoustic waves from leaving one side of
the grid and re-entering on the opposite side, 
an anisotropic absorbing boundary condition
called a perfectly matched layer (PML) was employed to
enclose the computational grids.
The performance of the PML was dependent on both the size and
attenuation of the layer. A PML thickness of 10 grid points,
together with a PML absorption coefficient of 2 nepers per meter,
were found to be sufficient to reduce boundary reflection and
transmission for normally incident waves 
\cite{TabeiJASA2002,KatsibasIEEE2004}
and were employed in this study. To accurately and stably model
wave propagation, the temporal and spatial steps
were related by the Courant-Friedrichs-Lewy (CFL) number as
\cite{MastIEEE2001,treeby2010k}
\begin{equation}
\label{eq:delta_t}
\Delta t \le \frac{\text{CFL} \Delta r_\text{min}}{c_\text{max}},
\end{equation}
where the $\Delta r_\text{min}$ is the minimum grid spacing,
and a CFL number of 0.3 typically provides a good compromise
between computation accuracy and speed \cite{TabeiJASA2002,treeby2010k}.
A more detailed description of the implementation of 
the k-space pseudospectral method 
can be found in Refs. \cite{TabeiJASA2002,treeby2010k}.

The action of the backprojection matrix on the 
measured pressure data $\hat{\mathbf{p}}$ was implemented 
according to Eqn. \eqref{eq:HT}. It can be verified that 
$\mathbf{p}^{\rm bp} = \mathbf{H}^{\rm T} \hat{\mathbf{p}}$ 
can be computed as 
\begin{gather}
\mathbf{v}^{M-1}  = 
\boldsymbol\Theta^{\rm T} \hat{\mathbf{p}}_{M-1} , \\
\mathbf{v}^{m-1}  = \boldsymbol\Theta^{\rm T} \hat{\mathbf{p}}_{m-1} +
\mathbf{W}^{\rm T} \mathbf{v}^m, \quad m=M-1,\cdots,1\\
\mathbf{p}^{\rm bp} = \boldsymbol\tau^{\rm T} \mathbf{v}^0.
\end{gather}

Since $\boldsymbol\Theta$ and $\boldsymbol\tau$ 
are both sparse matrices that can be stored and transposed,
$ \boldsymbol\Theta^{\rm T} \hat{\mathbf{p}}_m$
and $ \boldsymbol\tau^{\rm T} \mathbf{v}^1$
can be readily computed. Most of block matrices in 
the propagator matrix $\mathbf{W}$ are zero or identity matrices.
Therefore, to compute $\mathbf{W}^{\rm T} \mathbf{v}^m$, we only need to
compute the actions of transposed non-trivial block matrices in $\mathbf{W}$. 
To incorporate the PML boundary condition,
both $\mathbf{W}$ and $\mathbf{W}^{\rm T}$ 
should be modified as described in Ref. \cite{TabeiJASA2002}.

\subsection{Reconstruction algorithms}

\begin{figure*}[!t]
\centering
  \subfigure[]{\resizebox{2.3in}{!}{\includegraphics{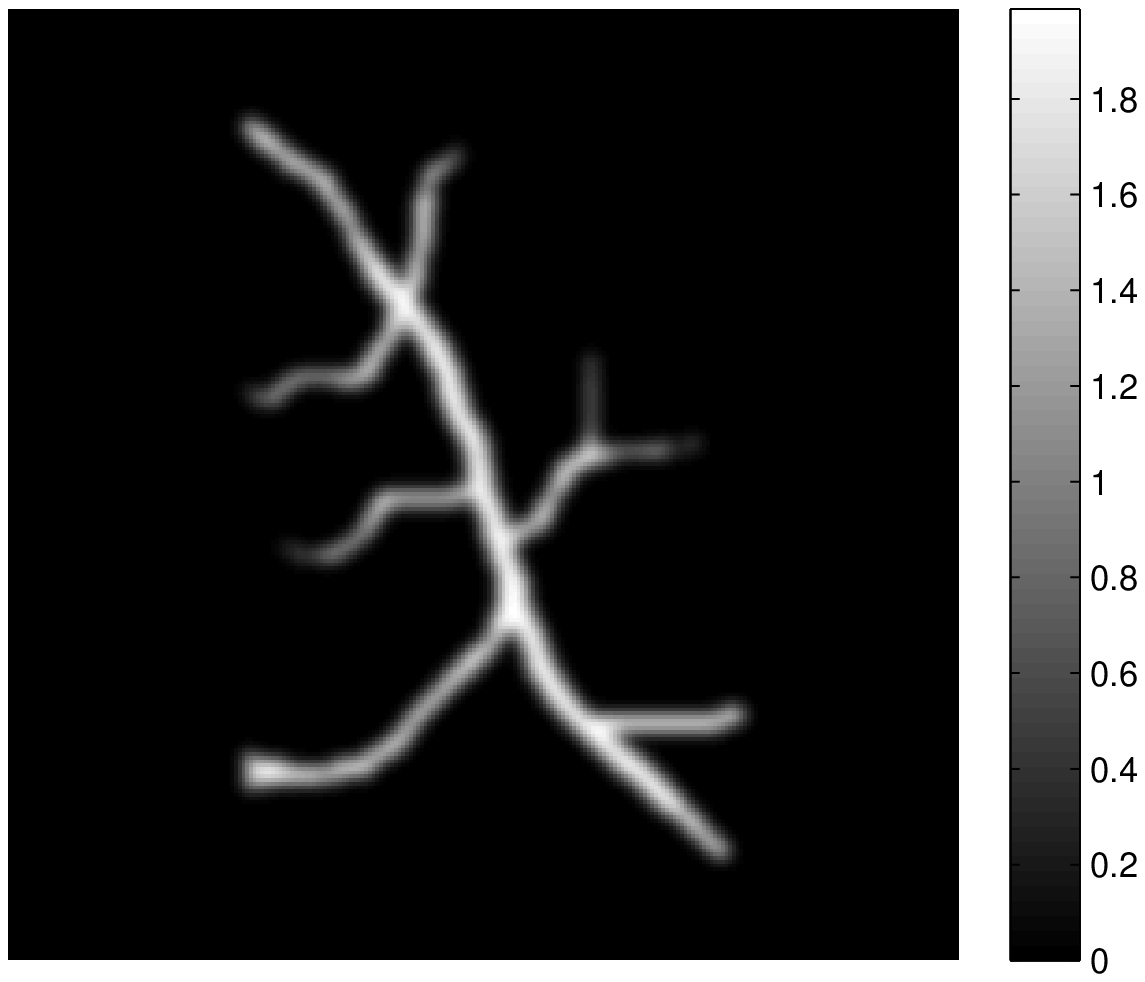}}}
  \subfigure[]{\resizebox{2.3in}{!}{\includegraphics{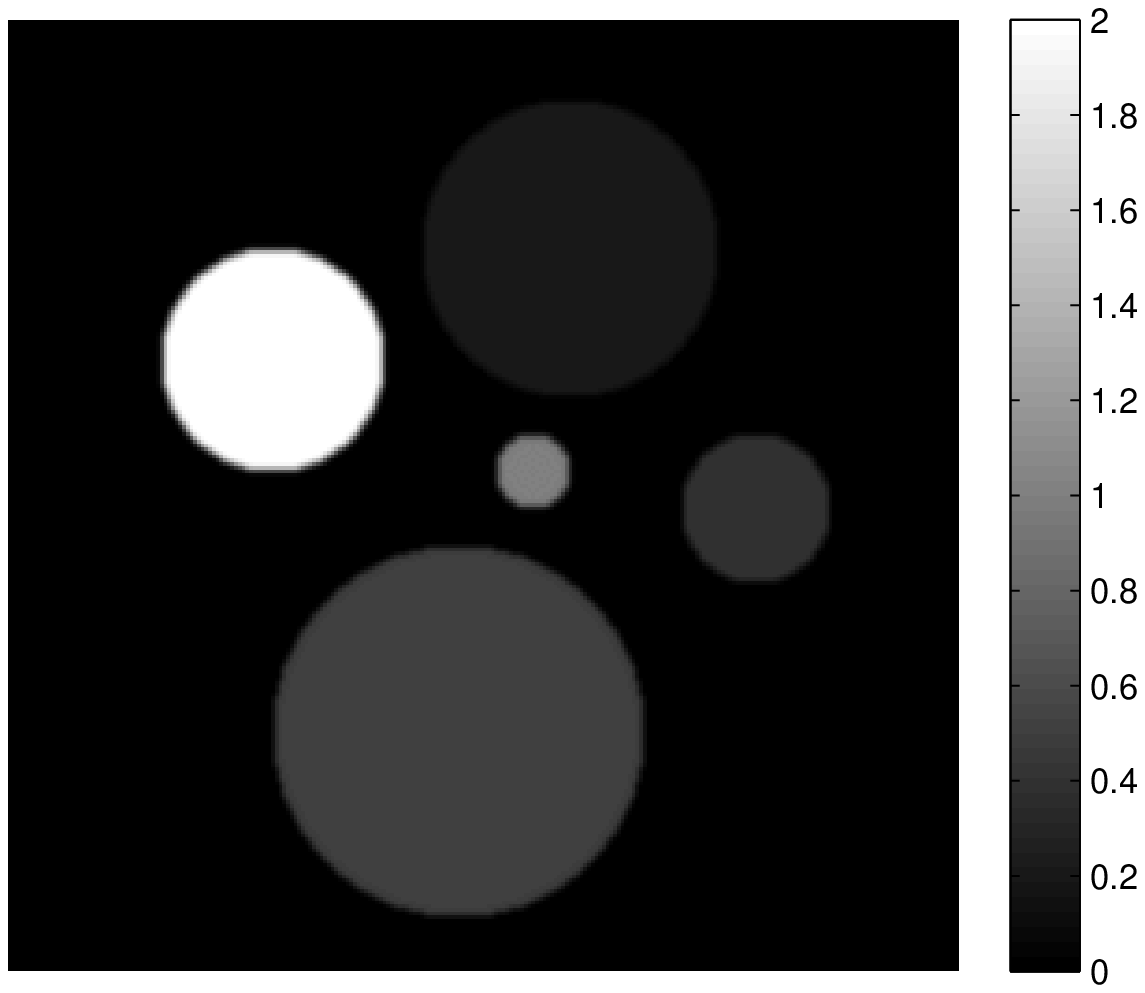}}}
  \subfigure[]{\resizebox{2.3in}{!}{\includegraphics{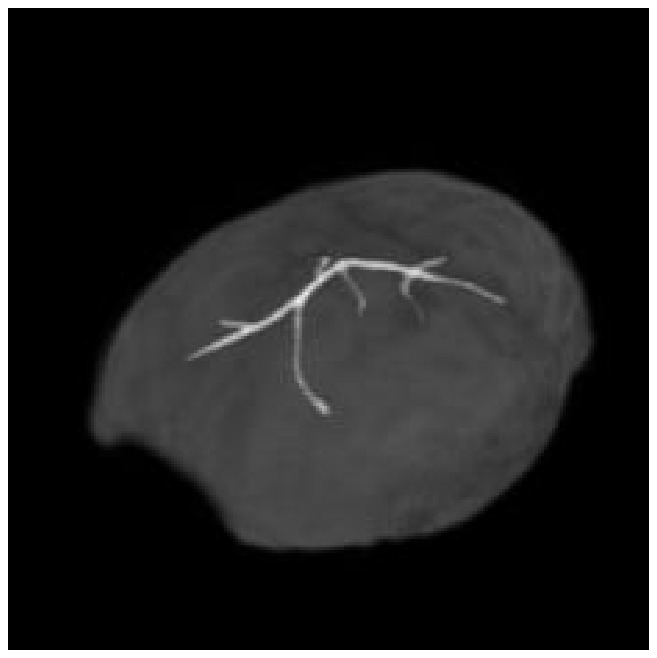}}}
\caption{\label{fig:phantoms}
The (a) blood vessel and (b) disc numerical phantoms employed
to represent $\mathbf p_0$
in the 2D computer-simulation studies.
Panel (c) is the overlapped image with 3D vessel phantom
and skull, which is only used to show the
relative position of the phantom to the skull.
}
\end{figure*}

By use of the proposed discrete imaging model
and methods for implementing $\mathbf H$ and $\mathbf H^{\rm T}$,
a wide variety of iterative image reconstruction algorithms
can be employed for determining estimates of $\mathbf p_0$.
In this work, we utilized an algorithm
that sought solutions of  the optimization problem
\begin{equation}
\label{eq:tv_solution}
\hat{\mathbf{p}}_0 =
\operatorname*{arg\,min}_{\mathbf{p}_0 \ge 0}
\| \mathbf{\hat{p}} - \mathbf{H} \mathbf{p}_0 \|^2
+ \lambda |\mathbf{p}_0|_{\text{TV}},
\end{equation}
where $\lambda$ is the regularization parameter, 
and a non-negativity constraint was employed.
For the 3D case, the TV-norm is defined as 
\begin{equation}
\label{eq:tv_norm}
\begin{split}
|\mathbf{p}_0|_{\text{TV}} = \sum_{n=1}^N
\big \{
& ([\mathbf{p}_0]_n - [\mathbf{p}_0]_{n_1^-})^2 + \\
& ([\mathbf{p}_0]_n - [\mathbf{p}_0]_{n_2^-})^2 +
([\mathbf{p}_0]_n - [\mathbf{p}_0]_{n_3^-})^2
\big \}^{\frac{1}{2}},
\end{split}
\end{equation}
where $[\mathbf{p}_0]_n$ denotes the $n$-th
grid node, and $[\mathbf{p}_0]_{n_1^-}$, 
$[\mathbf{p}_0]_{n_2^-}, [\mathbf{p}_0]_{n_3^-}$ 
are neighboring nodes before the $n$-th node
along the first, second and third dimension, respectively.
The fast iterative shrinkage/thresholding algorithm 
(FISTA) \cite{BeckIEEE2009, WangkunPMB2012} was employed 
to solve Eqn. \eqref{eq:tv_solution}, and 
its implementation is given in Appendix-B.
The regularization parameter $\lambda$ was 
empirically selected  to have a value of $0.001$ and was
fixed for all studies.


A TR image reconstruction algorithm based on the k-space
pseudospectral \cite{TreebyIP2010} method was also utilized in the studies
described below.
The TR reconstruction algorithm
solves the discretized acoustic
Eqns. \eqref{eq:d_motion} - \eqref{eq:d_loss}
backward in time subject to initial
and boundary conditions as described in reference \cite{TreebyIP2010}.
The parameters of the PML boundary condition
were the same with the ones employed in our system matrix construction.

For both algorithms, images were reconstructed on a uniform grid
of $512 \times 512$ pixels with a pitch of 0.2 mm for
the 2D simulation studies and on a $256\times 256\times 128$
grid with a pitch of $0.4$ mm for the 3D studies.
All simulations were computed in the MATLAB environment
on a workstation that contained dual hexa-core Intel(R) Xeon(R) 
E5645 CPUs and a NVIDIA Tesla C2075 GPU. The GPU was equiped with
448 1.15 GHz CUDA Cores and 5 GB global memory. 
The Jacket toolbox \cite{Jacket} was employed to perform the computation
of Eqns. \eqref{eq:d_motion} - \eqref{eq:d_loss}
and (40) - (42) on the GPU.

\subsection{Computer-simulation studies of 2D PACT}

%
%

\emph{Scanning geometries}:
Three different 2D scanning geometries 
were considered to investigate the robustness 
of the reconstruction methods
to different types and degrees of data incompleteness.
A `full-view' scanning geometry utilized 180 transducers that were
evenly distributed on a circle 
of radius 40 mm.
A `few-view'  scanning geometry utilized
60 transducers that were equally distributed on the
circle.
Finally, a `limited-view' scanning geometry
utilized 90 transducers that were evenly located on a semi-circle
of radius 40 mm.

\emph{Numerical phantoms}:
The two numerical phantoms shown in 
Fig. \ref{fig:phantoms}-(a) and (b) were chosen to 
represent the initial pressure distributions $\mathbf{p}_0$
in the 2D computer-simulation studies.
The blood vessel phantom shown in Fig. \ref{fig:phantoms}-(a)
was employed to investigate the robustness 
of the reconstruction methods with respect to  
different types and degrees of data incompleteness 
mentioned above. The low contrast disc phantom
displayed in Fig. \ref{fig:phantoms}-(b)
was employed to investigate the robustness of the
reconstruction methods with respect to
errors in the SOS and density maps introduced below.

\emph{Measurement data}:
Assuming ideal point-like transducer
and neglecting the transducer EIR and acoustic attenuation,
simulated pressure data corresponding 
to the numerical phantoms were computed
at the transducer locations by
use of the k-space pseudospectral method
for the 3 measurement geometries.
To avoid committing an `inverse crime'
\cite{KaipioJCAM2007},
a  $1024 \times 1024$ grid with a pitch of 0.1 mm
was employed in this computation.
A total of 20,000 temporal samples were computed
at each transducer location with time step
$\Delta t = 30 $ ns, all of which were
employed by the TR image reconstruction method.  However, only
the first 1,500 temporal samples were employed
by the iterative reconstruction method. The same procedure
was repeated for noisy pressure data,
where 3\% (with respect to maximum value of noiseless data)
additive white Gaussian noise (AWGN) was
added to the simulated pressure data.

\begin{figure}[!t]
\centering
  \subfigure[]{\resizebox{1.7in}{!}{\includegraphics{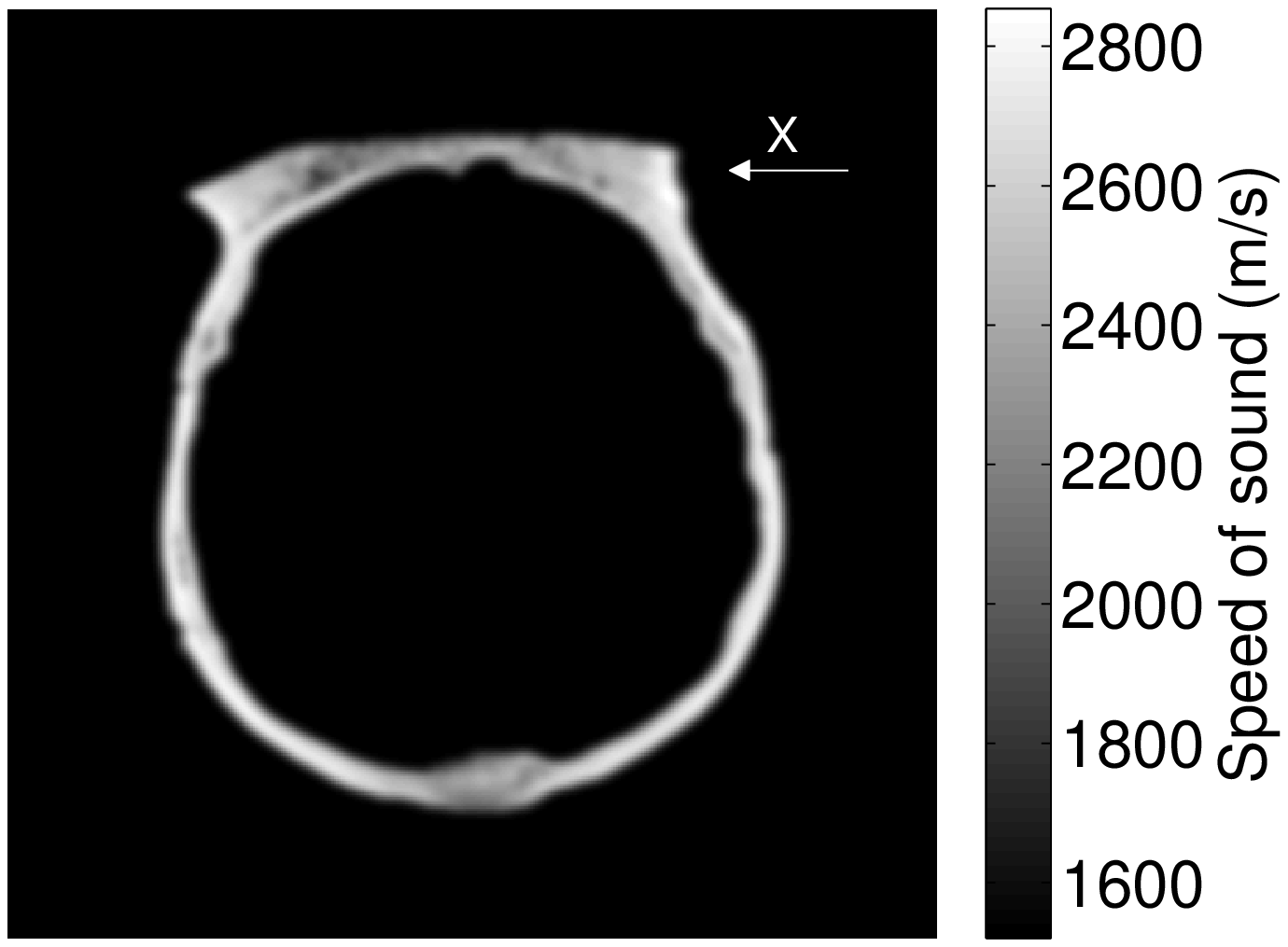}}}
  \subfigure[]{\resizebox{1.7in}{!}{\includegraphics{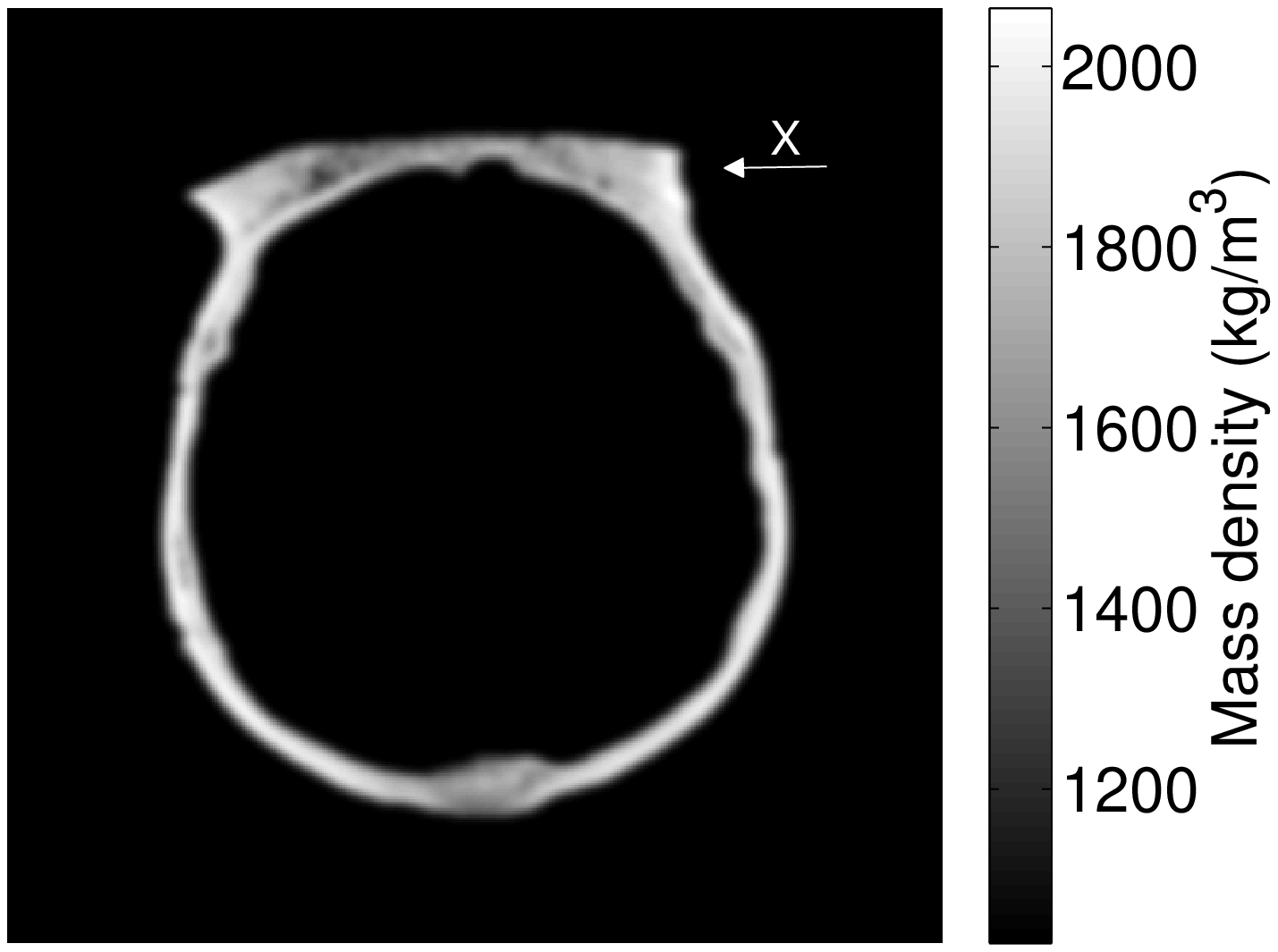}}}\\
  \subfigure[]{\resizebox{1.7in}{!}{\includegraphics{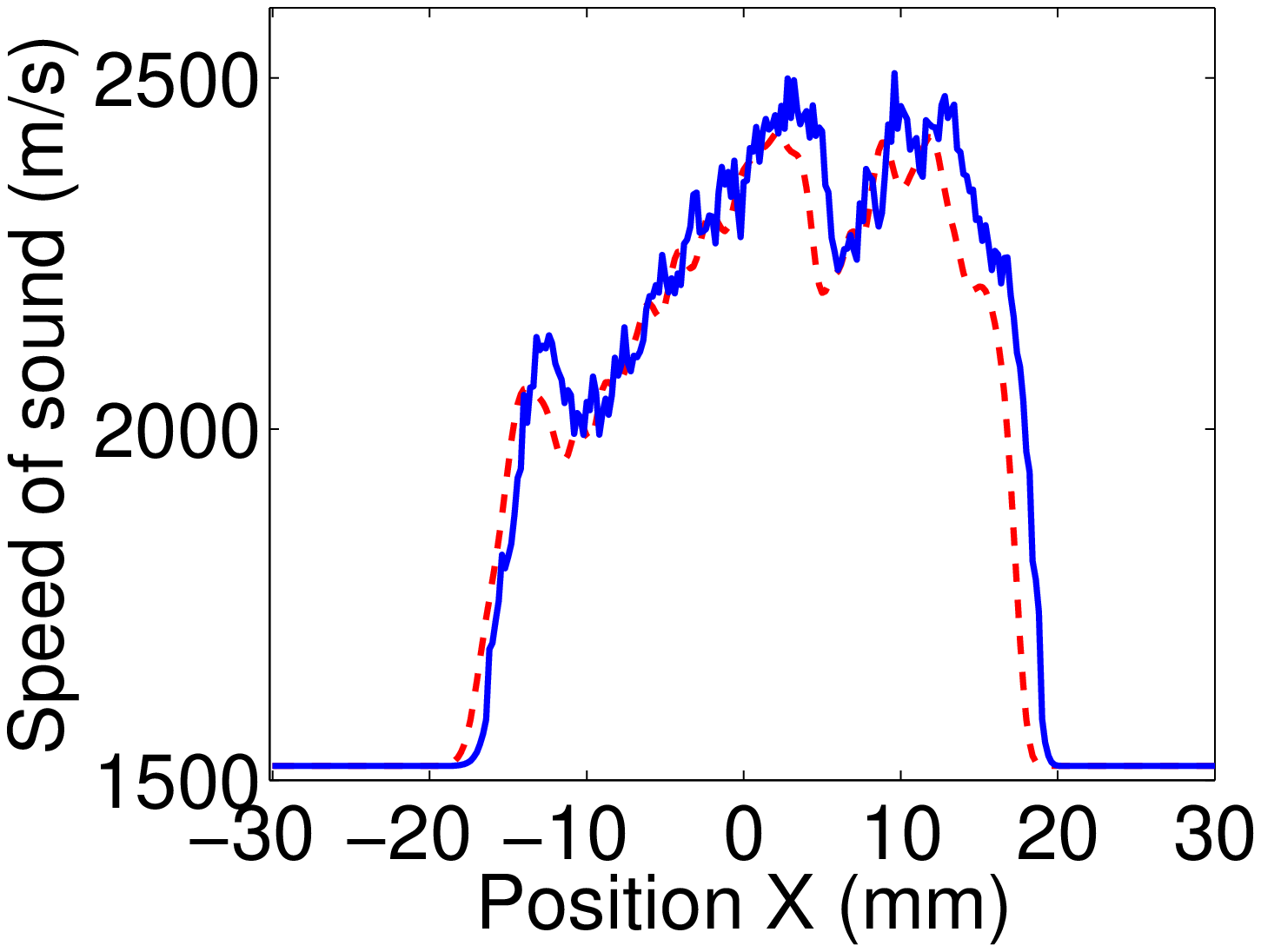}}}
  \subfigure[]{\resizebox{1.7in}{!}{\includegraphics{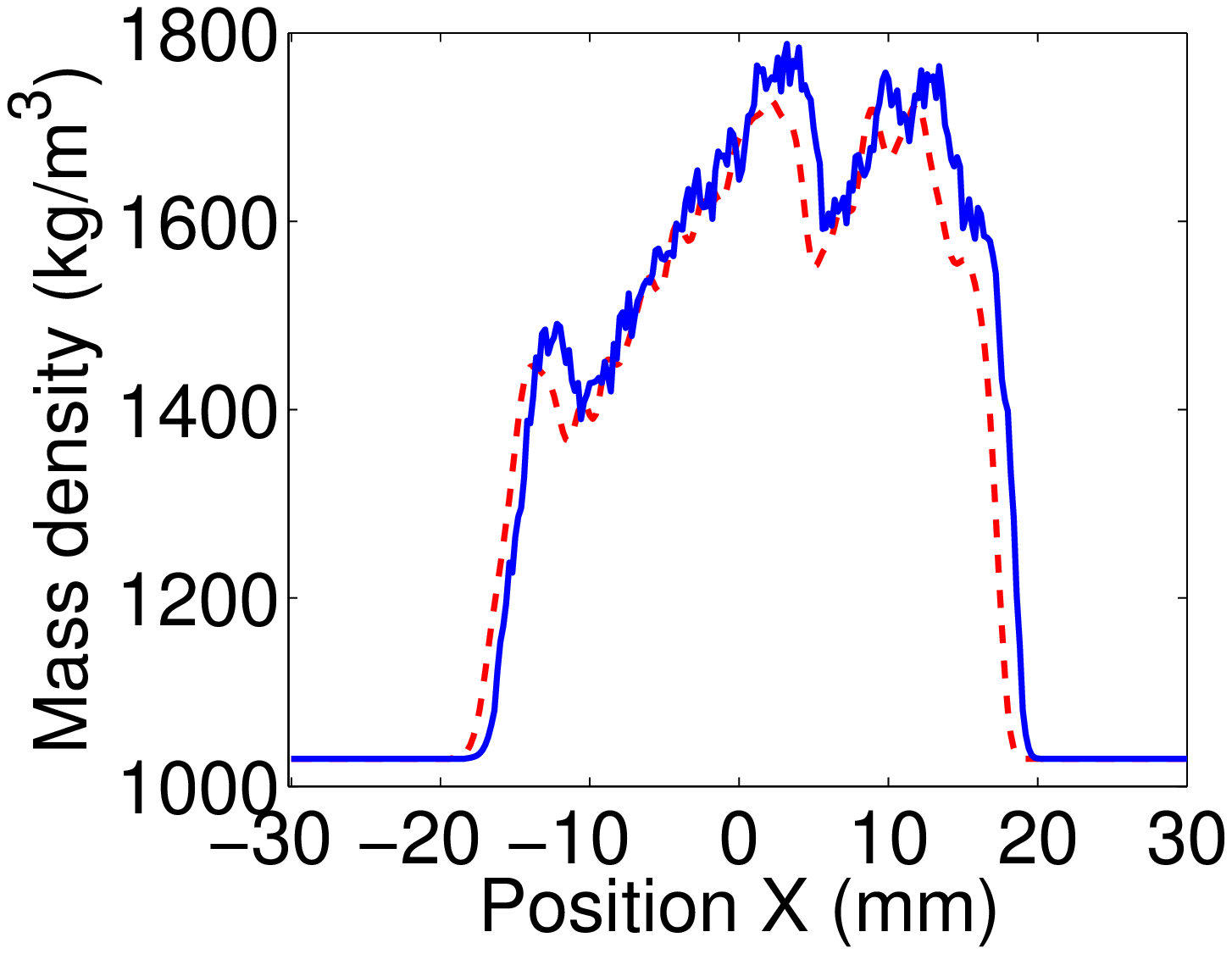}}}
\caption{\label{fig:maps}
A slice of the SOS (a) and density (b) map
deduced from the X-ray CT data of a monkey skull.
Panel (c) and (d) display profiles of the SOS
and density maps along the `X'-axis
indicated in Fig. \ref{fig:maps},
respectively. Red dashed lines are the profiles
of the assumed maps, whereas the blue solid lines
are the profiles of maps with errors.
}
\end{figure}

\emph{Investigation of systematic errors}:
The SOS and density maps employed in the simulation
studies were representative of a monkey skull
\cite{HuangchaoJBObrain}.
The dimensions of the skull were
approximately $7$ cm $\times$ $6$ cm, and its thickness
ranges from 2 to 4 mm. Figure \ref{fig:maps}-(a) and (b)
show a transverse slice of the SOS and density maps, 
which were used in the 2D simulations.

Since errors in the estimated SOS and density maps
are inevitable regardless in how they are determined, 
we investigated the robustness
of the reconstruction methods
with respect to the SOS and density map errors,
which were generated in two steps.
First, 1.3\% (with respect to maximum value)
uncorrelated Gaussian noise with mean value of
1.7\% of the maximum value was
added to the SOS and density maps to
simulate inaccuracy of the SOS and density values. 
Subsequently, the maps were shifted
by 7 pixels (1.4 mm) to simulate
a registration error.
Figure \ref{fig:maps}-(c) and (d) show profiles
of the SOS and density maps with those errors
along the `X'-axis indicated by the arrows
in Fig. \ref{fig:maps}-(a) and (b), respectively.

\subsection{Computer-simulation studies of 3D PACT}
Because PACT is inherently a 3D method,
we also conducted 3D simulation studies 
to evaluate and compare the iterative reconstruction method 
and the TR method. As in the 2D studies described above, 
the 3D SOS and density maps were representative of
a monkey skull. A 3D blood vessel phantom 
was positioned underneath the skull to mimic 
the blood vessels on the cortex surface.
To demonstrate this configuration,
Figure \ref{fig:phantoms}-(c) shows 
the overlapped images of the 3D  phantom and the skull.
The assumed scanning geometry was a hemispherical cap
with radius of 46 mm, and 484 transducers
were evenly distributed on the hemispherical
cap by use of the golden section spiral method \cite{CGAFaq}.
The pressure data were computed on a 
$512 \times 512 \times 256$ grid with a pitch of 0.2 mm
and a time step $\Delta t=30$ ns. The simulated pressure data
were then contaminated with 3\% AWGN.
The TR reconstruction method employed 2,000 temporal samples
at each transducer location, whereas the iterative method
employed 1,000 samples.

\subsection{Studies utilizing experimental data}

\begin{figure}[!t]
  \centering
    \resizebox{2.3in}{!}{
      \includegraphics{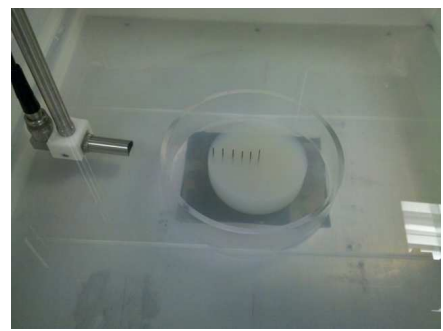}
  }
  \caption{\label{fig:setup}
A photograph of the pencil leads held in agar and
surrounded by an acrylic cylindrical shell.
}
\end{figure}

%
Since the acoustic absorption and dispersion were
modeled by the system matrix, the iterative method
can naturally compensate for absorption
and dispersion effects during reconstruction.
To demonstrate the compensation for those effects, 
images were reconstructed by use of the iterative method
with experimental data obtained from
a well-characterized phantom object
that is displayed in Fig. \ref{fig:setup}.
The phantom  contained 6 optically
absorbing structures (pencil leads with diameter 1 mm) 
embedded in agar. These structures were surrounded 
by an acrylic cylinder, which
represents the acoustic heterogeneities 
and absorption in the experiments.
The cylinder had inner and outer radii of
7.1 and 7.6 cm, respectively, and a height of 3 cm.
The density and SOS of the acrylic were
measured and found to be 1200 kg m$^{-3}$ and
3100 m s$^{-1}$, and the estimated
acoustic absorption parameters were found to be
$\alpha_0=1.3$ dB MHz$^{-y}$ cm$^{-1}$ and $y=0.9$ \cite{HuangchaoJBOatten}.
These values were assigned to the 
the annular region occupied by the acrylic
in the 2D SOS maps $c_0(\mathbf r)$,
density map $\rho_0(\mathbf r)$ and 
attenuation coefficient $\alpha_0(\mathbf r)$,
respectively. The SOS value 1480 m s$^{-1}$ and 
density value 1000 kg m$^{-3}$ of water
were assigned elsewhere. Since we neglected 
the relatively weak acoustic attenuation due to
the water bath and agar, $\alpha_0(\mathbf r)$
was also set to zero elsewhere.

The experimental data were acquired from
a cylindrically focused ultrasound transducer
that had a central frequency of 2.25 MHz with a bandwidth of 70\%  \cite{NielimingJBO2011}.
The transducer was scanned along a circular trajectory
of radius 95 mm, and 20,000 temporal samples
were measured at each transducer location 
at a sampling rate of 20 MHz. More details 
about the data acquisition can be found 
in Ref. \cite{HuangchaoJBOatten}.
In this study, images were reconstructed 
by use of PA signals recorded at 200, 100 (over 180 degrees), 
and 50 transducer locations, which correspond to 
the full-view, limited-view, and few-view scanning geometry,
respectively. The TR reconstruction method employed 
20,000 temporal samples at each transducer location, 
while the iterative method employed 2,000 samples.
The reference images were also reconstructed
by use of the data obtained at 200 transducer locations 
when the acrylic cylinder was absent. 
{Since the pencil lead phantom is expected to generate quasi-cylindrical waves
and the morphology of the acoustic heterogeneity 
(the acrylic shell) was a cylinder,
the cylindrical wave propagation can be approximated by 
the 2D PA wave equation. Accordingly, we employed
a 2D imaging model in the experimental study, }and 
all the reconstructions were performed on a grid of 
$512 \times 512$ pixels with a pitch of 0.5 mm.
{The effects of shear wave propagation in the
acrylic cylinder were neglected,
which we expected to be of 
 second-order importance compared to wavefield perturbations
that arise  from inhomogeneties in the
SOS and density distributions \cite{BobJBO2012}. 
}

\section{Simulation and experimental results}
\label{sect:results}
\subsection{Computer-simulations corresponding to different scanning geometries}

\begin{figure}[!t]
\centering
  \subfigure[]{\resizebox{1.7in}{!}{\includegraphics{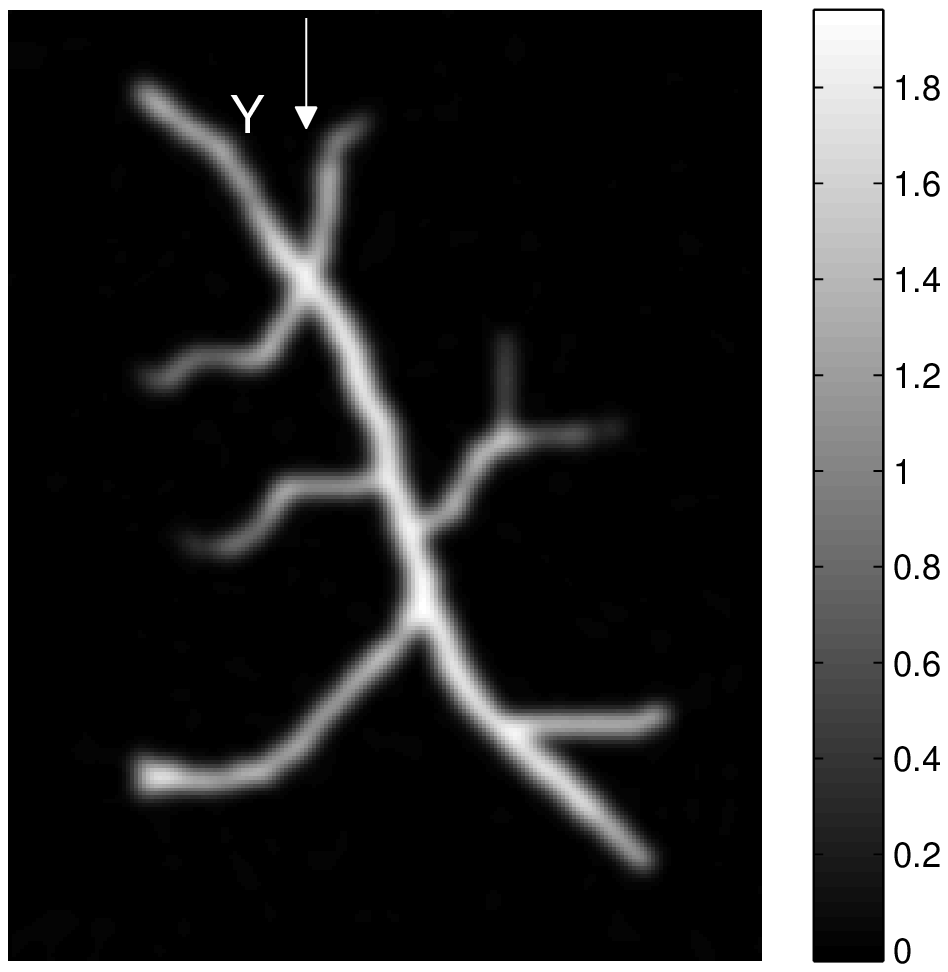}}}
  \subfigure[]{\resizebox{1.7in}{!}{\includegraphics{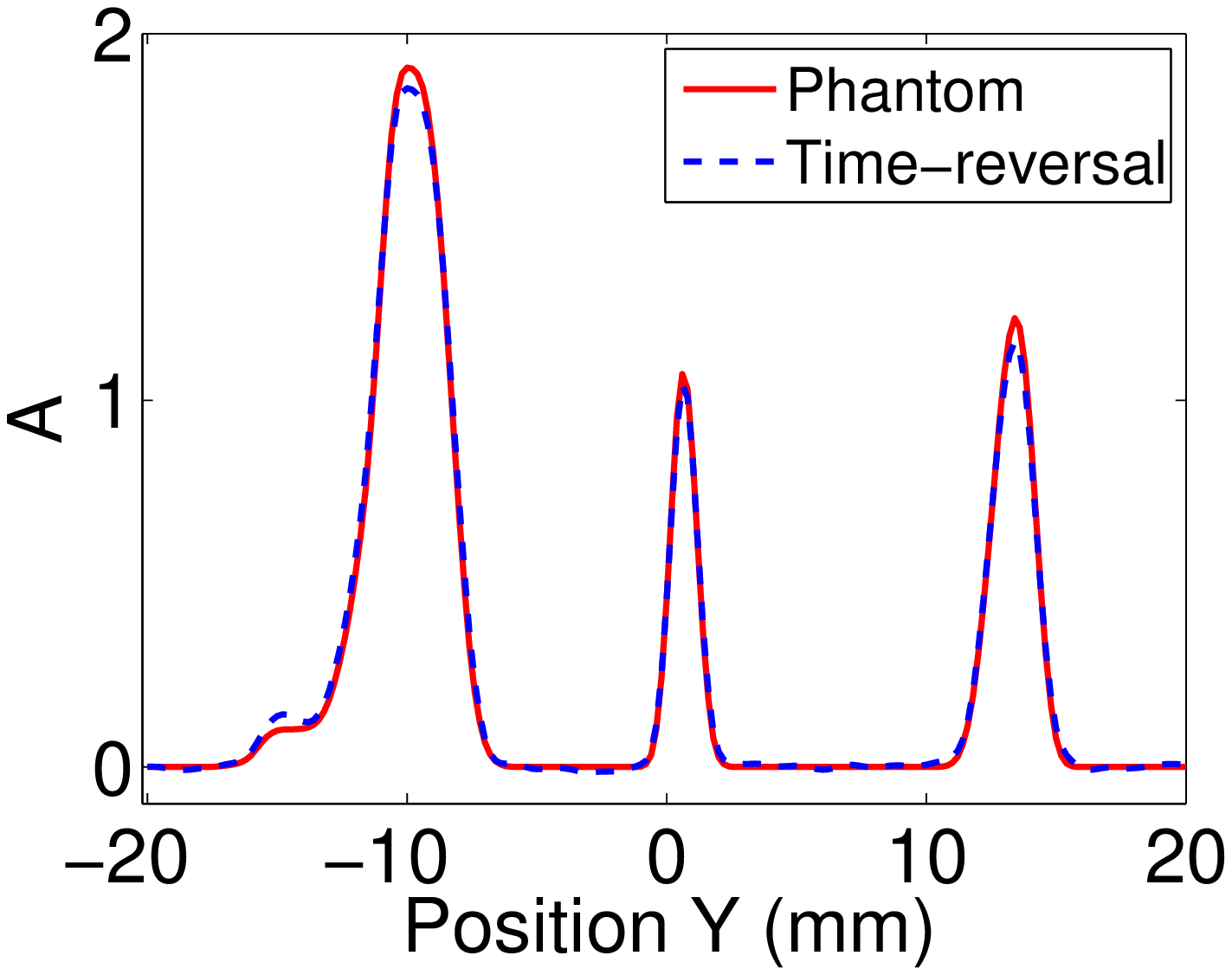}}}\\
  \subfigure[]{\resizebox{1.7in}{!}{\includegraphics{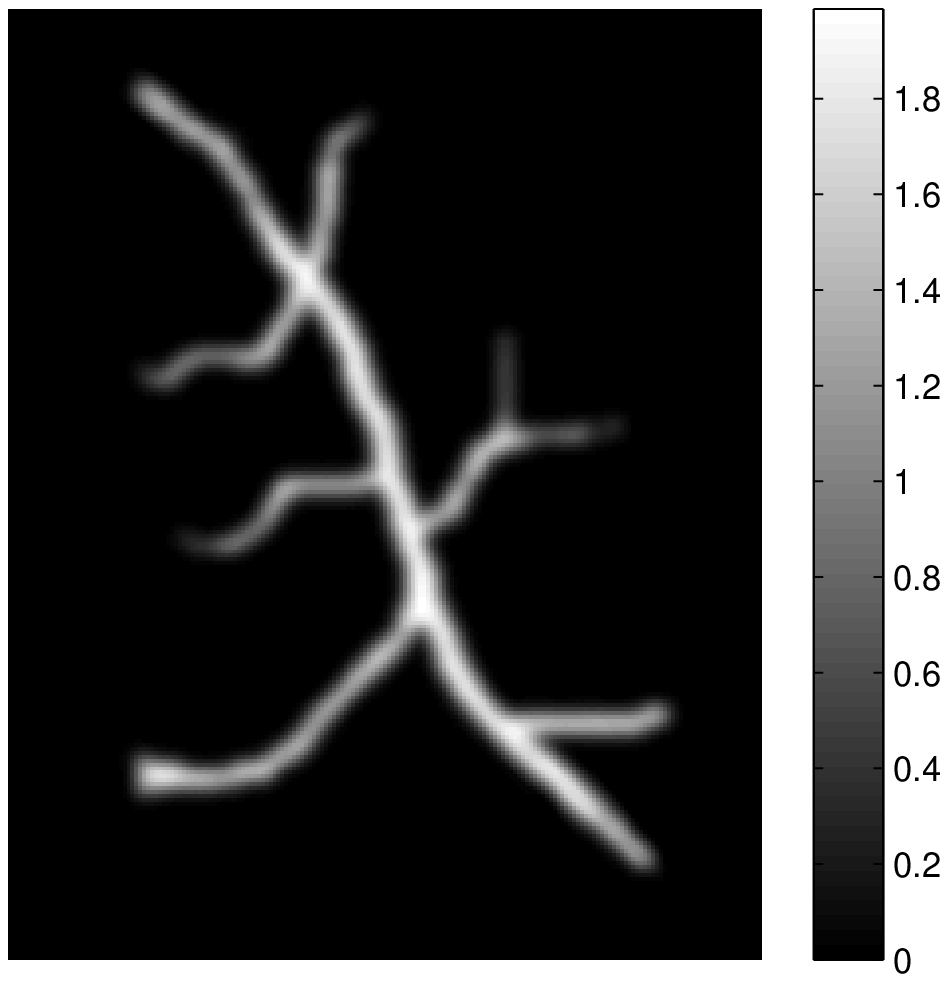}}}
  \subfigure[]{\resizebox{1.7in}{!}{\includegraphics{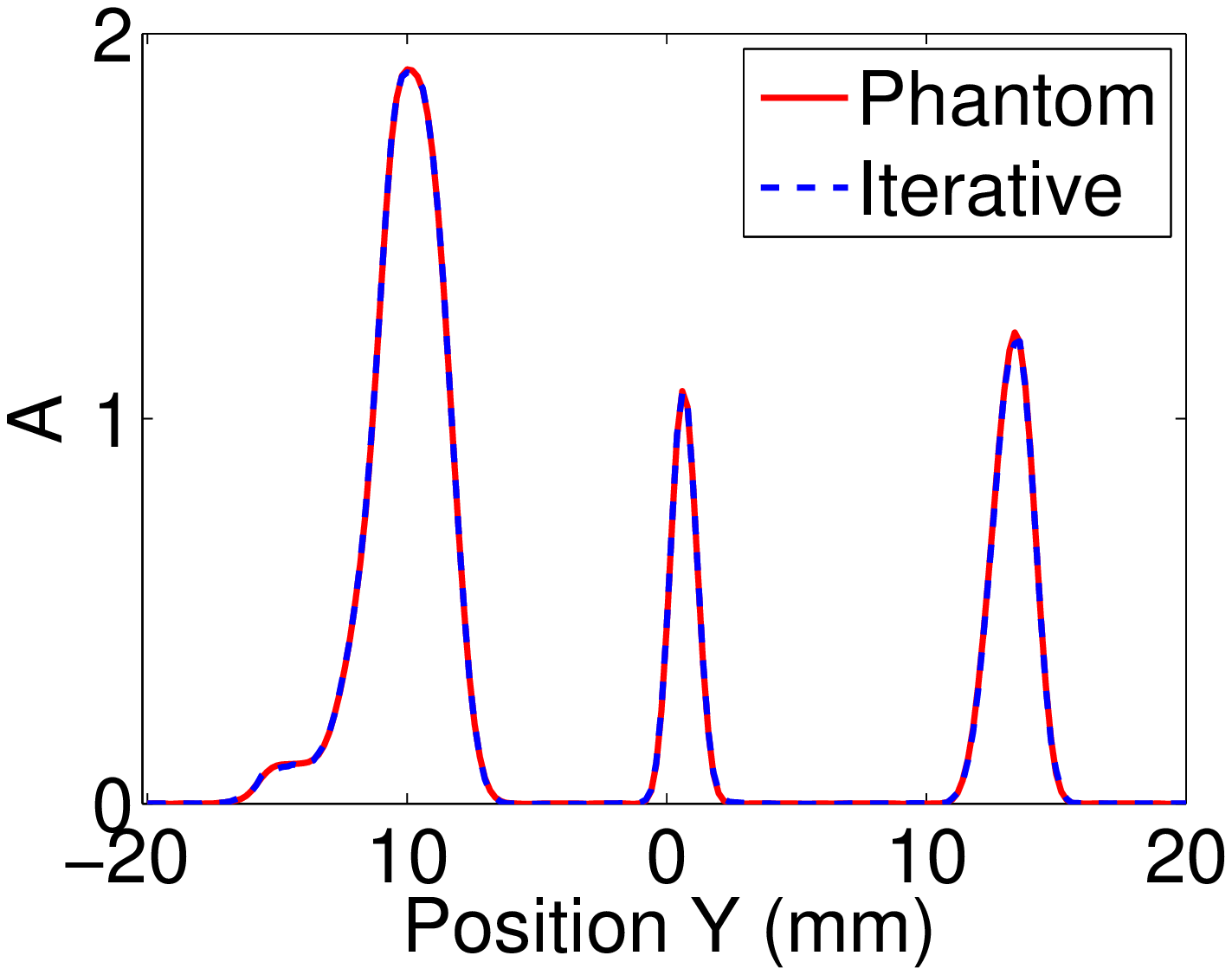}}}
\caption{\label{fig:vessel_full}
(a) and (c) are reconstructed images from noiseless
data with full-view scanning geometry by use of
the TR method and iterative method,
respectively. (b) and (d) are the corresponding profiles
along the `Y'-axis indicated in panel (a).
}
\end{figure}
\begin{figure}[!t]
\centering
  \subfigure[]{\resizebox{1.7in}{!}{\includegraphics{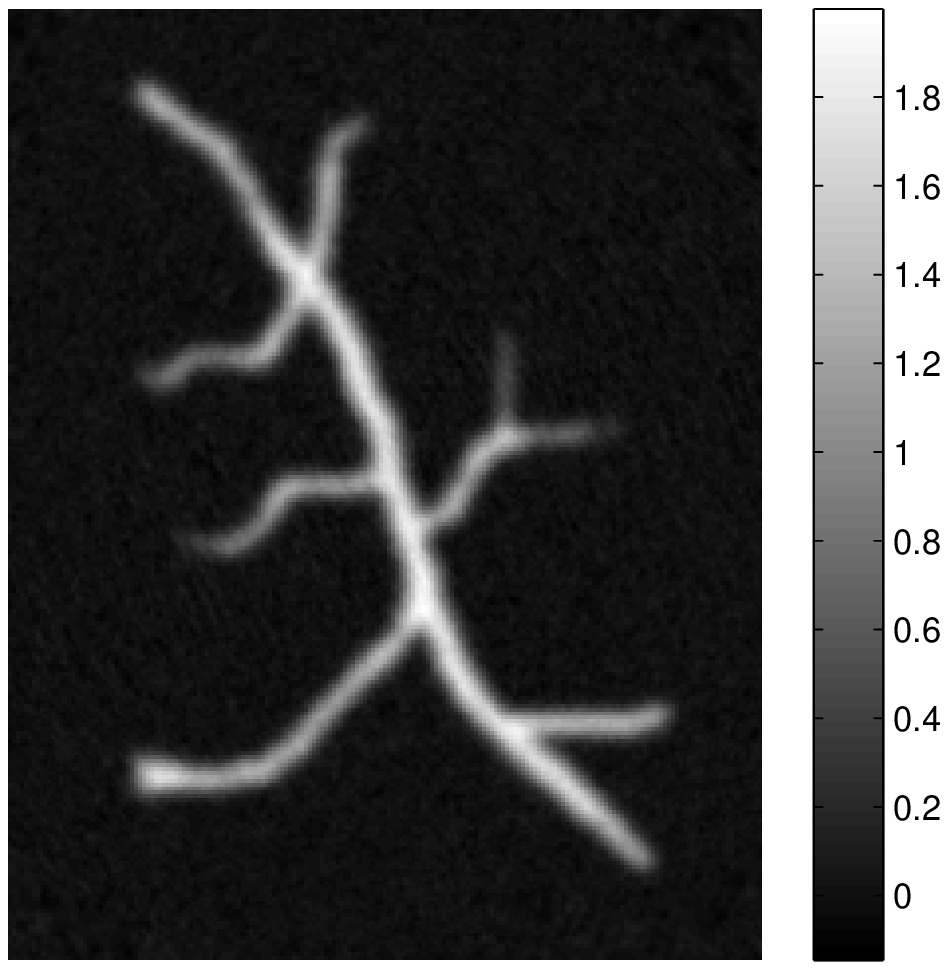}}}
  \subfigure[]{\resizebox{1.7in}{!}{\includegraphics{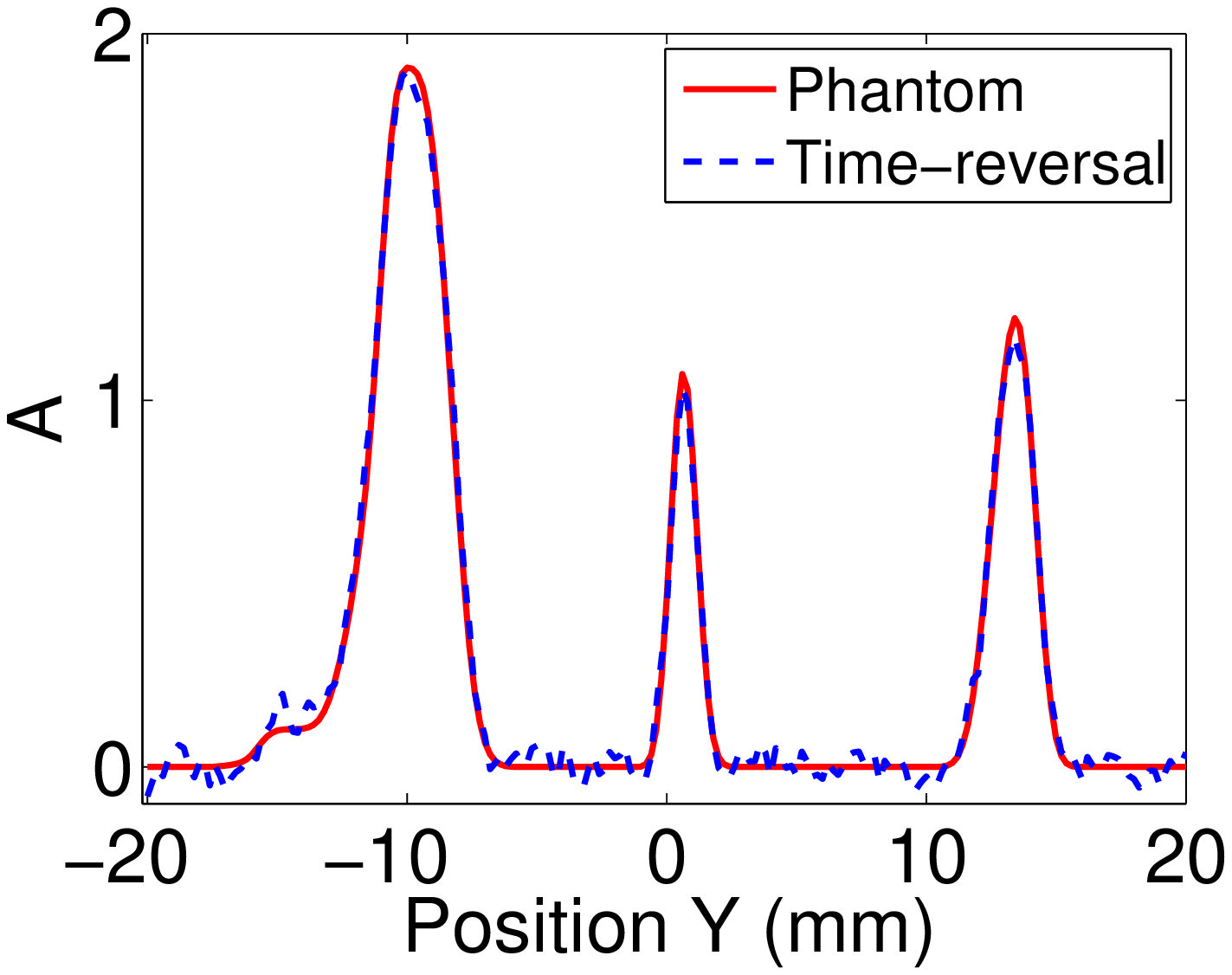}}}\\
  \subfigure[]{\resizebox{1.7in}{!}{\includegraphics{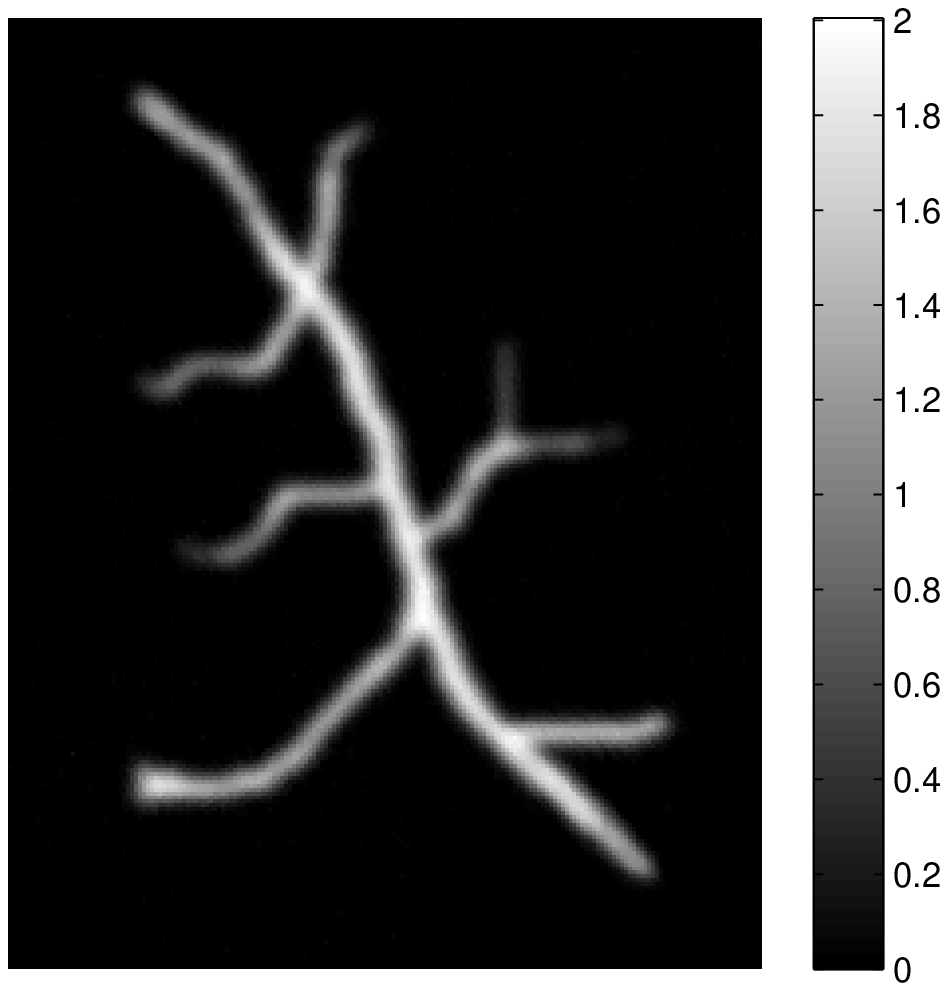}}}
  \subfigure[]{\resizebox{1.7in}{!}{\includegraphics{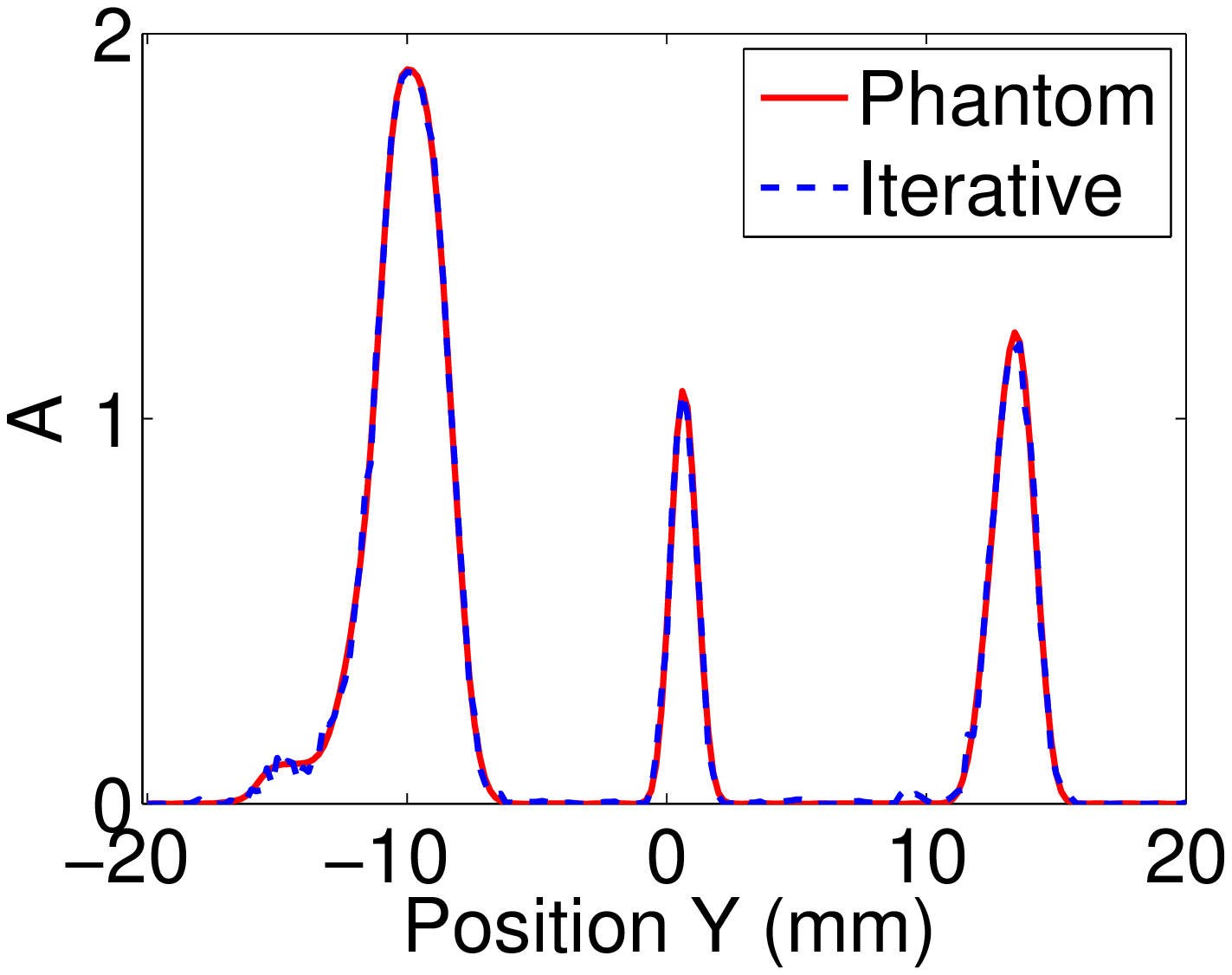}}}
\caption{\label{fig:vessel_fulln}
(a) and (c) are reconstructed images from the noisy pressure data
with 3\% AWGN corresponding to the full-view scanning geometry by use of
the TR method and iterative method,
respectively. (b) and (d) are the corresponding profiles.
}
\end{figure}
\begin{figure}[!t]
\centering
  \subfigure[]{\resizebox{1.7in}{!}{\includegraphics{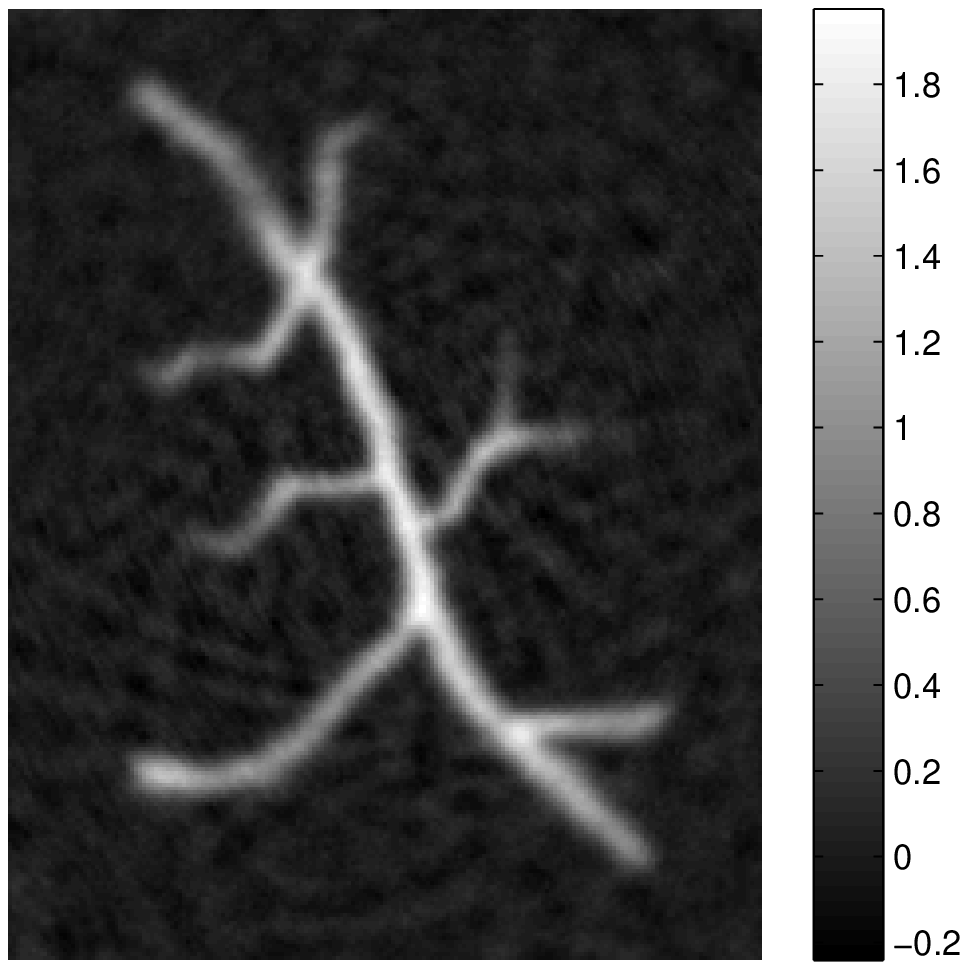}}}
  \subfigure[]{\resizebox{1.7in}{!}{\includegraphics{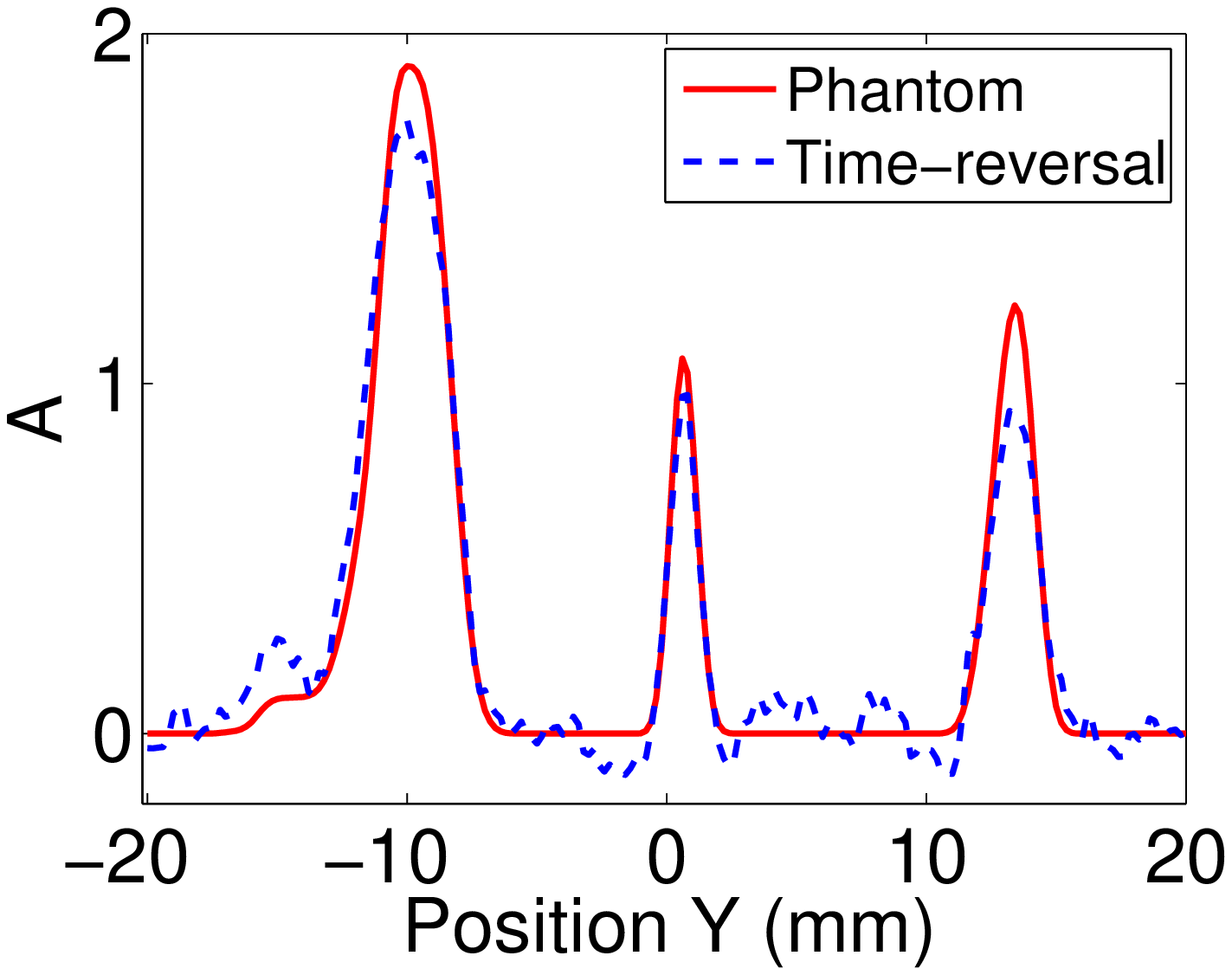}}}\\
  \subfigure[]{\resizebox{1.7in}{!}{\includegraphics{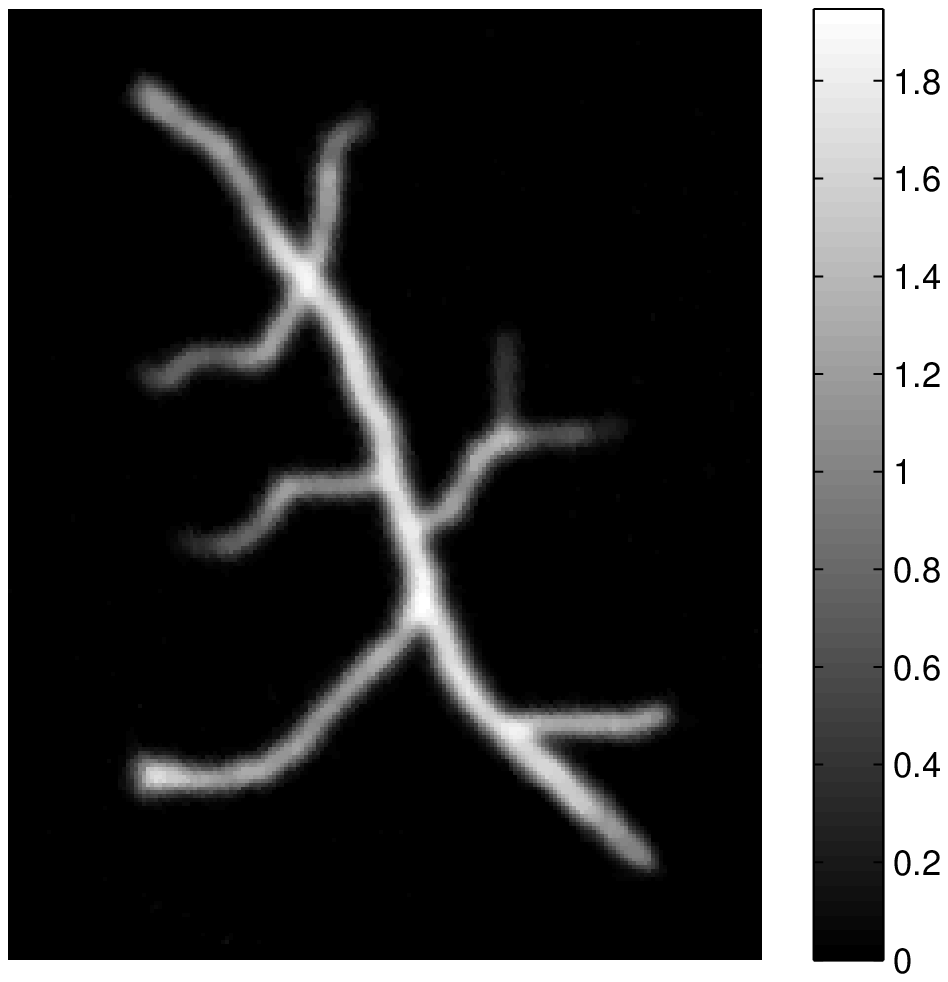}}}
  \subfigure[]{\resizebox{1.7in}{!}{\includegraphics{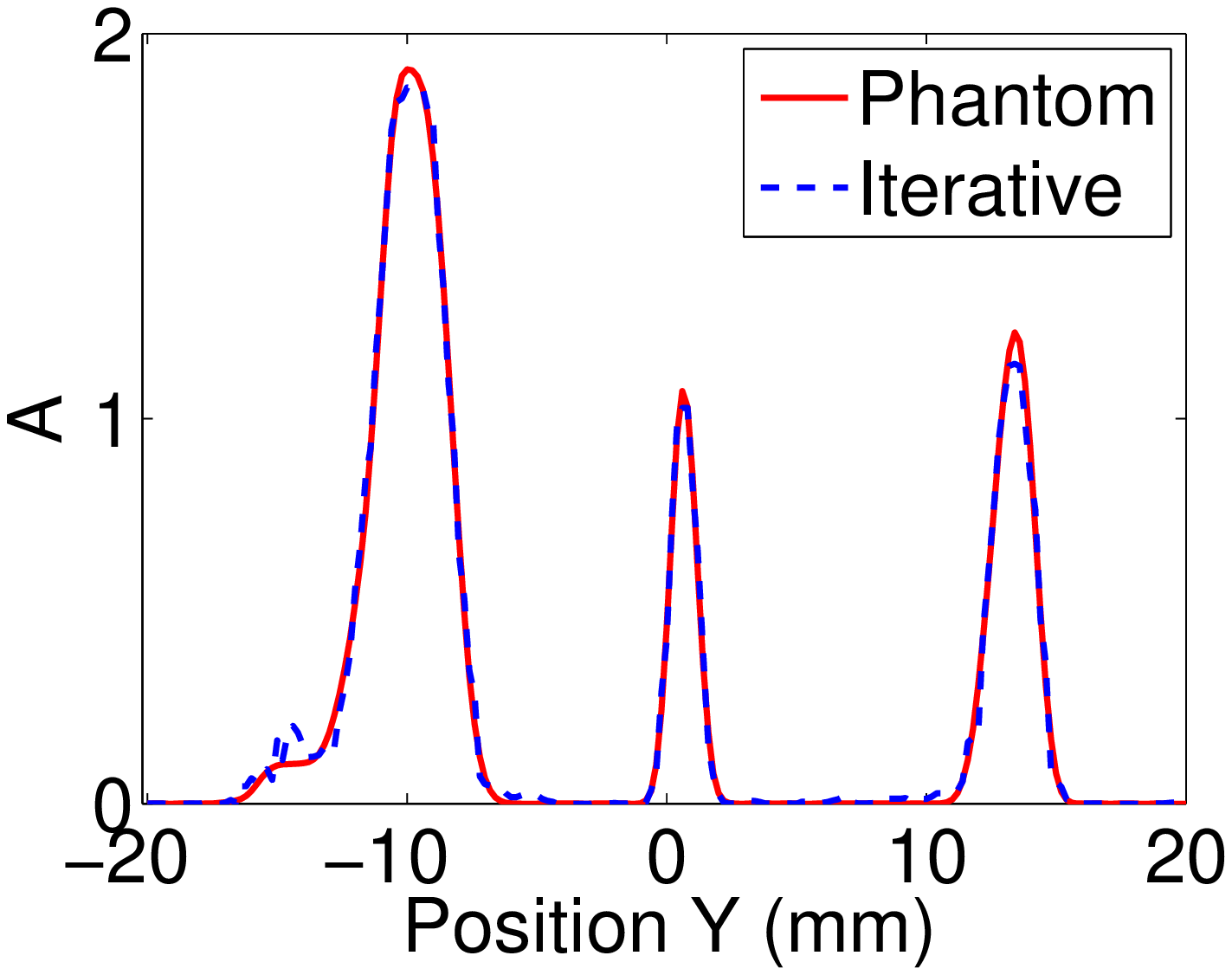}}}
\caption{\label{fig:vessel_fewn}
(a) and (c) are reconstructed images from the noisy pressure data
with 3\% AWGN corresponding to the few-view scanning geometry by use of
the TR method and iterative method,
respectively. (b) and (d) are the corresponding profiles.
}
\end{figure}
\begin{figure}[!t]
\centering
  \subfigure[]{\resizebox{1.7in}{!}{\includegraphics{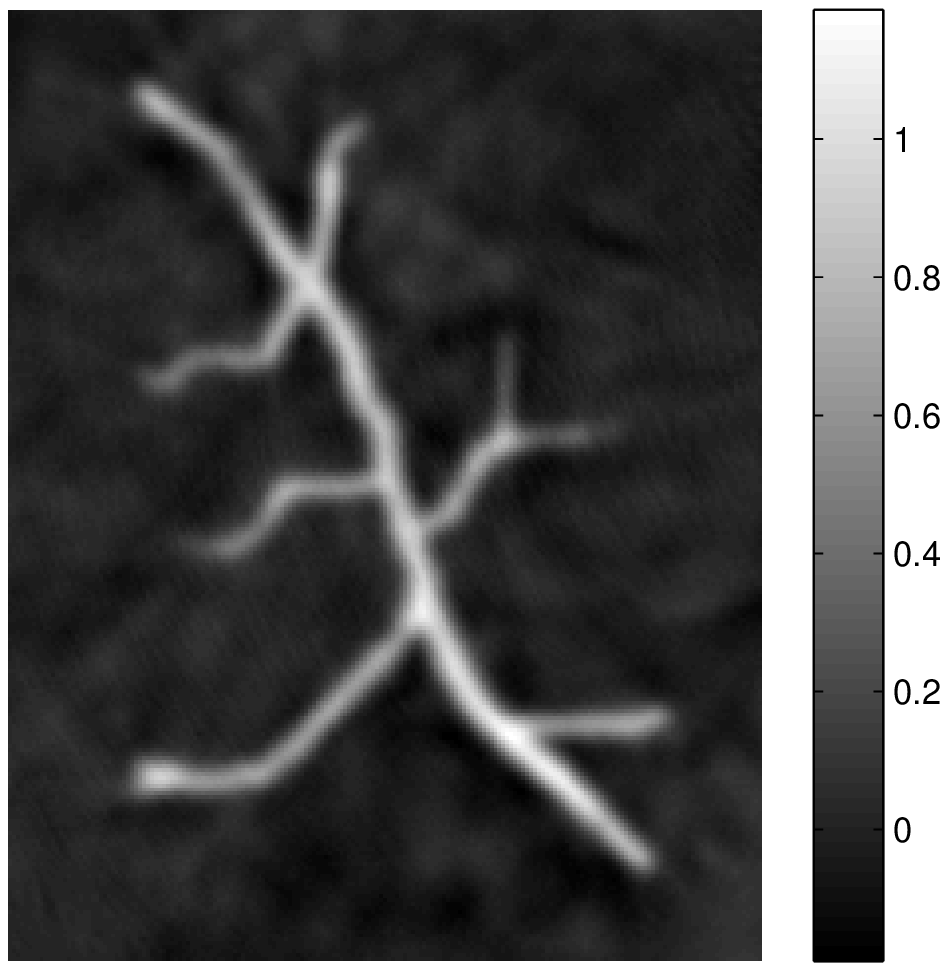}}}
  \subfigure[]{\resizebox{1.7in}{!}{\includegraphics{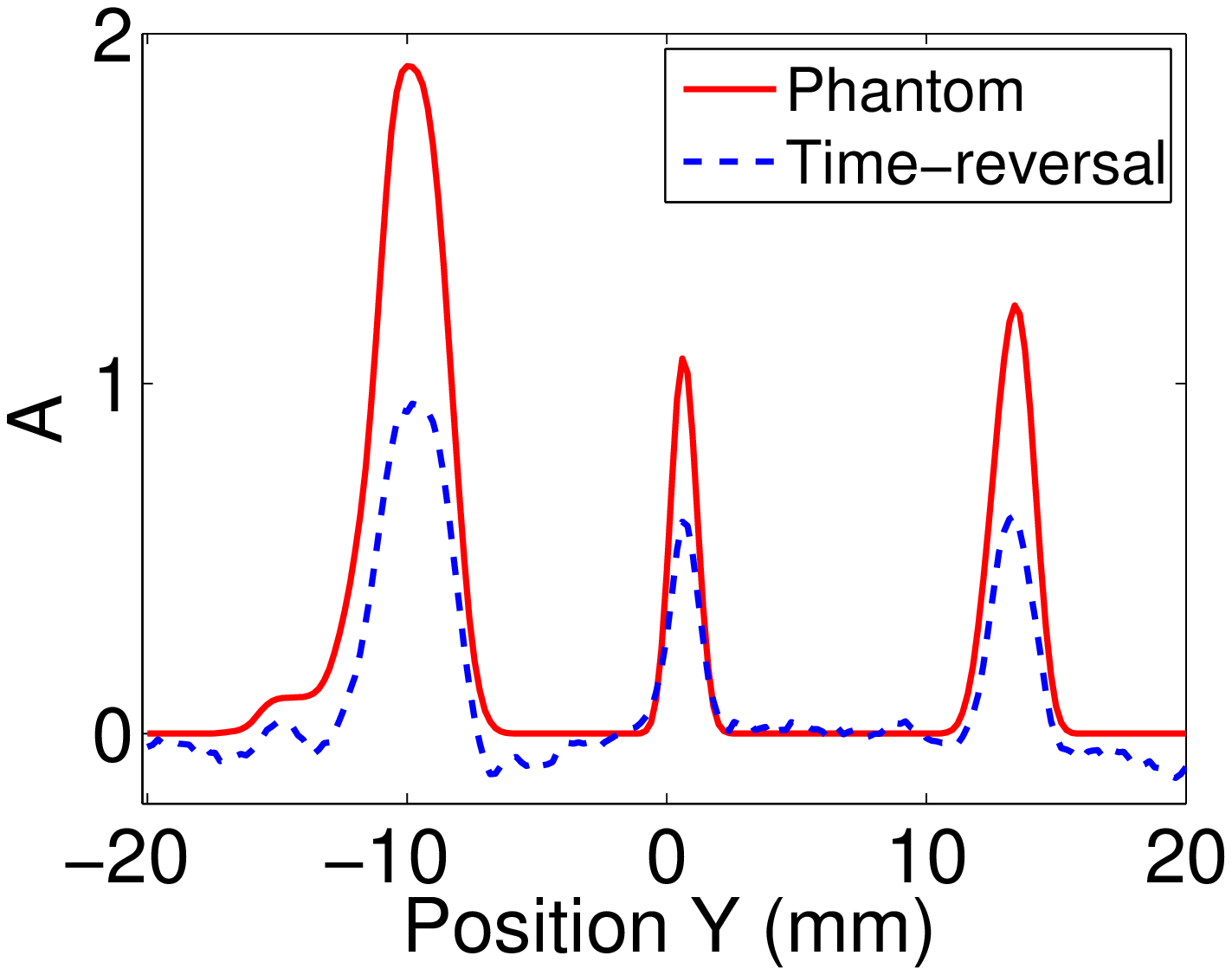}}}\\
  \subfigure[]{\resizebox{1.7in}{!}{\includegraphics{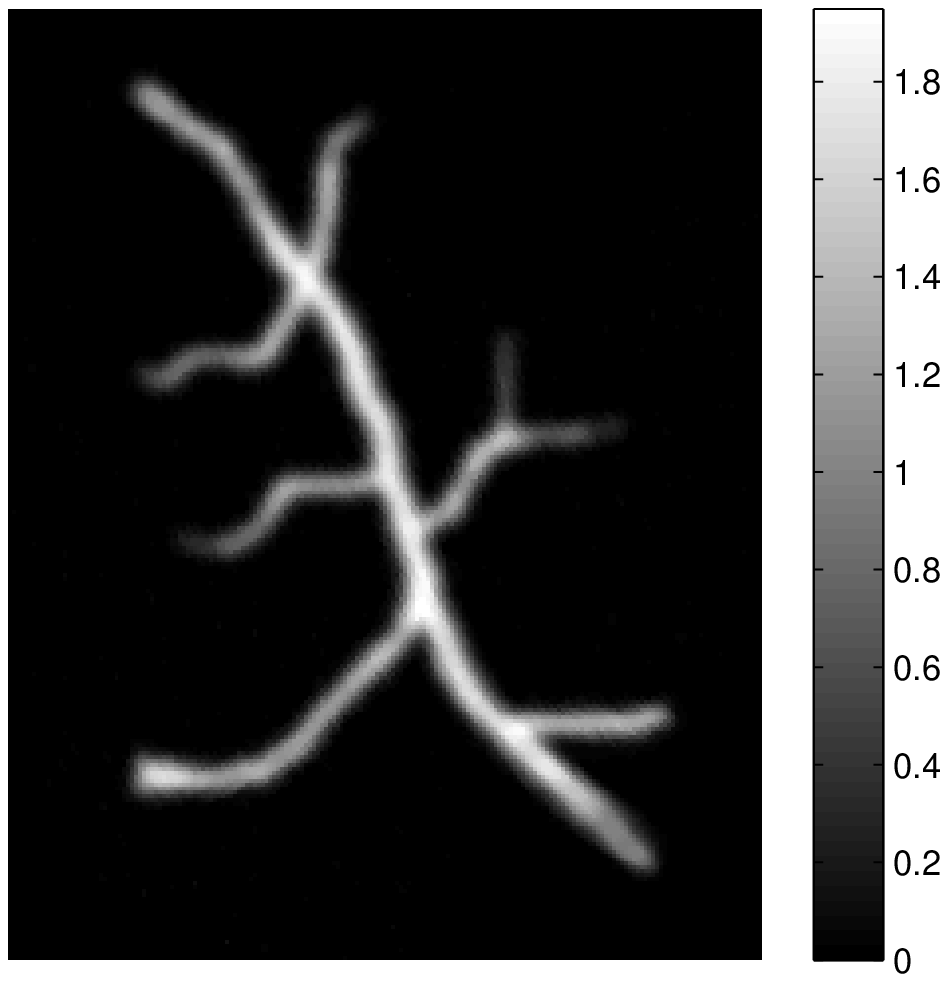}}}
  \subfigure[]{\resizebox{1.7in}{!}{\includegraphics{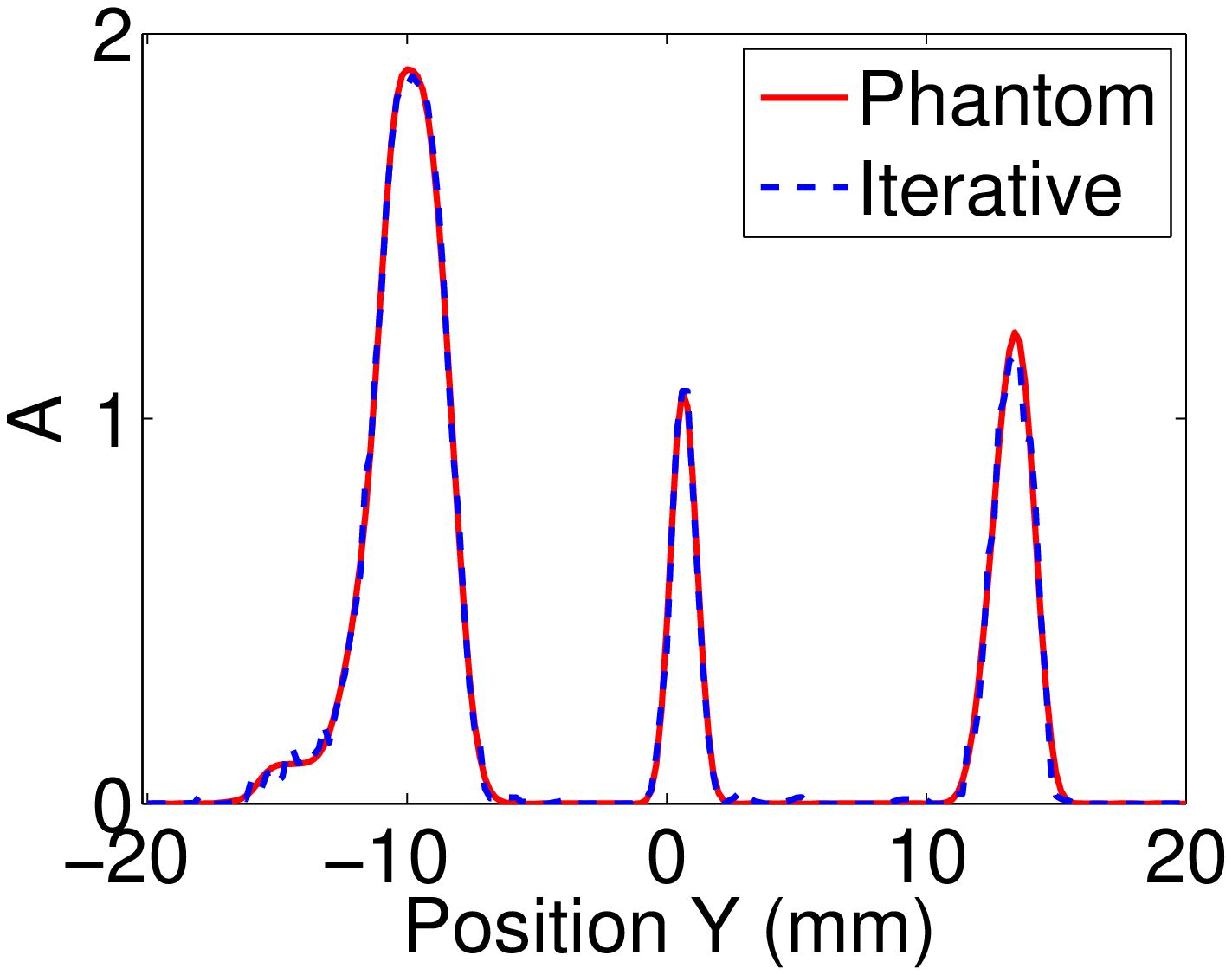}}}
\caption{\label{fig:vessel_limitedn}
(a) and (c) are reconstructed images from the noisy pressure data
with 3\% AWGN corresponding to the limited-view scanning geometry by use of
the TR method and iterative method,
respectively. (b) and (d) are the corresponding profiles.
}
\end{figure}

The reconstructed images corresponding to 
the three scanning geometries are displayed 
in Figs. \ref{fig:vessel_full} 
- \ref{fig:vessel_limitedn}.
In each figure, the results in the top row correspond to 
use of the TR reconstruction method, 
while the bottom row shows the corresponding results 
obtained by use of the iterative method. 
The profiles shown in each figure are
along the `Y'-axis indicated 
by the arrow in Fig. \ref{fig:vessel_full}-(a). 
The red solid lines and blue dashed lines correspond to
profiles through the phantom and reconstructed images, respectively. 
With the full-view scanning geometry, 
the TR method and the iterative method 
both produce accurate reconstructed images.
However, with the few-view and the limited-view 
scanning geometries, the images reconstructed 
from the iterative method contain fewer artifacts and less noise
than the TR results 
\footnote{
With the limited view scanning geometry, 
we also implemented the iterated TR method \cite{iterTR},
which produced images with fewer artifacts than
the ordinary TR results, but the background
was still not as clean as the iterative results.
Given the limited space, those results were not
included in this article.
}.
Also, the values of the images 
reconstructed from the iterative method are much closer to
the values of the phantom than those produced by the 
TR method. The root mean square error (RMSE) between the 
phantom and the reconstructed images were also 
computed. 
The RMSE of images reconstructed by use of the TR method 
and the iterative method corresponding to noisy pressure data
with the full-view, few-view, and limited-view
scanning geometries are 0.011, 0.042, 0.081 and
0.003, 0.007, 0.008, respectively. 
The computational time of the TR method
was 1.7 minutes, while the iterative method
took approximately 10 minutes to finish 20 iterations.

\subsection{Simulation results with errors in SOS and density maps}

\begin{figure}[!t]
\centering
   \subfigure[]{\resizebox{1.7in}{!}{\includegraphics{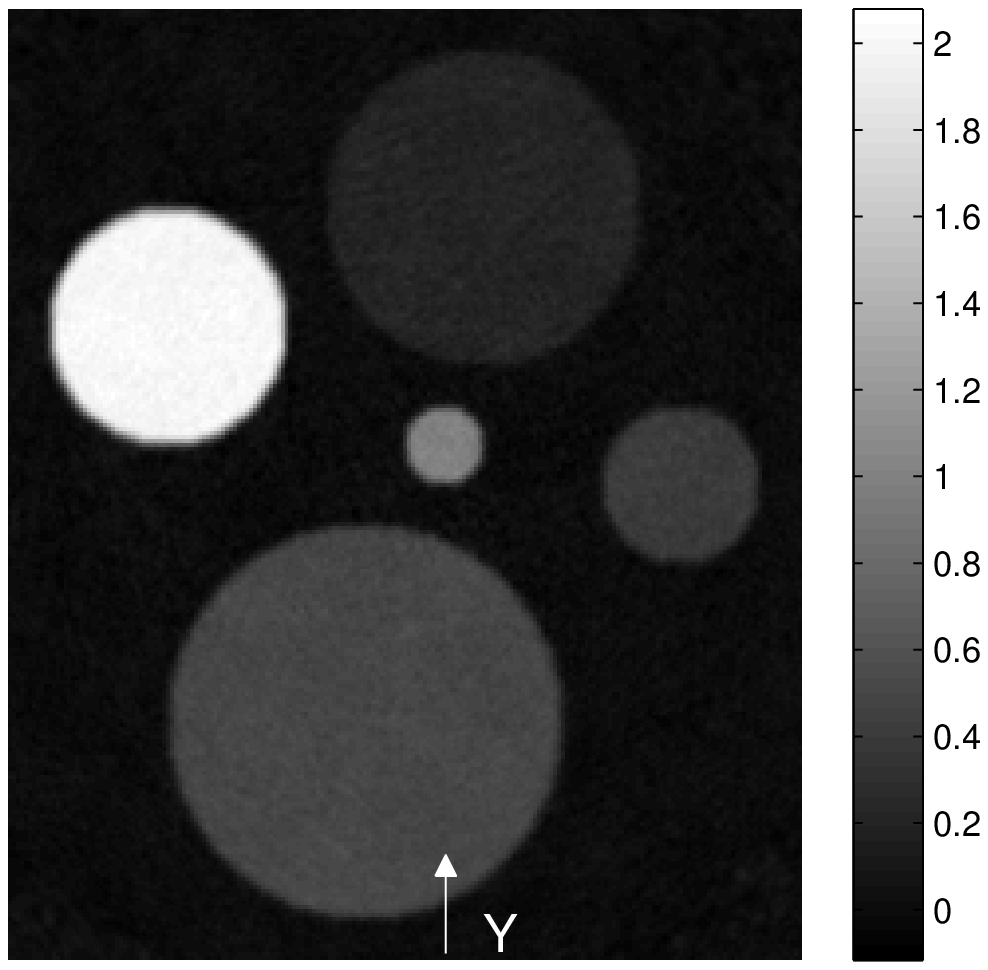}}}
   \subfigure[]{\resizebox{1.7in}{!}{\includegraphics{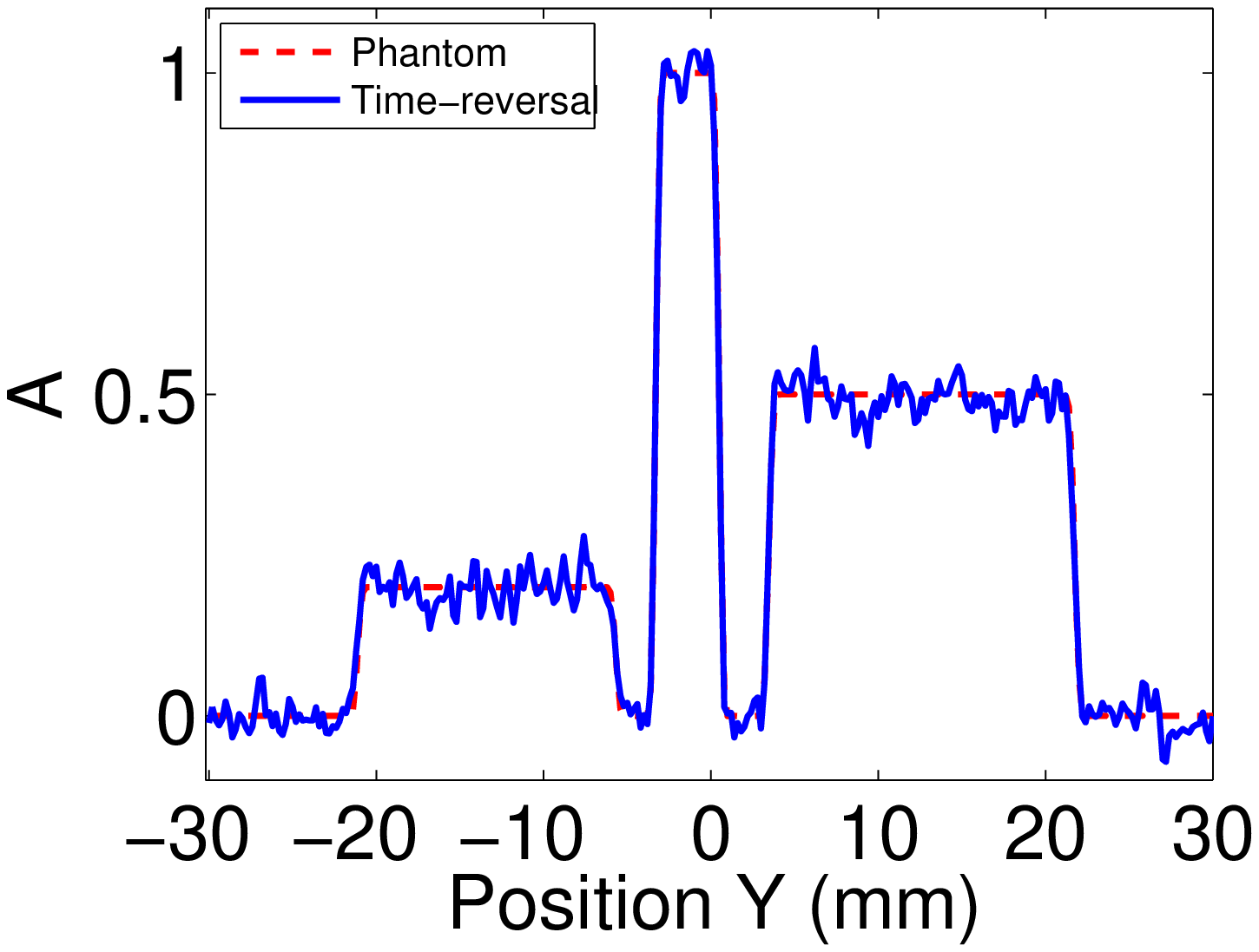}}}\\
   \subfigure[]{\resizebox{1.7in}{!}{\includegraphics{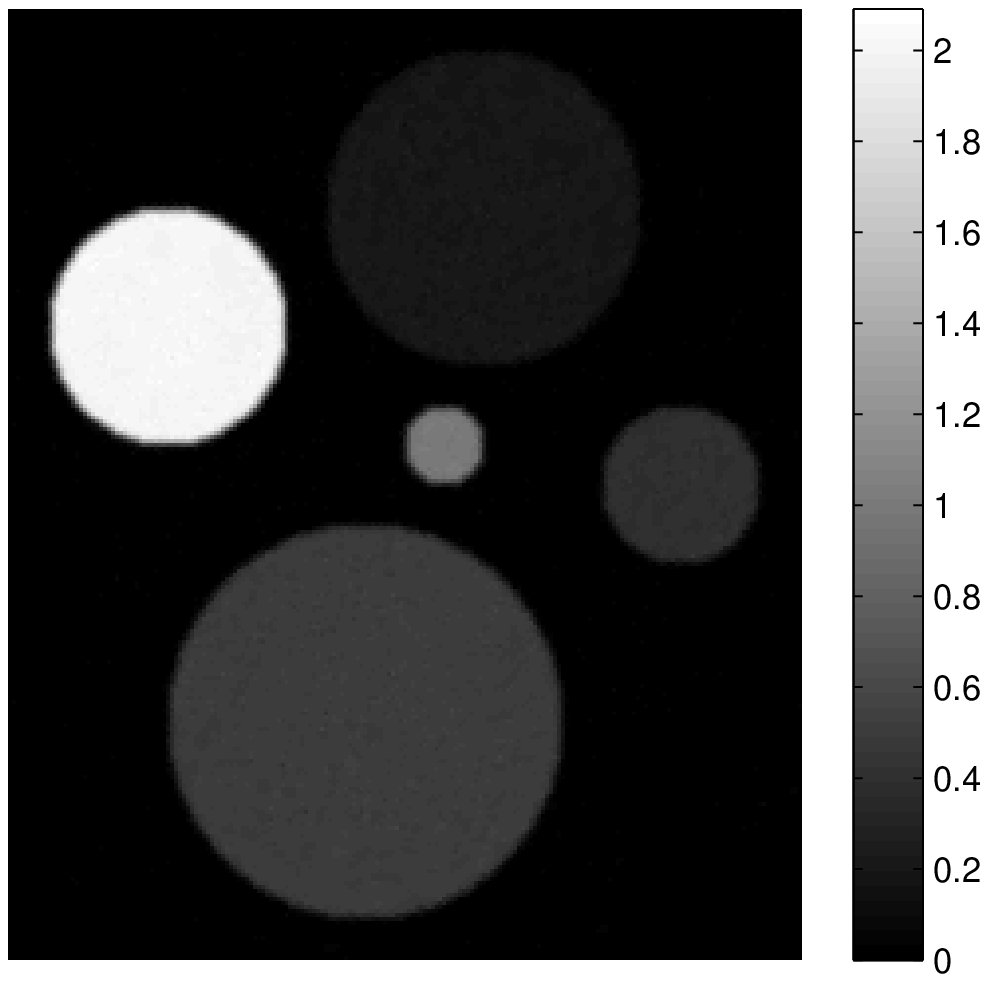}}}
   \subfigure[]{\resizebox{1.7in}{!}{\includegraphics{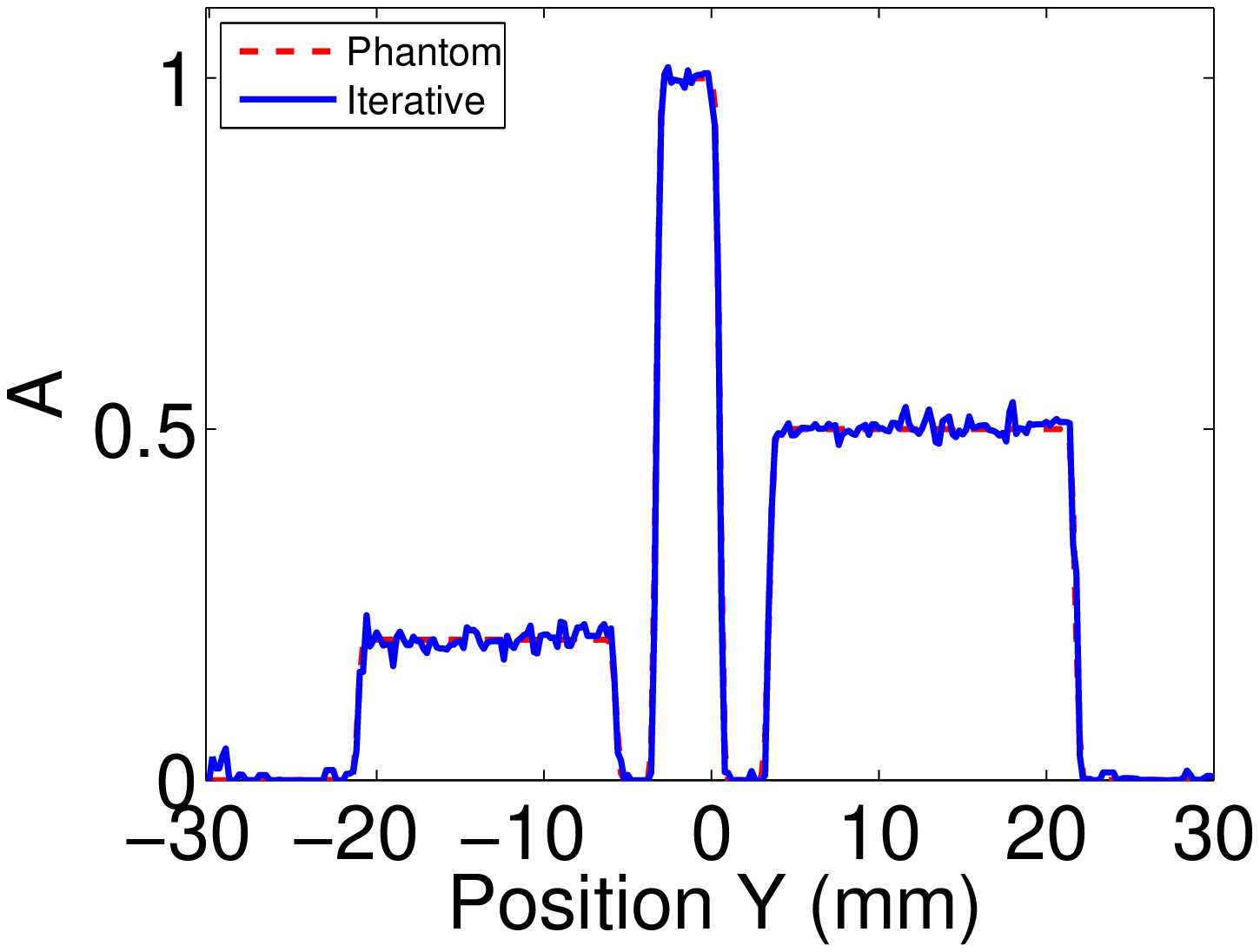}}}
\caption{\label{fig:disc_recon}
(a) and (c) are reconstructed images with
actual SOS and density maps by use of
the TR method and iterative method,
respectively. (b) and (d) are the corresponding profiles
along the `Y'-axis indicated in panel (a).
}
\end{figure}

Figure \ref{fig:disc_recon} shows the images 
reconstructed from noisy pressure data 
corresponding to the low contrast disc phantom
in the case where SOS and density maps have no error. 
The results corresponding to TR and iterative
image reconstruction algorithms are shown 
in the top and bottom row, respectively.
The RMSE corresponding to 
the time-reversal and the iterative results 
are 0.026 and 0.007, respectively. These results 
suggest that the iterative algorithm can more effectively
reduce the noise level in the reconstructed 
images than the time-reversal algorithm.

\begin{figure}[!t]
\centering
  \subfigure[]{\resizebox{1.7in}{!}{\includegraphics{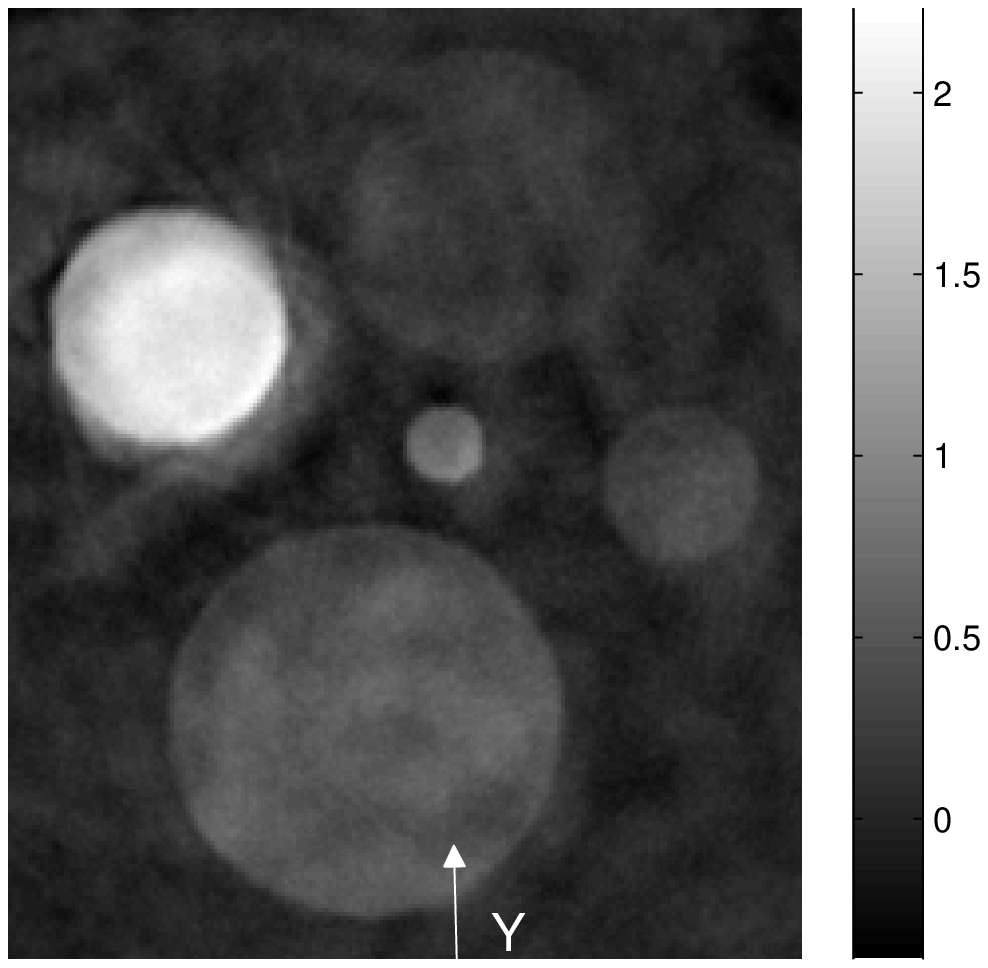}}}
  \subfigure[]{\resizebox{1.7in}{!}{\includegraphics{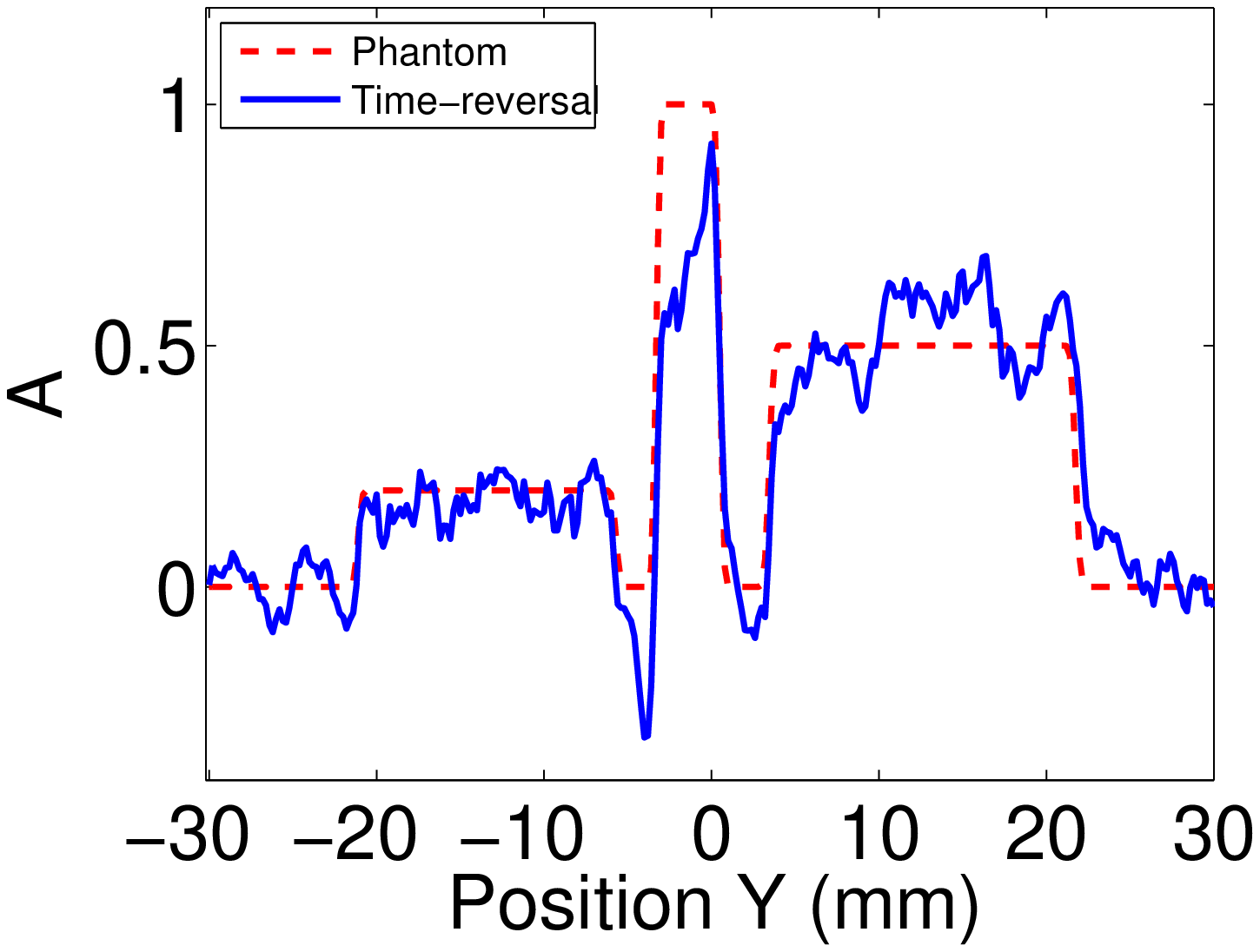}}}\\
  \subfigure[]{\resizebox{1.7in}{!}{\includegraphics{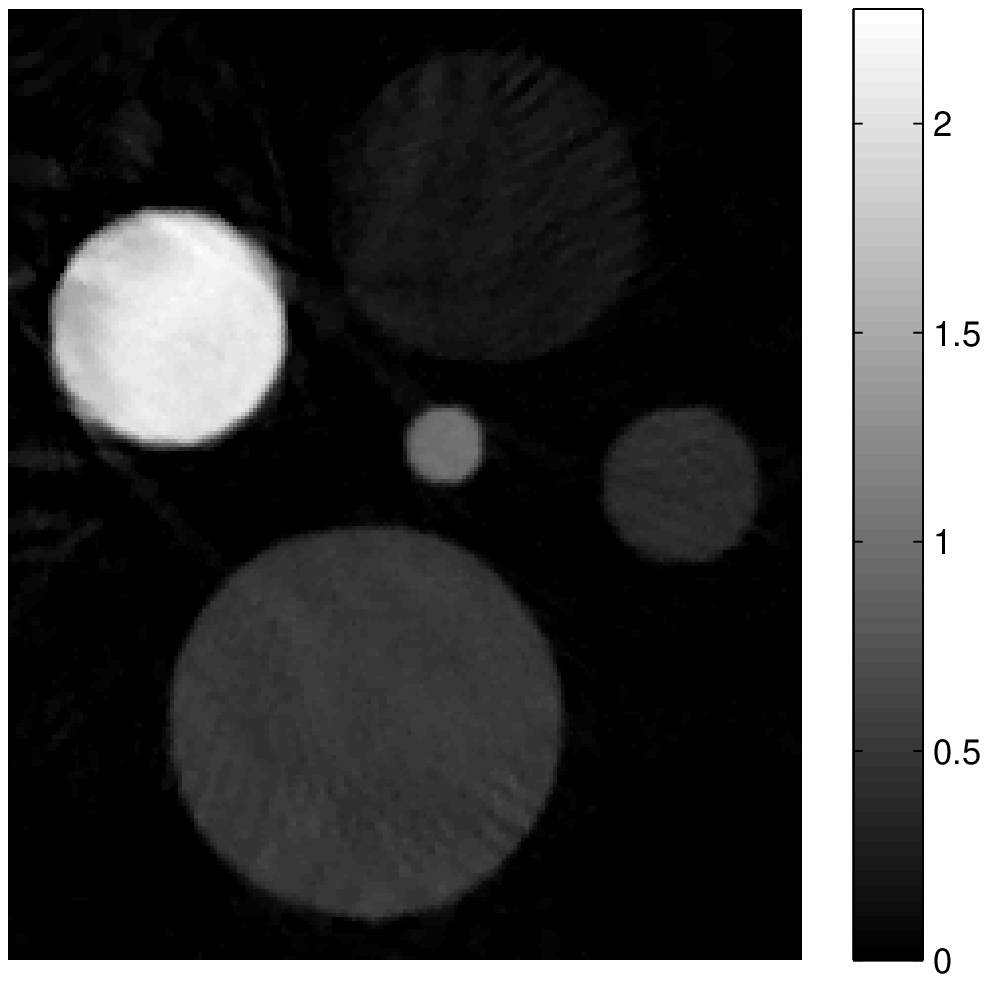}}}
  \subfigure[]{\resizebox{1.7in}{!}{\includegraphics{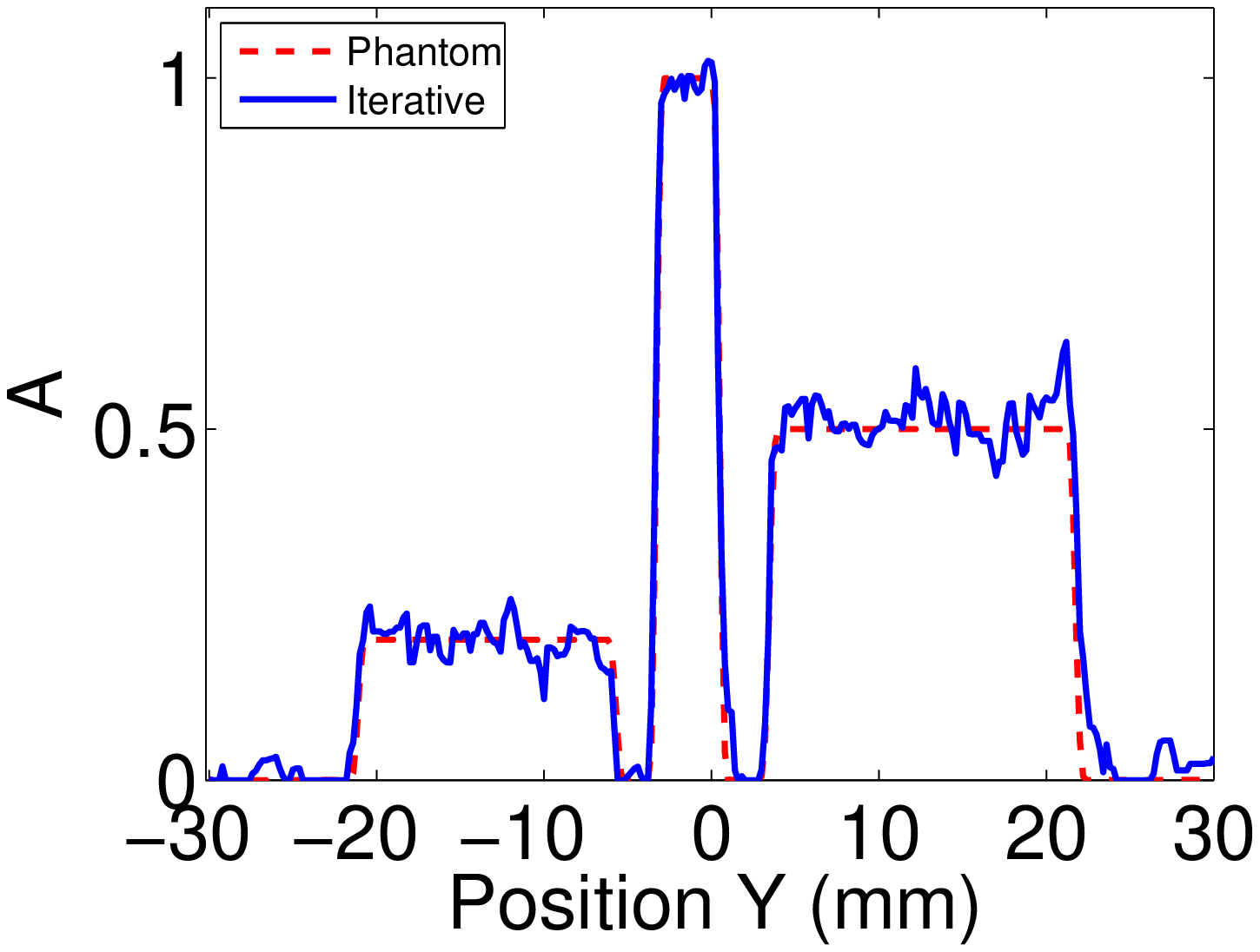}}}
\caption{\label{fig:error_recon}
(a) and (c) are reconstructed images with
SOS and density maps with errors by use of
the TR method and iterative method,
respectively. (b) and (d) are the corresponding profiles
along the `Y'-axis indicated in panel (a).
}
\end{figure}

The images reconstructed by use of the SOS and density maps with 
errors are shown in Fig. \ref{fig:error_recon}.
The image produced by the iterative method 
has cleaner background than the TR result, and 
the RMSE corresponding to the TR and 
the iterative results are 0.086 and 0.034, respectively. 
The boundaries of the disc phantoms also appear sharper in the image 
reconstructed by the iterative method as compared to 
the TR result. This can be attributed to
the TV regularization employed in the iterative method. 
These results suggest that appropriately regularized
iterative reconstruction methods can be more
robust to the errors in the SOS and density maps
than the TR method.

\subsection{3D simulation results}

\begin{figure*}[!t]
\centering
  \subfigure[]{\resizebox{2.3in}{!}{\includegraphics{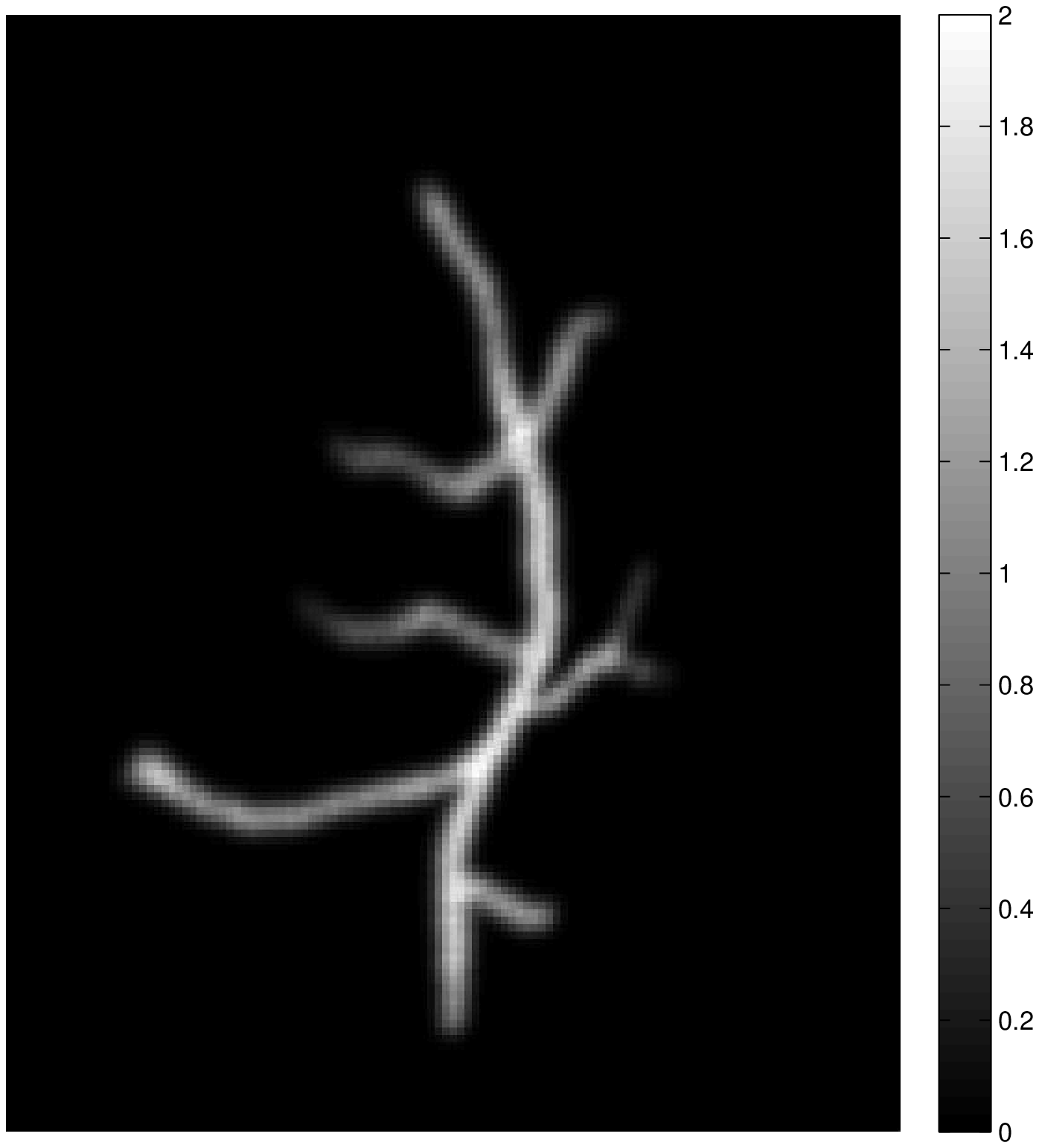}}}
  \subfigure[]{\resizebox{2.3in}{!}{\includegraphics{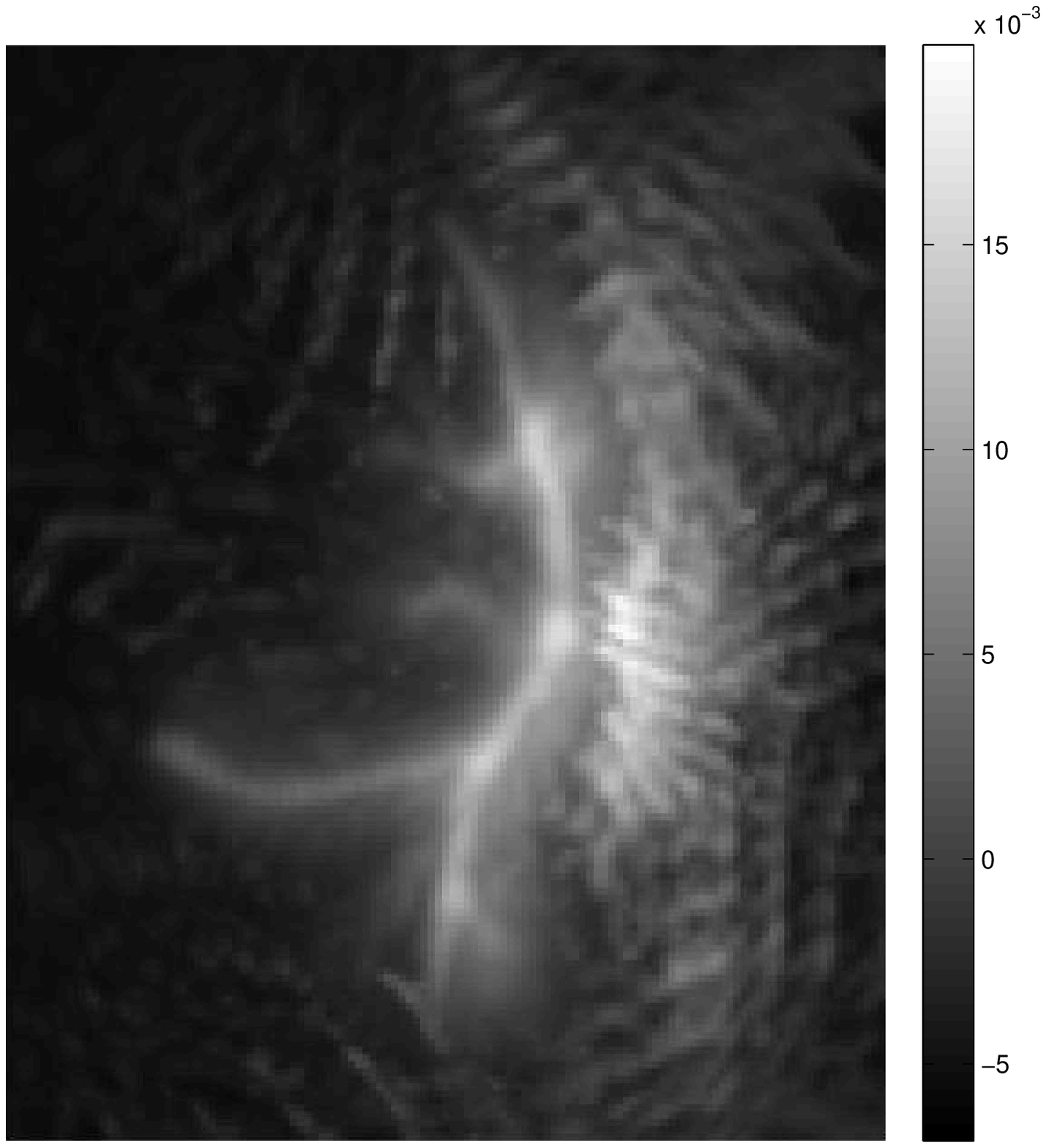}}}
  \subfigure[]{\resizebox{2.3in}{!}{\includegraphics{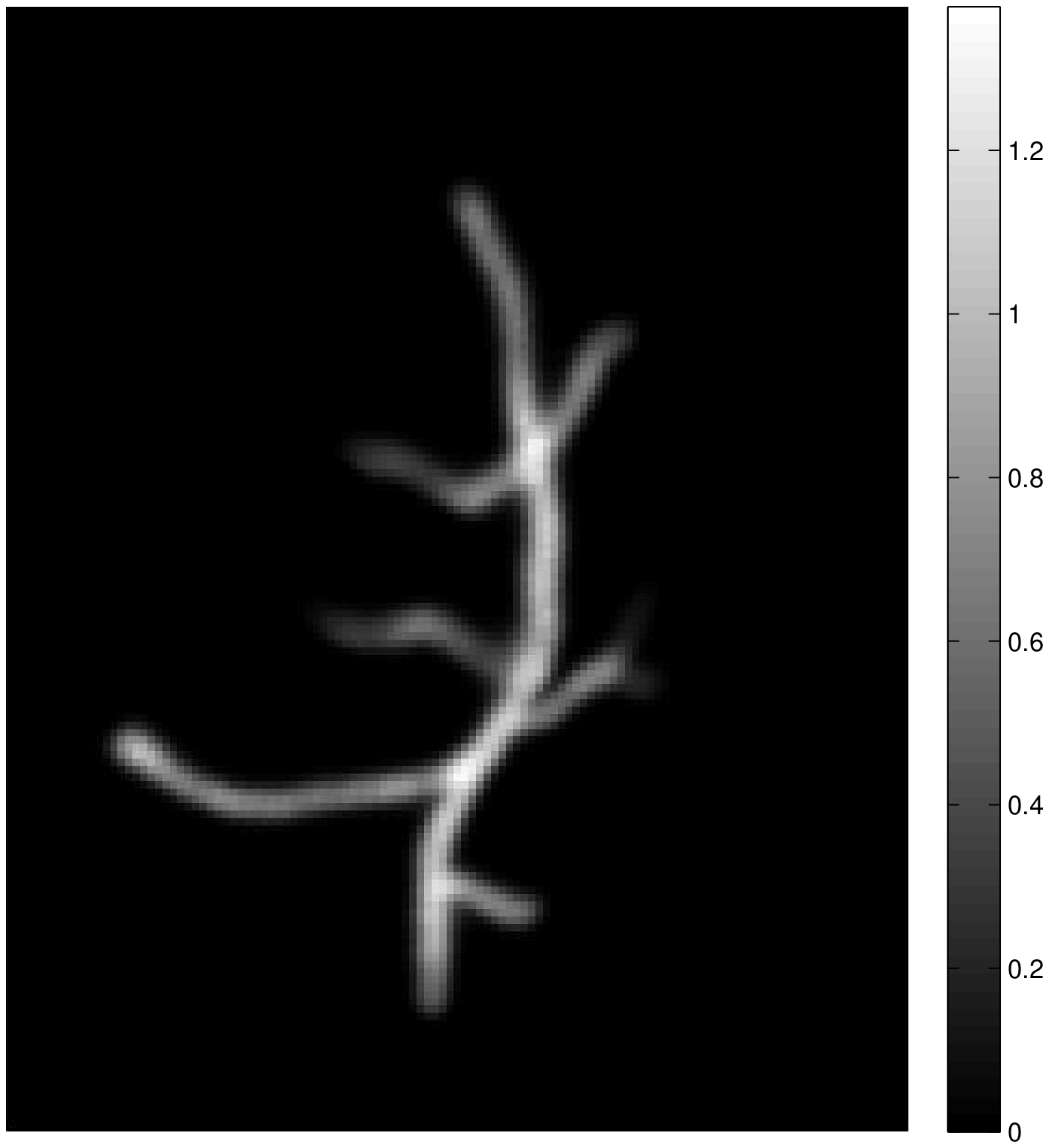}}}
\caption{\label{fig:3d}
Maximum intensity projection renderings of the 3D phantom (a),
and the reconstructed 3D images by use of the TR method (b)
and the iterative method (c).
}
\end{figure*}

The 3D blood vessel phantom and the reconstructed images
were visualized by the maximum intensity projection
(MIP) method. Figure \ref{fig:3d}-(a) shows the phantom image, and
Fig. \ref{fig:3d}-(b) and (c) display the images reconstructed
by use of the TR method and the iterative method, 
respectively. They are all displayed in the same grey scale window.
The RMSE corresponding to the TR and the iterative results 
are 0.018 and 0.003, respectively. These results suggest that 
the iterative method is robust to the data incompleteness 
and the noise in the pressure data.
The computational time of the TR method was approximately 6 minutes, 
while the iterative method with 10 iterations 
required 110 minutes.

\subsection{Experimental results}
%

\begin{figure}[!t]
\centering
  \subfigure[]{\resizebox{1.7in}{!}{\includegraphics{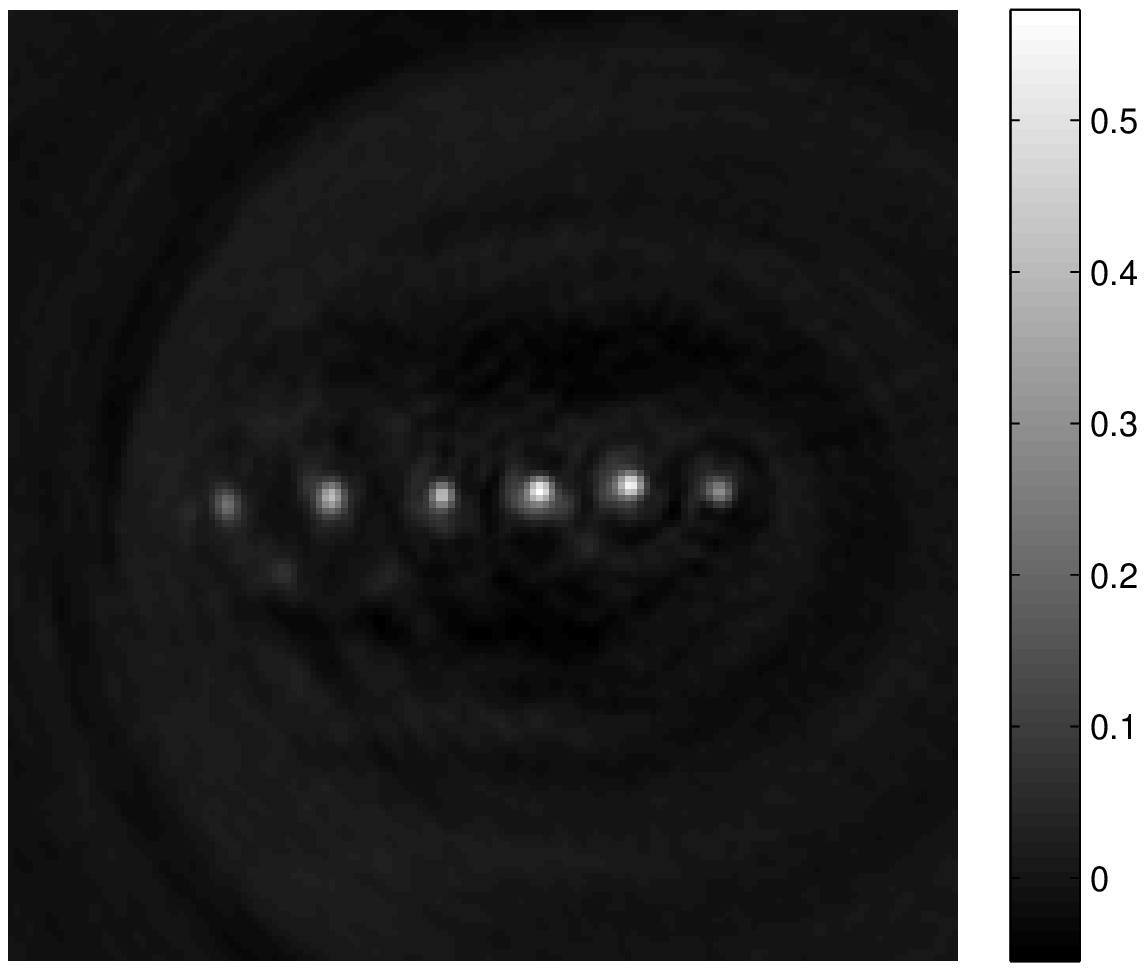}}}
  \subfigure[]{\resizebox{1.7in}{!}{\includegraphics{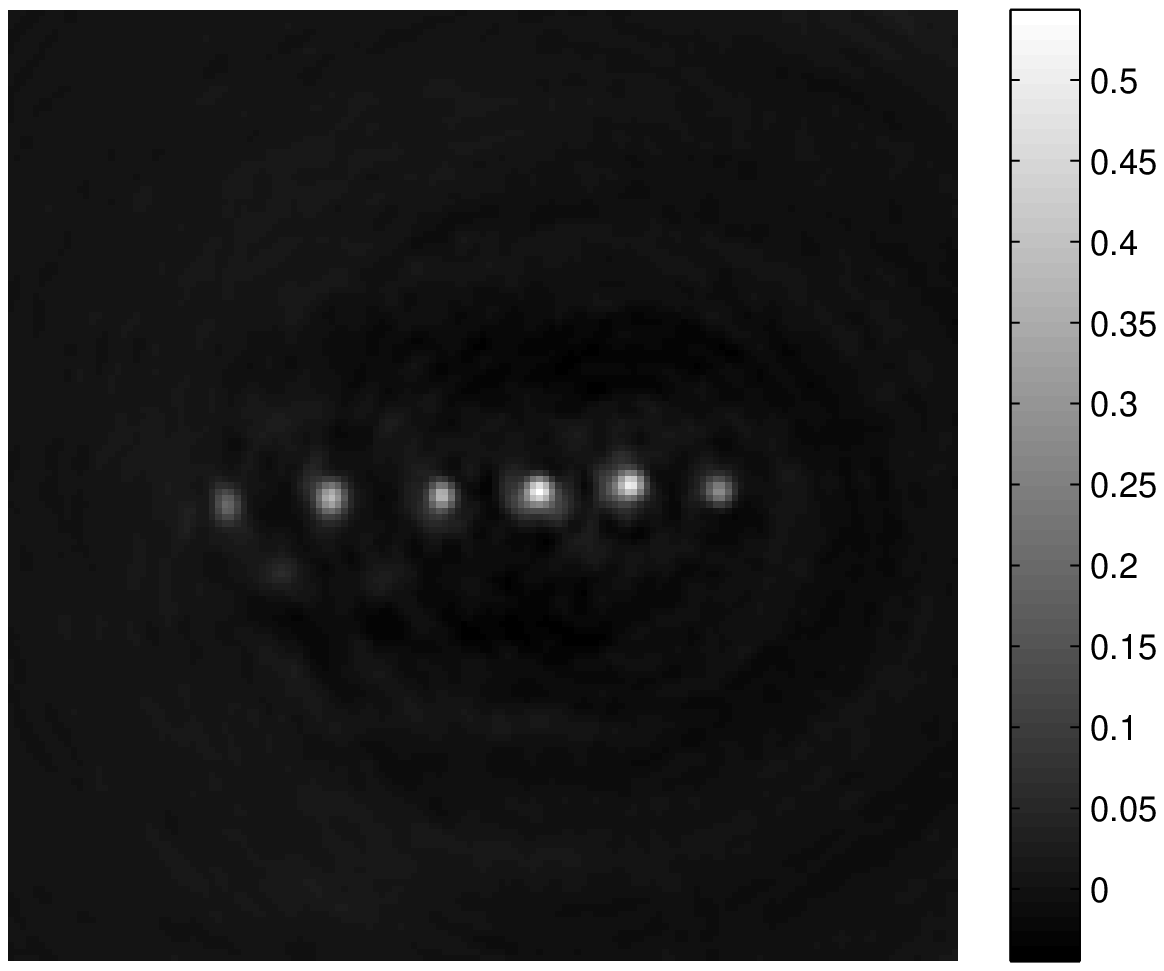}}}\\
  \subfigure[]{\resizebox{1.7in}{!}{\includegraphics{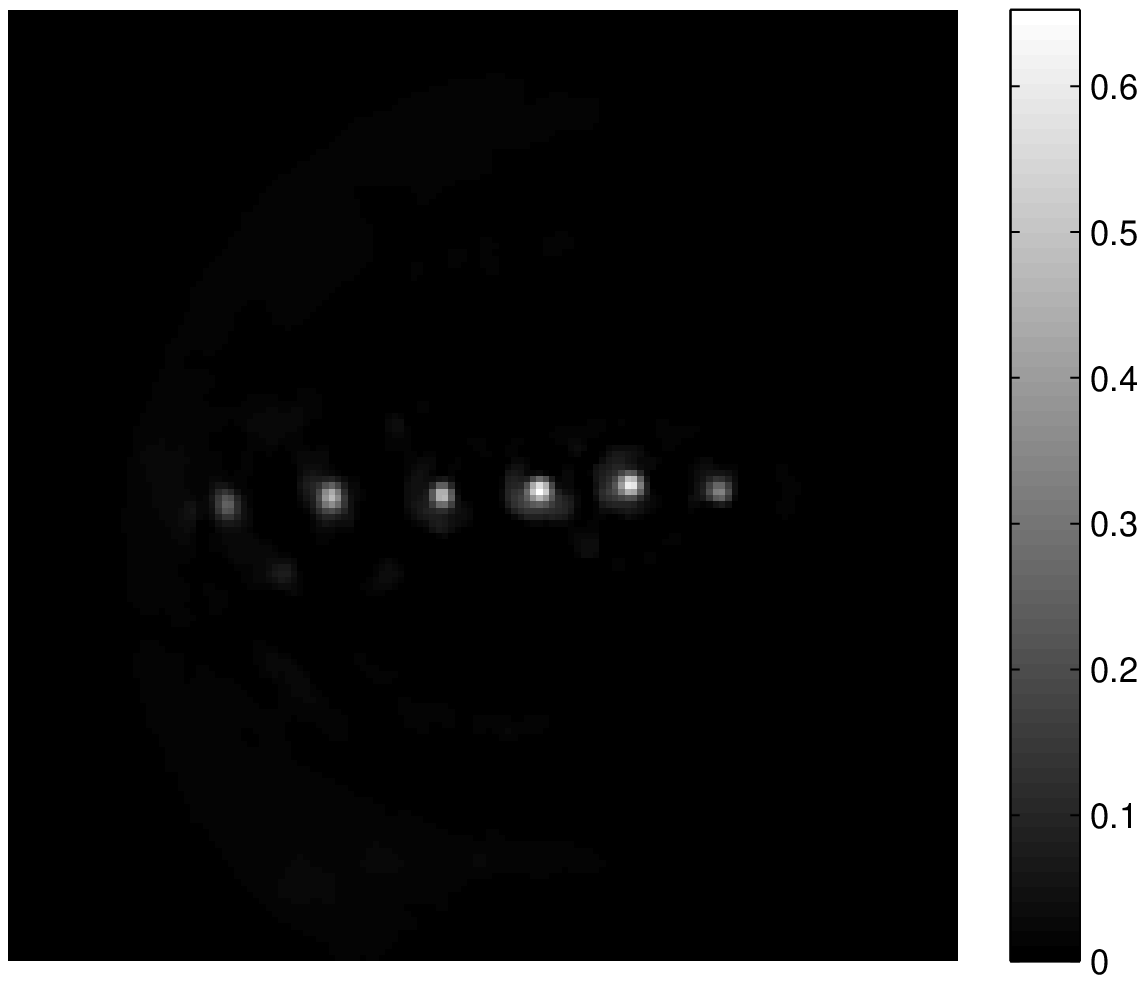}}}
  \subfigure[]{\resizebox{1.7in}{!}{\includegraphics{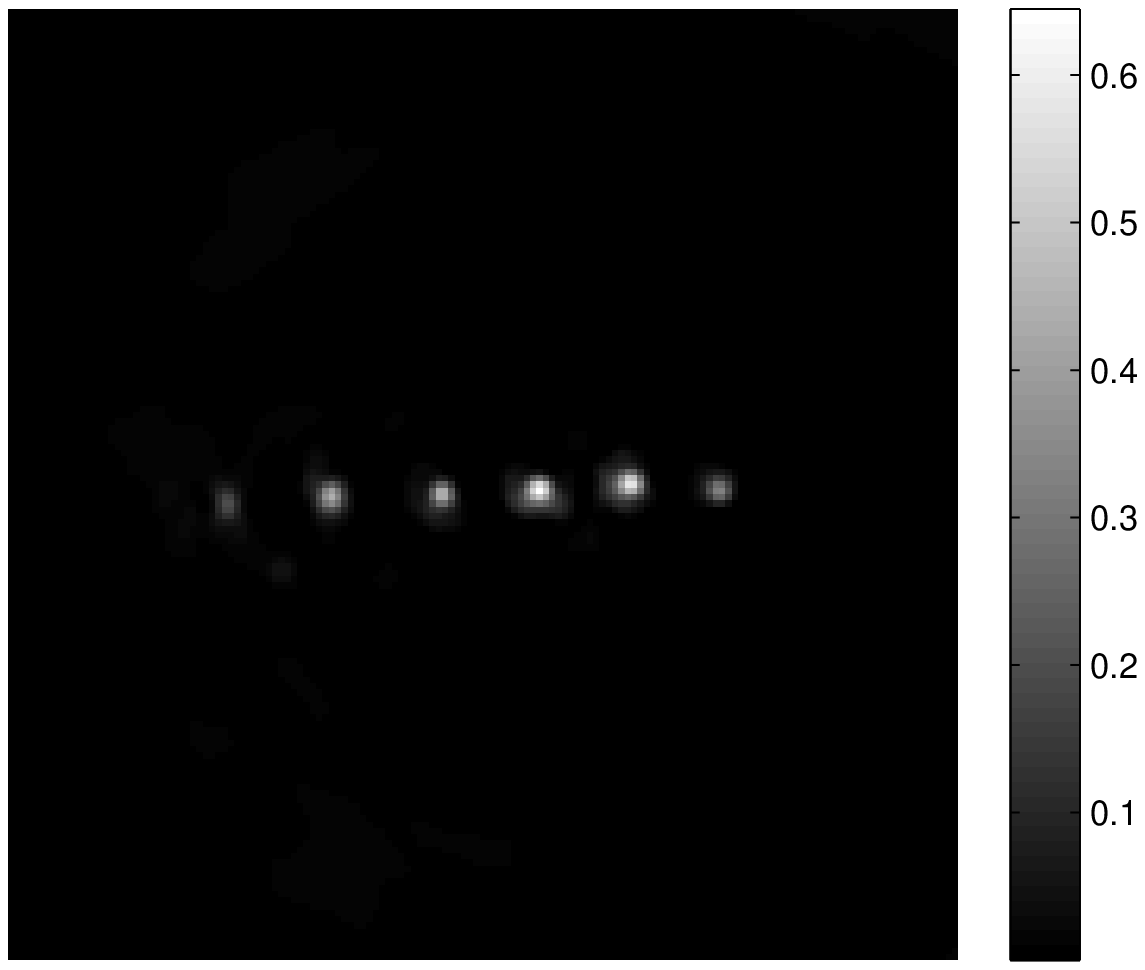}}}
\caption{\label{fig:acrylic_full}
(a) and (b) are reconstructed images
by use of the TR method from 200 views
with acrylic shell absent and present, respectively.
(c) and (d) are reconstructed images
by use of the iterative method
from 200 views with acrylic shell absent
and present, respectively.
}
\end{figure}

\begin{figure}[!t]
\centering
  \subfigure[]{\resizebox{1.7in}{!}{\includegraphics{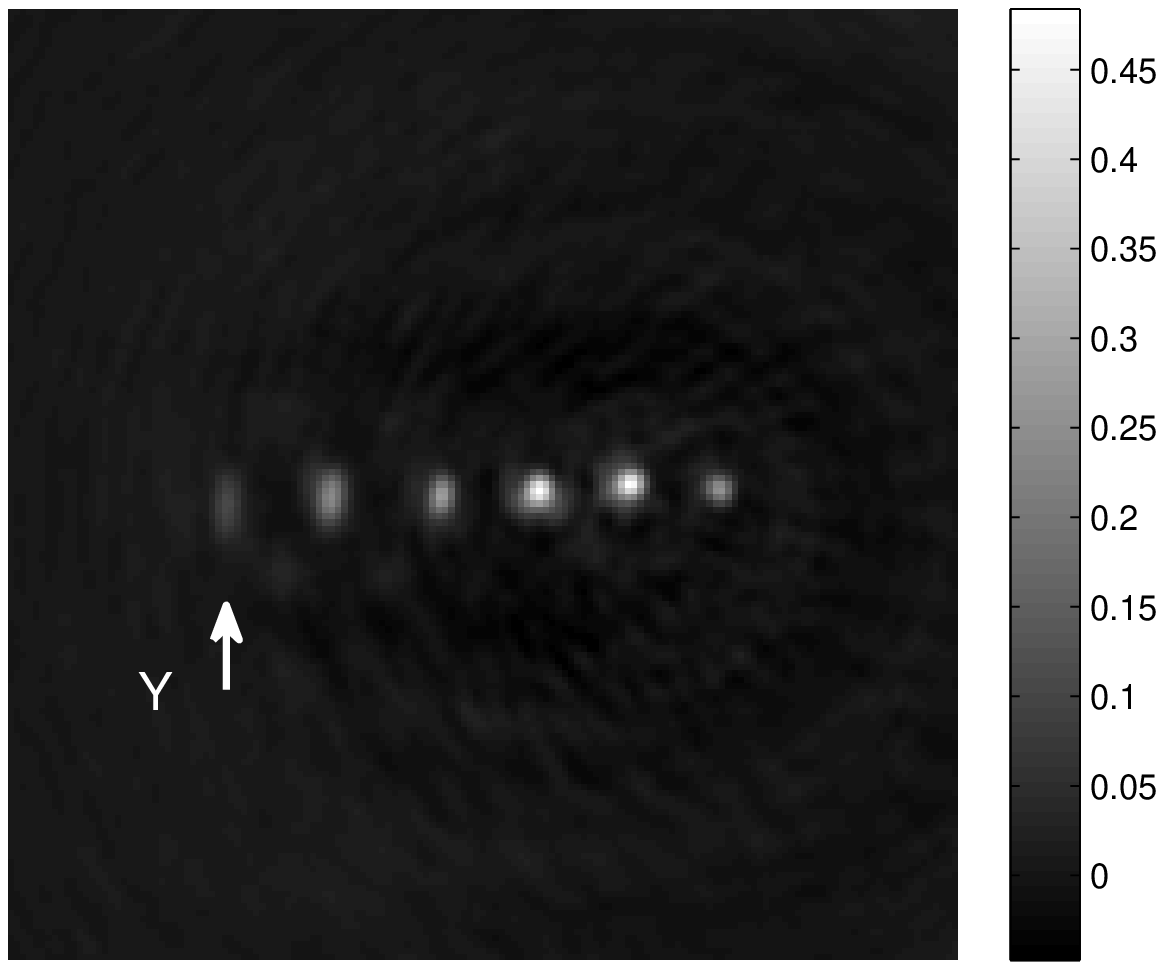}}}
  \subfigure[]{\resizebox{1.7in}{!}{\includegraphics{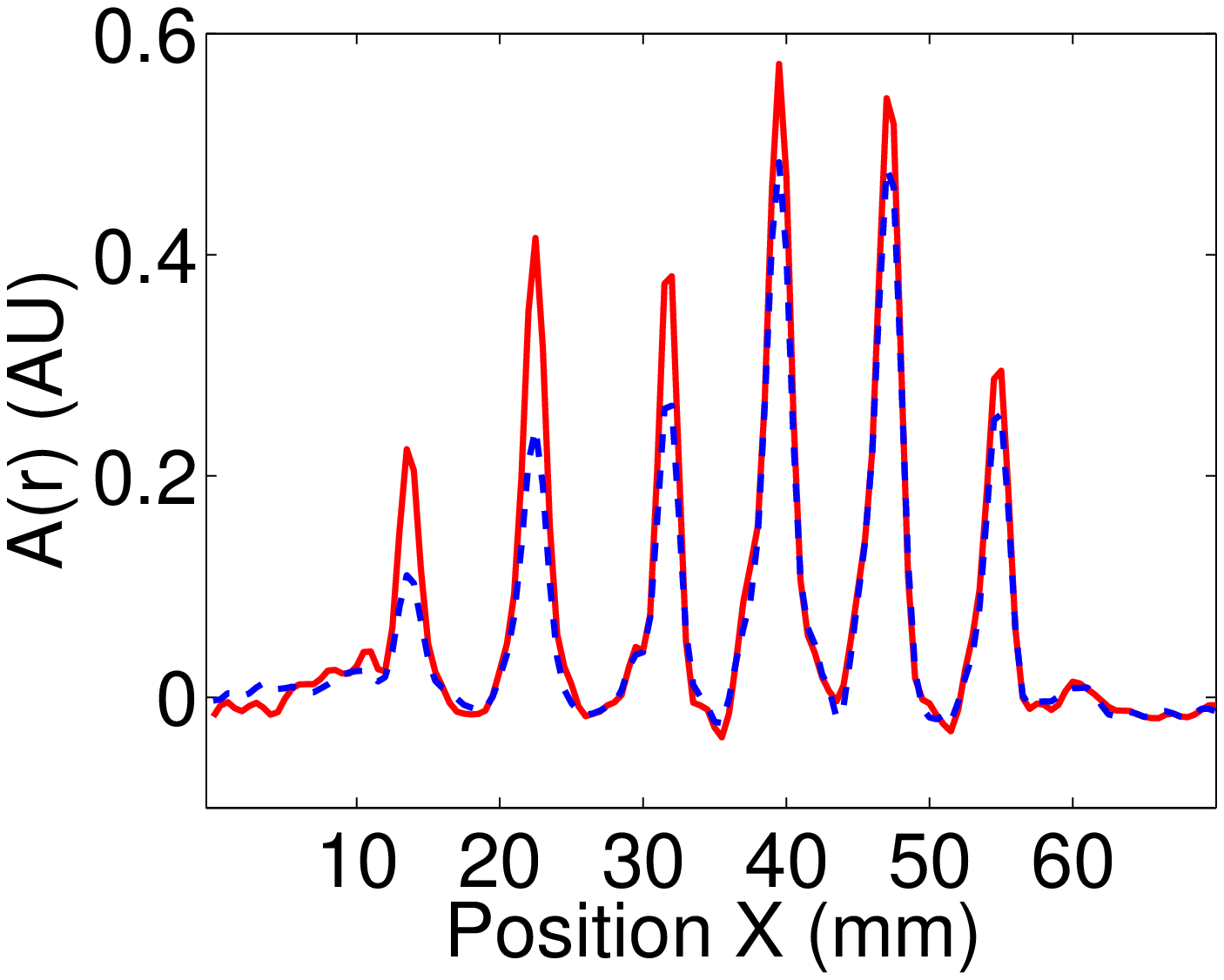}}}\\
  \subfigure[]{\resizebox{1.7in}{!}{\includegraphics{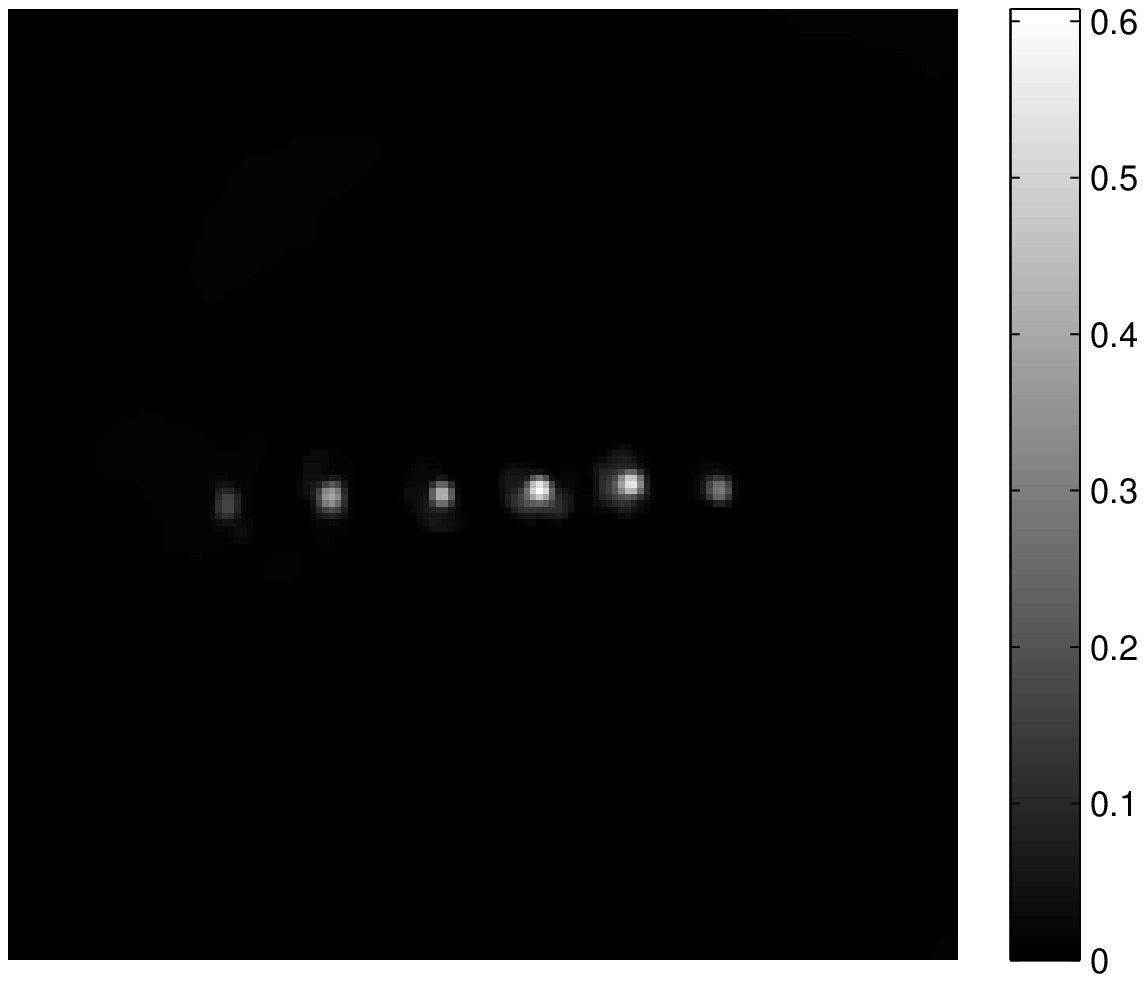}}}
  \subfigure[]{\resizebox{1.7in}{!}{\includegraphics{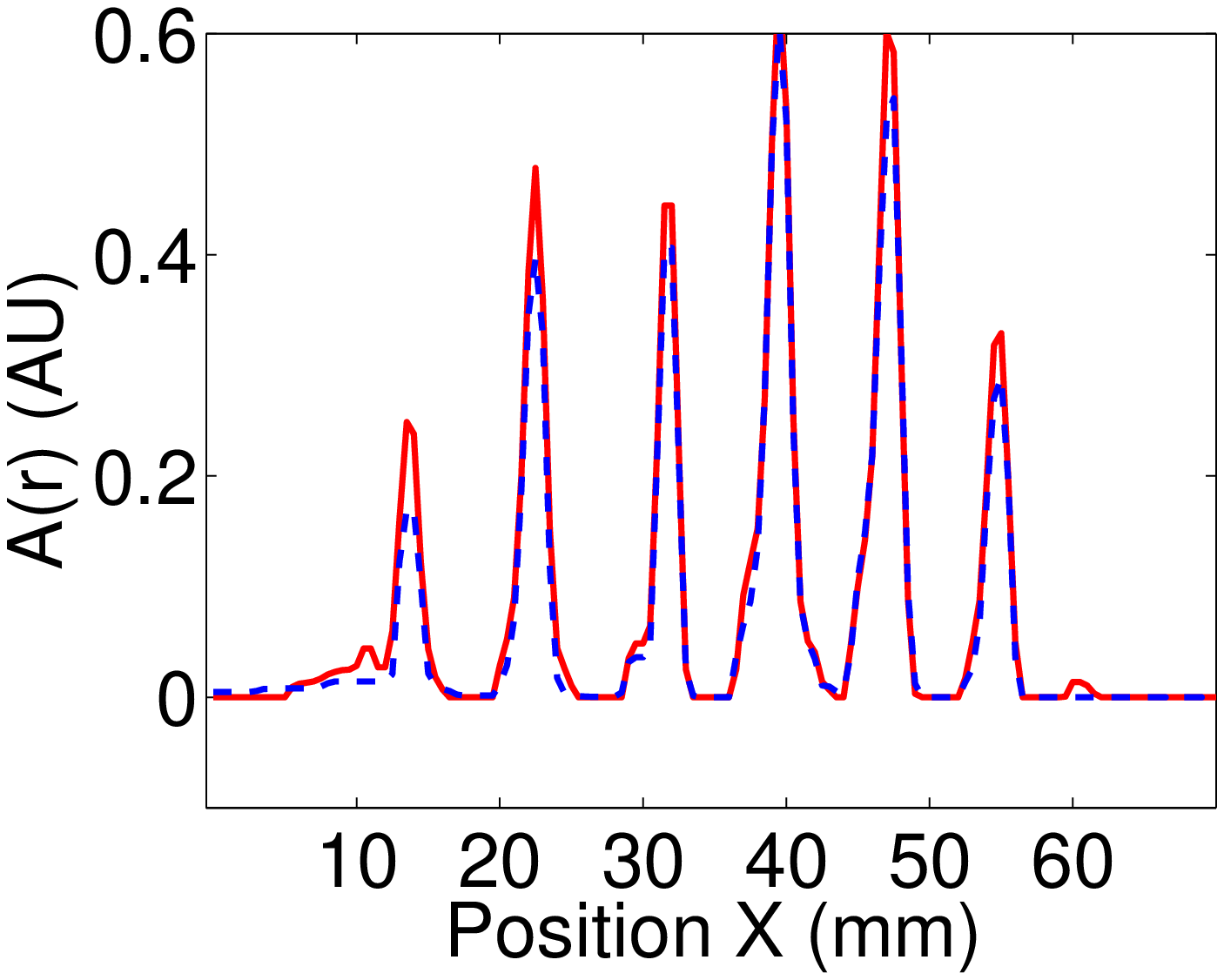}}}
\caption{\label{fig:acrylic_few}
(a) and (c) are reconstructed images with
data from 50 view angles over 360 degrees
(acrylic shell present) by use of the
TR method and iterative method, respectively.
(b) and (d) are their corresponding profiles
(dashed blue lines), where red solid lines
are the profiles of the reference images in Fig.
\ref{fig:acrylic_full} (a) and (c).
}
\end{figure}

\begin{figure}[!t]
  \centering
    \resizebox{2.3in}{!}{
      \includegraphics{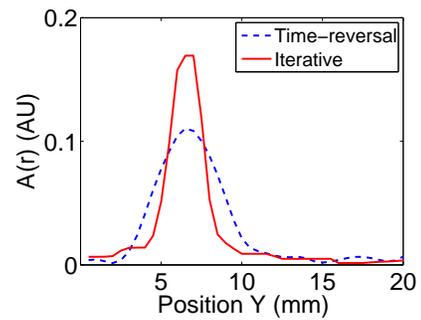}
  }
  \caption{\label{fig:acrylic_few_pfl}
The profiles of the reconstructed images
in Fig. \ref{fig:acrylic_few} along the
`Y'-axis indicated in Fig. \ref{fig:acrylic_few}(a).
}
\end{figure}

\begin{figure}[!t]
\centering
  \subfigure[]{\resizebox{1.7in}{!}{\includegraphics{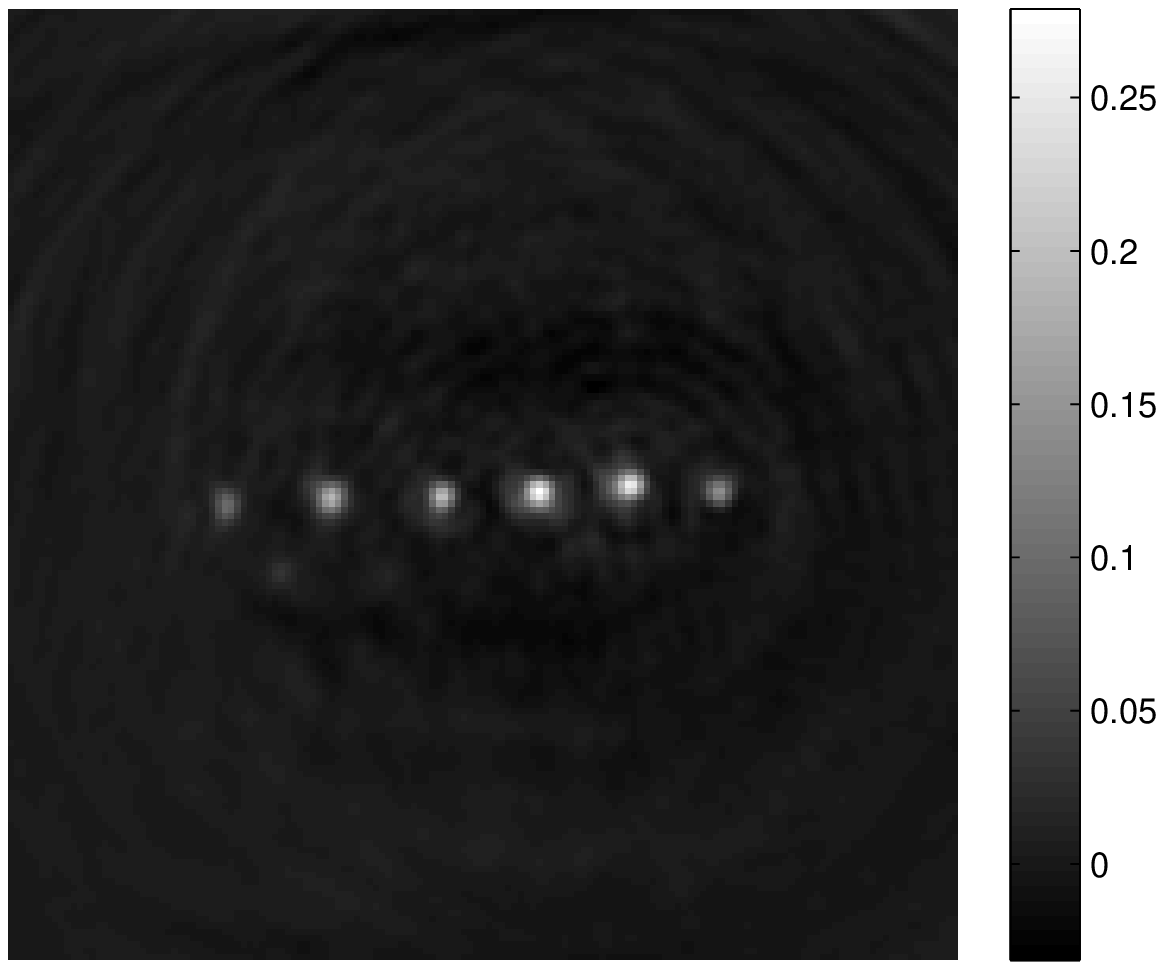}}}
  \subfigure[]{\resizebox{1.7in}{!}{\includegraphics{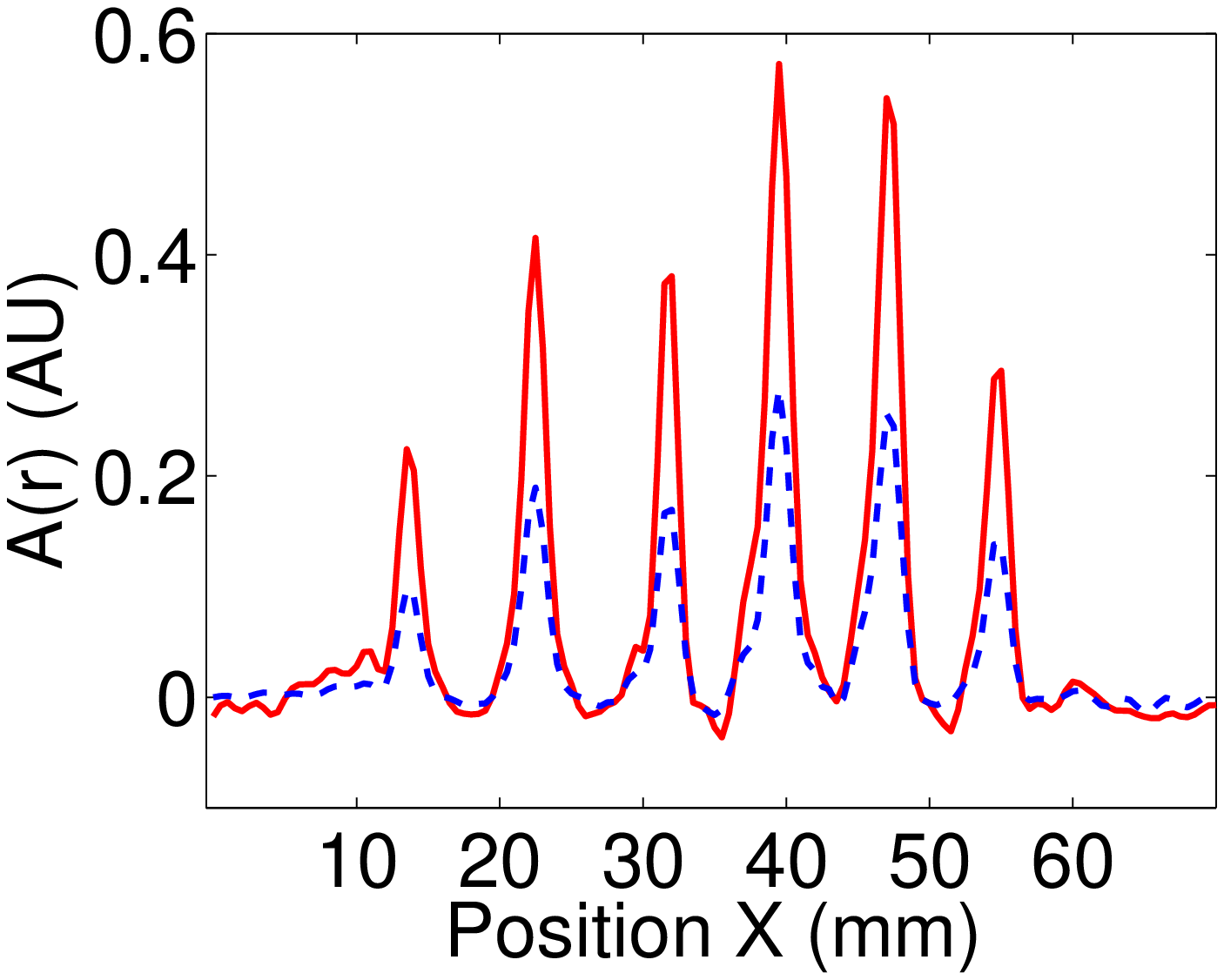}}}\\
  \subfigure[]{\resizebox{1.7in}{!}{\includegraphics{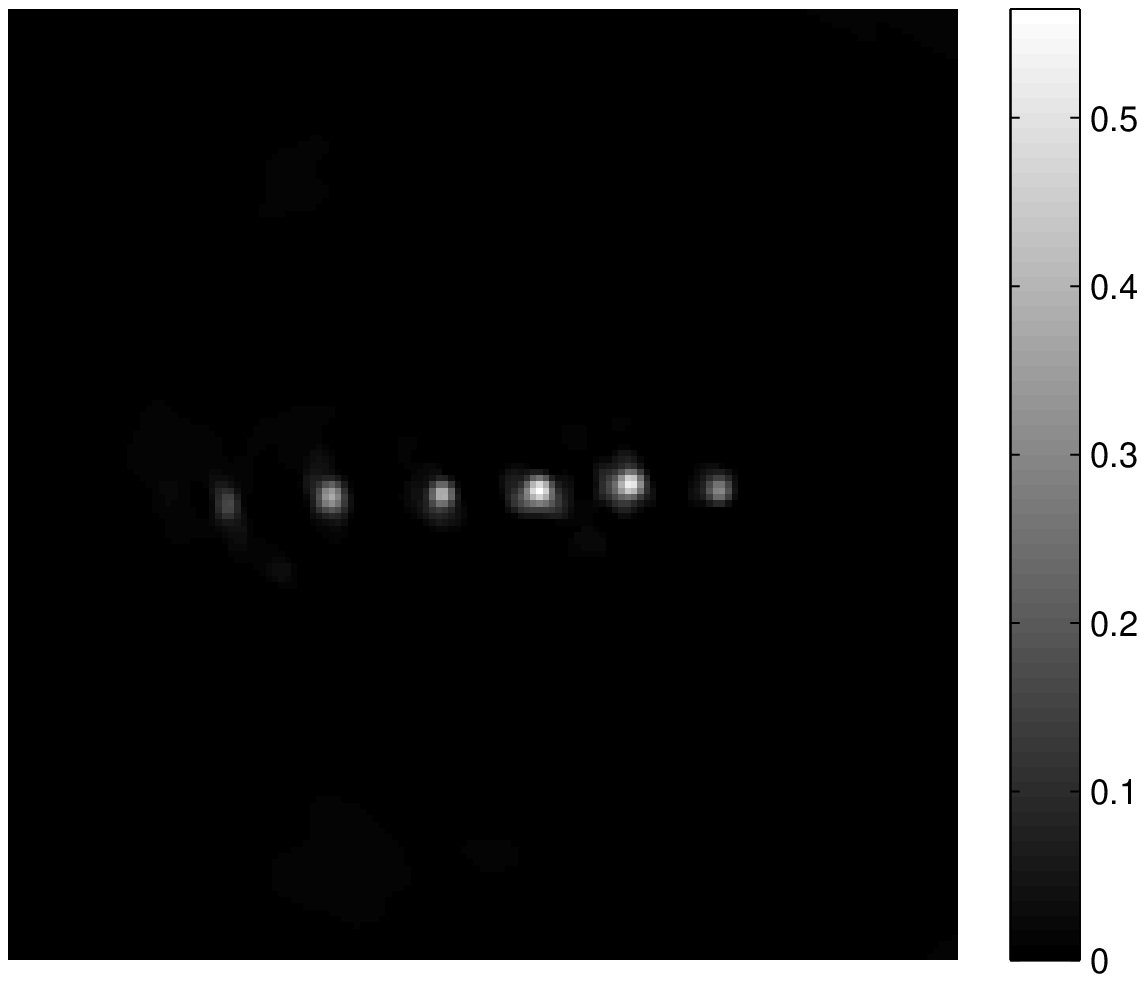}}}
  \subfigure[]{\resizebox{1.7in}{!}{\includegraphics{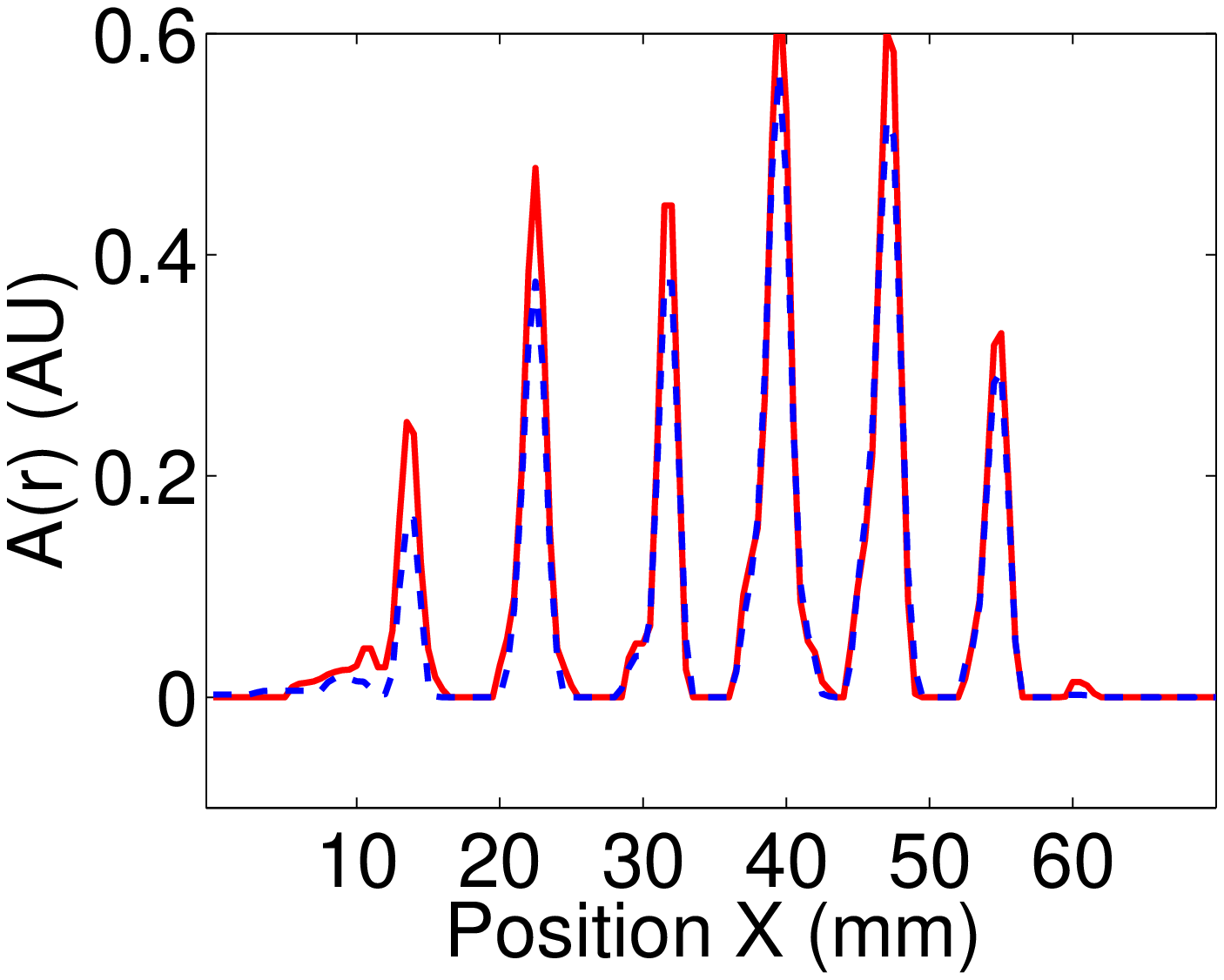}}}
\caption{\label{fig:acrylic_limited}
(a) and (c) are reconstructed images with
data from 100 view angles over 180 degrees
(acrylic shell present) by use of the
TR method  and iterative method, respectively.
(b) and (d) are their corresponding profiles
(dashed blue lines), where red solid lines
are the profiles of the reference images in Fig.
\ref{fig:acrylic_full} (a) and (c).
}
\end{figure}

The images reconstructed from the experimental data are 
shown in Figs. \ref{fig:acrylic_full} -
\ref{fig:acrylic_limited}. Figure \ref{fig:acrylic_full}
shows the image reconstructed with the 
full-view scanning geometry 
by use of the TR method (top row) and 
the iterative method (bottom row). 
Figure \ref{fig:acrylic_full}-(a) and (c) display
the reference images produced by each of the methods
when the acrylic shell was absent. 
Figure \ref{fig:acrylic_full}-(b) and (e) show 
the reconstructed images for the case
when the acrylic shell was present.
The RMSE between Fig. \ref{fig:acrylic_full}-(b), (d)
and the reference images  \ref{fig:acrylic_full}-(a), (c)
are 0.003 and 0.002, respectively.
Figure \ref{fig:acrylic_few}-(a) and (c) show the images
reconstructed with the few-view  scanning geometry
when the acrylic shell was present.
The corresponding image profiles are displayed in 
Figure \ref{fig:acrylic_few}-(b) and (d).
The profiles of Fig. \ref{fig:acrylic_few}-(a) and (c) 
along the `Y'-axis were shown in Fig. \ref{fig:acrylic_few_pfl},
which shows that the iterative method produced
higher resolution images than the TR method.
This can be attritubed to the TV regularization that 
mitigates model errors that arise, for example,
by neglecting the shear wave and finite transducer aperture effects.
The RMSE between Fig. \ref{fig:acrylic_few}-(b), (d)
and their reference images are 0.005 and 0.002, , respectively.
Figure \ref{fig:acrylic_limited} displays 
the images reconstructed with the limited-view
scanning geometry when the acrylic shell was present. 
The RMSE between Fig. \ref{fig:acrylic_limited}-(a), (c)
and their reference images are 0.007 and 0.003, respectively. 
These results show that the iterative algorithm can 
effectively compensate for the acoustic attenuation 
and mitigate artifacts and distortions 
due to incomplete measurement data.

\section{Conclusion and discussion}
\label{sect:summary}
We proposed and investigated a full-wave approach to 
iterative image reconstruction in PACT 
with acoustically inhomogeneous lossy media.
An explicit formulation of
the discrete imaging model based on the k-space 
pseudospectral method was described, and 
the details of implementing the forward and 
backprojection operators were provided. The matched
operator pair was employed in an iterative image
reconstruction algorithm that sought to minimize 
a TV-regularized PLS cost function. The developed 
reconstruction methodology was investigated by 
use of both computer-simulated and experimental 
PACT measurement data, and the results demonstrated that 
the reconstruction methodology can effectively mitigate 
image artifacts due to data incompleteness, noise, 
finite sampling, and modeling errors.
This suggests that the proposed image reconstruction method
has the potential to be adopted in preclinical and
clinical PACT applications.

There remain several important topics to further investigate
and validate the proposed iterative reconstruction method.
It has been shown \cite{HristovaIP2008,QianiJIS2011}
that the performance of reconstruction methods 
can be degraded when the SOS distribution 
satisfies a trapping condition 
\cite{HristovaIP2008,QianiJIS2011}.
Therefore, future studies may include the investigation
of numerical properties of the proposed image reconstruction method 
for cases in which the SOS distribution satisfies the trapping condition.
Also, because the signal detectability is affected by
the noise properties of an image reconstruction method, 
investigation of statistical properties 
of the iterative image reconstruction method
is another important topic for future studies.
Moreover, the proposed image reconstruction method 
can be further validated through additional experimental studies, 
and the quality of the produced images will be assessed 
by use of objective and quantitative measures.

\section*{acknowledments}
This research was supported by NIH award EB010049.

\section*{appendix-A: Modeling transducer impulse responses}
An important feature of the proposed 
discrete PACT imaging model is that the transducer's 
impulse responses, including the  spatial impulse response (SIR)
and the acousto-electrical impulse response (EIR),
can be readily incorporated into the system matrix.

The SIR accounts for the averaging effect
over the transducer surface \cite{GHarris1981,VAndreev2002,KunTMI2011},
which can be described as
\begin{equation}
\label{eq:CSIR}
\hat{p}^{\text{SIR}}(\mathbf{r}_l^d,m \Delta t)=
\frac{\int_{S(\mathbf{r}_l^d)} 
\mathrm{d}S(\mathbf{r}'_l)
p(\mathbf{r}'_l,m \Delta t)}
{S(\mathbf{r}_l^d)},
\end{equation}
where $\hat{p}^{\text{SIR}}(\mathbf{r}_l^d,m \Delta t)$ is the
averaged pressure at time $t=m \Delta t$ over the surface of
the $l$-th transducer, $S(\mathbf{r}_l^d)$
is the surface area of the $l$-th transducer
centered at $\mathbf{r}_l^d$.

In order to incorporate the SIR into the system matrix,
we can divide the transducer surface into $K$ small
patches with equal area $\Delta S$ that is much less than
the acoustic wavelength, so the integral in Eqn. \ref{eq:CSIR}
can be approximated by summation as
\begin{equation}
\label{eq:DSIR}
\hat{p}^{\text{SIR}}(\mathbf{r}_l^d,m \Delta t) \simeq
\sum\limits_{k=1}^K
p(\mathbf{r}^k_l,m \Delta t)
\frac{\Delta S}{S(\mathbf{r}_l^d)},
\end{equation}
or in the equivalent matrix form
\begin{equation}
\label{eq:MSIR}
\hat{p}^{\text{SIR}}(\mathbf{r}_l^d,m \Delta t) \simeq
\boldsymbol\gamma^{\text{SIR}} \hat{\mathbf{p}}_m^l
\end{equation}
where $\mathbf{r}^k_l$ denotes the center of
the $k$-th patch of the $l$-th transducer,
$\Delta S$ is the patch area,
$\boldsymbol\gamma^{\text{SIR}} \equiv \frac{\Delta S}{S(\mathbf{r}_l^d)}
(1,\cdots,1)$ is a $1 \times K$ vector, 
$\hat{\mathbf{p}}_m^l = (p(\mathbf{r}^1_l,m \Delta t),
\cdots, p(\mathbf{r}^K_l,m \Delta t))^{\rm T}$
denotes the acoustic pressure at patches of $l$-th transducer
at time $m \Delta t$. Here for simplicity, 
we assume all the transducers are divided into 
$K$ patches with equal area $\Delta S$, 
and it is readily to extend to general cases where 
$l$-th transducer is divided into $K_l$ patches 
with area of $\Delta S_{lk}$.

Recalling the measured pressure data
$\hat{\mathbf{p}}_m$ and $\hat{\mathbf{p}}$
defined for point-like transducer, we can 
redefine $\hat{\mathbf{p}}_m$ as a 
$KL \times 1$ vector that represents 
the acoustic pressure at patches 
of transducers with finite area at time 
$t=m \Delta t$ as
\begin{equation}
\label{eq:pm2}
\hat{\mathbf{p}}_m
\equiv
\begin{bmatrix}
\hat{\mathbf{p}}_m^1 \\
\vdots \\
\hat{\mathbf{p}}_m^L
\end{bmatrix}.
\end{equation}
The corresponding $\hat{\mathbf{p}}$ 
can be redefine as a $KLM \times 1$ vector 
denoting the measured pressure data 
corresponding to all transducer
and temporal samples as 
\begin{equation}
\label{eq:phat2}
\hat{\mathbf{p}}
\equiv
\begin{bmatrix}
\hat{\mathbf{p}}_0 \\
\vdots \\
\hat{\mathbf{p}}_{M-1}
\end{bmatrix}.
\end{equation}

The averaged pressure measured by all transducer
and temporal samples can be defined as the $LM \times 1$ vector
\begin{equation}
\label{eq:phatSIR}
\hat{\mathbf{p}}^{\text{SIR}}
\equiv
\begin{bmatrix}
\hat{\mathbf{p}}_0^{\text{SIR}} \\
\vdots \\
\hat{\mathbf{p}}_{M-1}^{\text{SIR}}
\end{bmatrix}.
\end{equation}
where the $L \times 1$ vector
\begin{equation}
\label{eq:pmSIR}
\hat{\mathbf{p}}_m^{\text{SIR}}
\equiv
\begin{bmatrix}
\hat{p}^{\text{SIR}}(\mathbf{r}_1^d,m \Delta t) \\
\vdots \\
\hat{p}^{\text{SIR}}(\mathbf{r}_L^d,m \Delta t)
\end{bmatrix}.
\end{equation}

According to Eqn. \ref{eq:MSIR},
$\hat{\mathbf{p}}$ and $\hat{\mathbf{p}}^{\text{SIR}}$
can be related as
\begin{equation}
\label{eq:phat_phatSIR}
\hat{\mathbf{p}}^{\text{SIR}}=\boldsymbol\Gamma^{\text{SIR}} \hat{\mathbf{p}}
\end{equation}
where the $KLM \times LM$ matrix
\begin{equation}
\label{eq:Gamma}
\boldsymbol\Gamma^{\text{SIR}} \equiv
\begin{bmatrix}
\boldsymbol\gamma^{\text{SIR}} & \mathbf{0}_{1 \times K} & \cdots & \mathbf{0}_{1 \times K}\\
\mathbf{0}_{1 \times K} & \boldsymbol\gamma^{\text{SIR}} & \cdots & \mathbf{0}_{1 \times K}\\
\vdots & \vdots & \ddots & \vdots \\
\mathbf{0}_{1 \times K} & \mathbf{0}_{1 \times K} & \cdots & \boldsymbol\gamma^{\text{SIR}}
\end{bmatrix}.
\end{equation}

The EIR models the electrical response of the 
piezoelectric transducer. With the assumption that 
the transducer is a linear shift invariant system 
with respect to the input averaged pressure time sequence, 
the output voltage signal is the convolution result 
of the input and the EIR. 

For simplicity, the transducers are assumed to 
process identical EIR, and let 
$\mathbf{h}^e = (h_1^e, \cdots, h_J^e)^{\rm T}$ 
be the discrete samples of the EIR. The input averaged pressure 
time sequence of the $l$-th transducer can be defined as a $L \times 1$ vector
$\hat{\mathbf{p}}^l_{\text{SIR}} \equiv 
(\hat{p}^{\text{SIR}}(\mathbf{r}_l^d, 0),\cdots,
\hat{p}^{\text{SIR}}(\mathbf{r}_l^d,(M-1) \Delta t) )^{\rm T}$.
Then the output voltage signal $\hat{\mathbf{p}}_l^{\text{IR}}$ of the 
$l$-th transducer can be expressed as a $(J+M-1) \times 1$ vector
\begin{equation}
\hat{\mathbf{p}}_l^{\text{IR}} = \mathbf{h}^e * \hat{\mathbf{p}}^l_{\text{SIR}},
\end{equation}
where $*$ denotes discrete linear convolution operation, 
which can be constructed as a matrix multiplication 
by converting one of the operands into 
the corresponding Toeplitz matrix.

The output voltage signals of all transducers 
$\hat{\mathbf{p}}^{\text{IR}} \equiv 
(\hat{\mathbf{p}}_1^{\text{IR}}, \cdots,
\hat{\mathbf{p}}_L^{\text{IR}} )^{\rm T}$
can then be computed as
\begin{equation}
\label{eq:peir}
\hat{\mathbf{p}}^{\text{IR}} = \boldsymbol\Gamma^{\text{EIR}} \hat{\mathbf{p}}^{\text{SIR}}
\end{equation}
where the $L(J+M-1) \times LM$ matrix
\begin{equation}
\boldsymbol\Gamma^{\text{EIR}} \equiv
\begin{bmatrix}
\boldsymbol\gamma^{\text{EIR}} \\
\vdots \\
\boldsymbol\gamma^{\text{EIR}}
\end{bmatrix}
\end{equation}
and $\boldsymbol\gamma^{\text{EIR}}$ 
is a $(J+M-1) \times LM$ Toeplitz-like matrix defined as
\begin{equation}
\boldsymbol\gamma^{\text{EIR}} \equiv
\begin{bmatrix}
h_1^e & \mathbf{0}_{1\times(L-1)} & 0 & \cdots & 0 & \mathbf{0}_{1\times(L-1)} & 0\\
\vdots & \vdots & h_1^e & \vdots & \vdots & \vdots & \vdots
\\
h_J^e & \vdots & \vdots & \vdots & 0 & \mathbf{0}_{1\times(L-1)} & 0\\
0 &  \mathbf{0}_{1\times(L-1)} & h_J^e & \cdots & h_1^e & \mathbf{0}_{1\times(L-1)} & 0\\
0 &  \mathbf{0}_{1\times(L-1)} & 0 & \cdots & \vdots & \vdots & h_1^e\\
\vdots & \vdots & \vdots & \vdots & h_J^e & \vdots & \vdots \\
0 &  \mathbf{0}_{1\times(L-1)} & 0 & \cdots & 0 & \mathbf{0}_{1\times(L-1)} & h_J^e
\end{bmatrix}
\end{equation}

By use of Eqns. \eqref{eq:H}, \eqref{eq:phat_phatSIR}, and \eqref{eq:peir},
it is readily found that
\begin{equation}
\label{eq:pir}
\hat{\mathbf{p}}^{\text{IR}} = 
\boldsymbol\Gamma^{\text{EIR}} \boldsymbol\Gamma^{\text{SIR}}
\mathbf{S} \mathbf{T}_{M-1} \cdots \mathbf{T}_1 \mathbf{T}_0
\mathbf{p}_0,
\end{equation}
and the corresponding system matrix that incorporates
the transducer impulse responses is found to be
\begin{equation}
\label{eq:pir}
\mathbf{H}^{\text{IR}} \equiv 
\boldsymbol\Gamma^{\text{EIR}} \boldsymbol\Gamma^{\text{SIR}}
\mathbf{S} \mathbf{T}_{M-1} \cdots \mathbf{T}_1 \mathbf{T}_0.
\end{equation}

\section*{appendix-B: Implementation of the FISTA algorithm for PACT}

Equation \eqref{eq:tv_solution} was solved iteratively whose pseudocodes are
provided in Alg.\ \ref{alg:PLSTV}, where `Lip' is the Lipschitz constant 
of the operator $2\mathbf H^{\rm T}\mathbf H$ \cite{BeckIEEE2009}. 

\begin{algorithm}[H]
\caption{\label{alg:PLSTV}
Solver of the optimization problem defined by Eqn.\ \eqref{eq:tv_solution}
}
\algsetup{indent=2em}
\begin{algorithmic}[1]
\REQUIRE $\hat{\mathbf p}$, $\mathbf p_0^{(0)}$, $\lambda$, $\rm Lip$
\ENSURE $ \hat{\mathbf p}_0$
  \STATE $t^{(0)} \leftarrow 1$;
         $\boldsymbol \sigma_0^{(1)} \leftarrow \mathbf p_0^{(0)} $
    \COMMENT{Set the initial guess (The zero initial guess was employed in all the studies in this article)}
  \FOR{$\zeta=1$ \TO $Z$}
    \STATE $\mathbf p_0^{(\zeta)}  
        \leftarrow \rm {F\_Dnoise}
          \big( \boldsymbol \sigma_0^{(\zeta)}-\frac{2}{\rm Lip}\mathbf H^T(\mathbf H 
                \boldsymbol \sigma_0^{(\zeta)}
               - \hat{\mathbf p}), 
          {2\lambda}/{Lip}\big)$
    \STATE $t^{(\zeta+1)} \leftarrow 0.5 + 0.5\sqrt{1+4(t^{(\zeta)})^2}$
    \STATE $\boldsymbol \sigma_0^{(\zeta+1)} \leftarrow
            \mathbf p_0^{(\zeta)} + 
            (t^{(\zeta)} -1)(\mathbf p_0^{(\zeta)} 
            - \mathbf p_0^{(\zeta-1)})/t^{(\zeta+1)}$
  \ENDFOR
  \STATE $\hat{\mathbf p}_0 \leftarrow \mathbf p_0^{(Z)}$
\end{algorithmic}
\end{algorithm}

Note that we extended the FISTA algorithm described in 
Ref. \cite{BeckIEEE2009} to 3D. 
The function `F\_Dnoise' in Alg.\ \ref{alg:PLSTV}-Line 3   
solves a de-noising problem defined as: 
\begin{equation}\label{eqn:Dnoise}
  \hat{\mathbf x} = \arg\min_{\mathbf x\geq 0} 
   \lVert \mathbf y - \mathbf x \rVert^2
   +\beta |\mathbf x|_{\rm TV},
\end{equation}
where $\beta = 2\lambda/{\rm Lip}$ and 
\begin{equation}
  \mathbf y = \hat{\mathbf p}
   -\frac{2}{\rm Lip}
   \mathbf H^{\rm T} (\mathbf H \boldsymbol\sigma_0^{(\zeta)}-\hat{\mathbf p}).
\end{equation} 
It has been demonstrated that Eqn.\ \eqref{eqn:Dnoise} can be solved 
efficiently \cite{BeckIEEE2009}, and the pseudocodes 
are provided in Alg.\ \ref{alg:Dnoise}.

\begin{algorithm}[H]
\caption{\label{alg:Dnoise}
Solver of the de-noising problem defined by Eqn.\ \eqref{eqn:Dnoise}
}
\algsetup{indent=2em}
\begin{algorithmic}[1]
\REQUIRE $\mathbf y$, $\beta$
\ENSURE $ \hat{\mathbf x}$
\STATE
       $\big[\mathbf a^{(1)}, \mathbf b^{(1)}, \mathbf c^{(1)}\big]\leftarrow$ \\
       $\big[\mathbf 0_{(N_1-1)\times N_2\times N_3}, 
         \mathbf 0_{N_1\times (N_2-1)\times N_3},
         \mathbf 0_{N_1\times N_2\times (N_3-1)}\big]$\\
       $\big[\mathbf d^{(0)}, \mathbf e^{(0)}, \mathbf f^{(0)}\big]\leftarrow$ \\
       $\big[\mathbf 0_{(N_1-1)\times N_2\times N_3},
         \mathbf 0_{N_1\times (N_2-1)\times N_3},
         \mathbf 0_{N_1\times N_2\times (N_3-1)}\big]$\\
       $t^{(1)}=1$
\FOR {$\zeta=1$ \TO $Z$}
  \STATE
     $\big[\mathbf d^{(\zeta)}, \mathbf e^{(\zeta)}, \mathbf f^{(\zeta)}\big]
     \leftarrow
     \mathcal P_p \Big\lbrace 
        [\mathbf a^{(\zeta)}, \mathbf b^{(\zeta)}, \mathbf c^{(\zeta)}]
     +(6\beta)^{-1} 
     \mathcal P_l^T \big\lbrace
     \mathcal P_c   \lbrace
              \mathbf y - 0.5\beta
          \mathcal P_l
                \lbrace\mathbf a^{(\zeta)}, \mathbf b^{(\zeta)}, \mathbf c^{(\zeta)}\rbrace
                    \rbrace
                  \big\rbrace
                  \Big\rbrace$
  \STATE $t^{(\zeta+1)} \leftarrow 1 + 0.5\sqrt{1+4(t^{(\zeta)})^2}$
  \STATE
  $\big[\mathbf a^{(\zeta+1)}, \mathbf b^{(\zeta+1)}, \mathbf c^{(\zeta+1)}\big]\leftarrow
    (t^{(\zeta)}-1)/t^{(\zeta+1)}
   \big[ \mathbf d^{(\zeta)}-\mathbf d^{(\zeta-1)}
    ,\mathbf e^{(\zeta)}-\mathbf e^{(\zeta-1)}
    ,\mathbf f^{(\zeta)}-\mathbf f^{(\zeta-1)}\big]
  $
\ENDFOR
\STATE $\hat{\mathbf x}
  \leftarrow \mathcal P_c \big\lbrace
         \mathbf y - \lambda \mathcal P_l 
     \lbrace\mathbf d^{(Z)}, \mathbf e^{(Z)}, \mathbf f^{(Z)}\rbrace
                 \big\rbrace
$
\end{algorithmic}
\end{algorithm}

The four operators 
$\mathcal P_l$
$\mathcal P_c$, 
$\mathcal P_l^T$ and $\mathcal P_p$ in  Alg.\ \ref{alg:Dnoise} 
are defined as follows: 

\noindent
$\mathcal P_l: \mathbb R^{(N_1-1)\times N_2\times N_3}
              \times\mathbb R^{N_1\times (N_2-1)\times N_3}
              \times\mathbb R^{N_1\times N_2\times (N_3-1)}
\rightarrow 
            \mathbb R^{ N_1 \times N_2\times N_3}
$.
\begin{equation}
\begin{split}
  &\big[\mathcal P_l\lbrace \mathbf a, \mathbf b, \mathbf c
      \rbrace\big]_{n_1,n_2,n_3}=\\
  & [\mathbf a]_{n_1,n_2,n_3} 
   +[\mathbf b]_{n_1,n_2,n_3}
   +[\mathbf c]_{n_1,n_2,n_3} -\\
  & [\mathbf a]_{n_1-1,n_2,n_3} 
   -[\mathbf b]_{n_1,n_2-1,n_3}
   -[\mathbf c]_{n_1,n_2,n_3-1}\\
   & \text{for}\quad
    n_1 = 1, \cdots, N_1, \,
    n_2 = 1, \cdots, N_2, \,
    n_3 = 1, \cdots, N_3, 
\end{split}
\end{equation}
where we assume $[\mathbf a]_{0,n_2,n_3} = [\mathbf a]_{N_1,n_2,n_3}
=[\mathbf b]_{n_1,0,n_3} = [\mathbf b]_{n_1,N_2,n_3}
=[\mathbf c]_{n_1,n_2,0} = [\mathbf c]_{n_1,n_2,N_3}\equiv 0$. 

\noindent
$\mathcal P_c: \mathbb R^{ N_1 \times N_2\times N_3}
     \rightarrow
        \mathbb R^{ N_1 \times N_2\times N_3}$. 
\begin{equation}
\big[\mathcal P_c \lbrace \mathbf x \rbrace \big]_{n_1,n_2,n_3}
  =\max
   \big\lbrace
     0, [\mathbf x ]_{n_1,n_2,n_3}. 
   \big\rbrace
\end{equation}

\noindent
$\mathcal P_l^T: 
            \mathbb R^{ N_1 \times N_2\times N_3}
\rightarrow 
            \mathbb R^{(N_1-1)\times N_2\times N_3}
              \times\mathbb R^{N_1\times (N_2-1)\times N_3}
              \times\mathbb R^{N_1\times N_2\times (N_3-1)}
$. 
If we denote the input and output matrices by $\mathbf y$
and $(\mathbf a, \mathbf b, \mathbf c)$ respectively, 
we have 
\begin{equation}
\begin{split}
  &[\mathbf a]_{n_1,n_2,n_3} = [\mathbf y]_{n_1,n_2,n_3}
                             -[\mathbf y]_{n_1+1,n_2,n_3}, \\
  &\text{for}\quad n_1=1,\cdots, N_1-1,  n_2=1,\cdots, N_2, 
                     n_3=1,\cdots, N_3\\
  &[\mathbf b]_{n_1,n_2,n_3} = [\mathbf y]_{n_1,n_2,n_3}
                             -[\mathbf y]_{n_1,n_2+1,n_3},\\
  &\text{for}\quad n_1=1,\cdots, N_1,  n_2=1,\cdots, N_2-1, 
                     n_3=1,\cdots, N_3\\
  &[\mathbf c]_{n_1,n_2,n_3} = [\mathbf y]_{n_1,n_2,n_3}
                             -[\mathbf y]_{n_1,n_2,n_3+1},\\
  &\text{for}\quad n_1=1,\cdots, N_1,  n_2=1,\cdots, N_2, 
                     n_3=1,\cdots, N_3-1. 
\end{split}
\end{equation}

\noindent
$\mathcal P_p: \mathbb R^{(N_1-1)\times N_2\times N_3}
              \times\mathbb R^{N_1\times (N_2-1)\times N_3}
              \times\mathbb R^{N_1\times N_2\times (N_3-1)}
\rightarrow
          \mathbb R^{(N_1-1)\times N_2\times N_3}
              \times\mathbb R^{N_1\times (N_2-1)\times N_3}
              \times\mathbb R^{N_1\times N_2\times (N_3-1)}
$. 
If we denote the input and output matrices by 
$(\mathbf a, \mathbf b, \mathbf c)$
and $(\mathbf d, \mathbf e, \mathbf f)$ respectively, we have
\begin{equation}
  \begin{split}
    [\mathbf d]_{n_1,n_2,n_3} = 
      \frac{[\mathbf a]_{n_1,n_2,n_3}}
           {\max\big\lbrace 1, \sqrt{[\mathbf a]^2_{n_1,n_2,n_3}
                       + [\mathbf b]^2_{n_1,n_2,n_3}
                       + [\mathbf c]^2_{n_1,n_2,n_3}}
                \big\rbrace
           } \\
    [\mathbf e]_{n_1,n_2,n_3} = 
      \frac{[\mathbf b]_{n_1,n_2,n_3}}
           {\max\big\lbrace 1, \sqrt{[\mathbf a]^2_{n_1,n_2,n_3}
                       + [\mathbf b]^2_{n_1,n_2,n_3}
                       + [\mathbf c]^2_{n_1,n_2,n_3}}
                \big\rbrace
           } \\
    [\mathbf f]_{n_1,n_2,n_3} = 
      \frac{[\mathbf c]_{n_1,n_2,n_3}}
           {\max\big\lbrace 1, \sqrt{[\mathbf a]^2_{n_1,n_2,n_3}
                       + [\mathbf b]^2_{n_1,n_2,n_3}
                       + [\mathbf c]^2_{n_1,n_2,n_3}}
                \big\rbrace
           },  
  \end{split}
\end{equation}
where $ n_1=1,\cdots, N_1$, $n_2=1,\cdots, N_2$, $n_3=1,\cdots, N_3$,
and we assume 
$[\mathbf a]_{0,n_2,n_3} = [\mathbf a]_{N_1,n_2,n_3}
=[\mathbf b]_{n_1,0,n_3} = [\mathbf b]_{n_1,N_2,n_3}
=[\mathbf c]_{n_1,n_2,0} = [\mathbf c]_{n_1,n_2,N_3}\equiv 0$. 

\bibliographystyle{IEEEtran}

\bibliography{reference}

\begin{thebibliography}{10}
\providecommand{\url}[1]{#1}
\csname url@samestyle\endcsname
\providecommand{\newblock}{\relax}
\providecommand{\bibinfo}[2]{#2}
\providecommand{\BIBentrySTDinterwordspacing}{\spaceskip=0pt\relax}
\providecommand{\BIBentryALTinterwordstretchfactor}{4}
\providecommand{\BIBentryALTinterwordspacing}{\spaceskip=\fontdimen2\font plus
\BIBentryALTinterwordstretchfactor\fontdimen3\font minus
  \fontdimen4\font\relax}
\providecommand{\BIBforeignlanguage}[2]{{%
\expandafter\ifx\csname l@#1\endcsname\relax
\typeout{** WARNING: IEEEtran.bst: No hyphenation pattern has been}%
\typeout{** loaded for the language `#1'. Using the pattern for}%
\typeout{** the default language instead.}%
\else
\language=\csname l@#1\endcsname
\fi
#2}}
\providecommand{\BIBdecl}{\relax}
\BIBdecl

\bibitem{WanglihongBook2009}
L.~V. Wang, ``Photoacoustic imaging and spectroscopy,'' in \emph{Photoacoustic
  Imaging and Spectroscopy}.\hskip 1em plus 0.5em minus 0.4em\relax CRC, 2009.

\bibitem{OraevskyBook2003}
A.~A. $\mathrm {Oraevsky}$ and A.~A. Karabutov, ``Optoacoustic tomography,'' in
  \emph{Biomedical Photonics Handbook}, T.~Vo-Dinh, Ed.\hskip 1em plus 0.5em
  minus 0.4em\relax CRC Press LLC, 2003.

\bibitem{XuminghuaRSI2006}
M.~Xu and L.~V. Wang, ``Photoacoustic imaging in biomedicine,'' \emph{Review of
  Scientific Instruments}, vol.~77, no. 041101, 2006.

\bibitem{XuzhunJBO2010}
\BIBentryALTinterwordspacing
Z.~Xu, C.~Li, and L.~V. Wang, ``Photoacoustic tomography of water in phantoms
  and tissue,'' \emph{Journal of Biomedical Optics}, vol.~15, no.~3, pp.
  036\,019--036\,019--6, 2010. [Online]. Available: \url{+
  http://dx.doi.org/10.1117/1.3443793}
\BIBentrySTDinterwordspacing

\bibitem{GusevBook1993}
V.~E. Gusev and A.~A. Karabutov, ``Laser optoacoustic,'' in \emph{Laser
  Optoacoustic}.\hskip 1em plus 0.5em minus 0.4em\relax AIP, 1993.

\bibitem{KrugerMP1999}
R.~Kruger, D.~Reinecke, and G.~Kruger, ``Thermoacoustic computed tomography-
  technical considerations,'' \emph{Medical Physics}, vol.~26, pp. 1832--1837,
  1999.

\bibitem{HaltmeierIP2004}
\BIBentryALTinterwordspacing
M.~Haltmeier, O.~Scherzer, P.~Burgholzer, and G.~Paltauf, ``Thermoacoustic
  computed tomography with large planar receivers,'' \emph{Inverse Problems},
  vol.~20, no.~5, pp. 1663--1673, 2004. [Online]. Available:
  \url{http://stacks.iop.org/0266-5611/20/1663}
\BIBentrySTDinterwordspacing

\bibitem{CoxOSA2006}
\BIBentryALTinterwordspacing
B.~T. Cox, S.~R. Arridge, K.~P. K\"{o}stli, and P.~C. Beard, ``Two-dimensional
  quantitative photoacoustic image reconstruction of absorption distributions
  in scattering media by use of a simple iterative method,'' \emph{Appl. Opt.},
  vol.~45, no.~8, pp. 1866--1875, Mar 2006. [Online]. Available:
  \url{http://ao.osa.org/abstract.cfm?URI=ao-45-8-1866}
\BIBentrySTDinterwordspacing

\bibitem{XuzhunJBO2011}
\BIBentryALTinterwordspacing
Z.~Xu, Q.~Zhu, and L.~V. Wang, ``In vivo photoacoustic tomography ofÂ mouse
  cerebral edema induced byÂ coldÂ injury,'' \emph{Journal of Biomedical
  Optics}, vol.~16, no.~6, pp. 066\,020--066\,020--4, 2011. [Online].
  Available: \url{+ http://dx.doi.org/10.1117/1.3584847}
\BIBentrySTDinterwordspacing

\bibitem{HuangchaoJBObrain}
\BIBentryALTinterwordspacing
C.~Huang, L.~Nie, R.~W. Schoonover, Z.~Guo, C.~O. Schirra, M.~A. Anastasio, and
  L.~V. Wang, ``Aberration correction for transcranial photoacoustic tomography
  of primates employing adjunct image data,'' \emph{Journal of Biomedical
  Optics}, vol.~17, no.~6, p. 066016, 2012. [Online]. Available:
  \url{http://link.aip.org/link/?JBO/17/066016/1}
\BIBentrySTDinterwordspacing

\bibitem{FryJASA1978}
\BIBentryALTinterwordspacing
F.~J. Fry and J.~E. Barger, ``Acoustical properties of the human skull,''
  \emph{The Journal of the Acoustical Society of America}, vol.~63, no.~5, pp.
  1576--1590, 1978. [Online]. Available:
  \url{http://dx.doi.org/doi/10.1121/1.381852}
\BIBentrySTDinterwordspacing

\bibitem{XingjinMP2008}
\BIBentryALTinterwordspacing
X.~Jin, C.~Li, and L.~V. Wang, ``Effects of acoustic heterogeneities on
  transcranial brain imaging with microwave-induced thermoacoustic
  tomography,'' \emph{Medical Physics}, vol.~35, no.~7, pp. 3205--3214, 2008.
  [Online]. Available: \url{http://dx.doi.org/10.1118/1.2938731}
\BIBentrySTDinterwordspacing

\bibitem{HuangchaoJBOatten}
\BIBentryALTinterwordspacing
C.~Huang, L.~Nie, R.~W. Schoonover, L.~V. Wang, and M.~A. Anastasio,
  ``Photoacoustic computed tomography correcting for heterogeneity and
  attenuation,'' \emph{Journal of Biomedical Optics}, vol.~17, no.~6, p.
  061211, 2012. [Online]. Available:
  \url{http://link.aip.org/link/?JBO/17/061211/1}
\BIBentrySTDinterwordspacing

\bibitem{XuyuanIEEE2003}
Y.~Xu and L.~Wang, ``Effects of acoustic heterogeneity in breast thermoacoustic
  tomography,'' \emph{Ultrasonics, Ferroelectrics and Frequency Control, IEEE
  Transactions on}, vol.~50, no.~9, pp. 1134 --1146, sept. 2003.

\bibitem{DimpleJBO2010}
\BIBentryALTinterwordspacing
D.~Modgil, M.~A. Anastasio, and P.~J. La~RiviÃ¨re, ``Image reconstruction in
  photoacoustic tomography with variable speed of sound using a higher-order
  geometrical acoustics approximation,'' \emph{Journal of Biomedical Optics},
  vol.~15, no.~2, pp. 021\,308--021\,308--9, 2010. [Online]. Available: \url{+
  http://dx.doi.org/10.1117/1.3333550}
\BIBentrySTDinterwordspacing

\bibitem{JoseOE2011}
\BIBentryALTinterwordspacing
J.~Jose, R.~G.~H. Willemink, S.~Resink, D.~Piras, J.~C.~G. van Hespen, C.~H.
  Slump, W.~Steenbergen, T.~G. van Leeuwen, and S.~Manohar, ``Passive element
  enriched photoacoustic computed tomography (per pact) for simultaneous
  imaging of acoustic propagation properties and light absorption,'' \emph{Opt.
  Express}, vol.~19, no.~3, pp. 2093--2104, Jan 2011. [Online]. Available:
  \url{http://www.opticsexpress.org/abstract.cfm?URI=oe-19-3-2093}
\BIBentrySTDinterwordspacing

\bibitem{DeanBenAPL2011}
\BIBentryALTinterwordspacing
X.~L. De\'{a}n-Ben, V.~Ntziachristos, and D.~Razansky, ``Statistical
  optoacoustic image reconstruction using a-priori knowledge on the location of
  acoustic distortions,'' \emph{Applied Physics Letters}, vol.~98, no.~17, p.
  171110, 2011. [Online]. Available:
  \url{http://link.aip.org/link/?APL/98/171110/1}
\BIBentrySTDinterwordspacing

\bibitem{Yuanzhen2007MP}
\BIBentryALTinterwordspacing
Z.~Yuan and H.~Jiang, ``Three-dimensional finite-element-based photoacoustic
  tomography: Reconstruction algorithm and simulations,'' \emph{Medical
  Physics}, vol.~34, no.~2, pp. 538--546, 2007. [Online]. Available:
  \url{http://link.aip.org/link/?MPH/34/538/1}
\BIBentrySTDinterwordspacing

\bibitem{Yaolei2011AO}
\BIBentryALTinterwordspacing
L.~Yao and H.~Jiang, ``Enhancing finite element-based photoacoustic tomography
  using total variation minimization,'' \emph{Appl. Opt.}, vol.~50, no.~25, pp.
  5031--5041, Sep 2011. [Online]. Available:
  \url{http://ao.osa.org/abstract.cfm?URI=ao-50-25-5031}
\BIBentrySTDinterwordspacing

\bibitem{HristovaIP2008}
\BIBentryALTinterwordspacing
Y.~Hristova, P.~Kuchment, and L.~Nguyen, ``Reconstruction and time reversal in
  thermoacoustic tomography in acoustically homogeneous and inhomogeneous
  media,'' \emph{Inverse Problems}, vol.~24, no.~5, p. 055006, 2008. [Online].
  Available: \url{http://stacks.iop.org/0266-5611/24/i=5/a=055006}
\BIBentrySTDinterwordspacing

\bibitem{TreebyIP2010}
\BIBentryALTinterwordspacing
B.~E. Treeby, E.~Z. Zhang, and B.~T. Cox, ``Photoacoustic tomography in
  absorbing acoustic media using time reversal,'' \emph{Inverse Problems},
  vol.~26, no.~11, p. 115003, 2010. [Online]. Available:
  \url{http://stacks.iop.org/0266-5611/26/i=11/a=115003}
\BIBentrySTDinterwordspacing

\bibitem{StefanovIP2009}
\BIBentryALTinterwordspacing
P.~Stefanov and G.~Uhlmann, ``Thermoacoustic tomography with variable sound
  speed,'' \emph{Inverse Problems}, vol.~25, no.~7, p. 075011, 2009. [Online].
  Available: \url{http://stacks.iop.org/0266-5611/25/i=7/a=075011}
\BIBentrySTDinterwordspacing

\bibitem{QianiJIS2011}
\BIBentryALTinterwordspacing
J.~Qian, P.~Stefanov, G.~Uhlmann, and H.~Zhao, ``An efficient neumann
  series-based algorithm for thermoacoustic and photoacoustic tomography with
  variable sound speed,'' \emph{SIAM J. Img. Sci.}, vol.~4, no.~3, pp.
  850--883, Sep. 2011. [Online]. Available:
  \url{http://dx.doi.org/10.1137/100817280}
\BIBentrySTDinterwordspacing

\bibitem{CoxJASA2007}
\BIBentryALTinterwordspacing
B.~T. Cox, S.~Kara, S.~R. Arridge, and P.~C. Beard, ``k-space propagation
  models for acoustically heterogeneous media: Application to biomedical
  photoacoustics,'' \emph{The Journal of the Acoustical Society of America},
  vol. 121, no.~6, pp. 3453--3464, 2007. [Online]. Available:
  \url{http://link.aip.org/link/?JAS/121/3453/1}
\BIBentrySTDinterwordspacing

\bibitem{MastIEEE2001}
T.~Mast, L.~Souriau, D.-L. Liu, M.~Tabei, A.~Nachman, and R.~Waag, ``A k-space
  method for large-scale models of wave propagation in tissue,''
  \emph{Ultrasonics, Ferroelectrics and Frequency Control, IEEE Transactions
  on}, vol.~48, no.~2, pp. 341 --354, march 2001.

\bibitem{PatrickOL2006}
\BIBentryALTinterwordspacing
P.~J.~L. Rivi\`{e}re, J.~Zhang, and M.~A. Anastasio, ``Image reconstruction in
  optoacoustic tomography for dispersive acoustic media,'' \emph{Opt. Lett.},
  vol.~31, no.~6, pp. 781--783, Mar 2006. [Online]. Available:
  \url{http://ol.osa.org/abstract.cfm?URI=ol-31-6-781}
\BIBentrySTDinterwordspacing

\bibitem{BurgholzerSPIE2007}
\BIBentryALTinterwordspacing
P.~Burgholzer, H.~Grün, M.~Haltmeier, R.~Nuster, and G.~Paltauf,
  ``Compensation of acoustic attenuation for high-resolution photoacoustic
  imaging with line detectors,'' \emph{Proceedings of the SPIE}, vol. 6437,
  no.~1, p. 643724, 2007. [Online]. Available:
  \url{http://dx.doi.org/doi/10.1117/12.700723}
\BIBentrySTDinterwordspacing

\bibitem{DimpleSPIE2009}
\BIBentryALTinterwordspacing
D.~Modgil, M.~A. Anastasio, and P.~J.~L. Riviere, ``Photoacoustic image
  reconstruction in an attenuating medium using singular value decomposition,''
  \emph{Proceedings of the SPIE}, vol. 7177, no.~1, p. 71771B, 2009. [Online].
  Available: \url{http://dx.doi.org/doi/10.1117/12.809030}
\BIBentrySTDinterwordspacing

\bibitem{BenPMB2011}
\BIBentryALTinterwordspacing
X.~L. Deán-Ben, D.~Razansky, and V.~Ntziachristos, ``The effects of acoustic
  attenuation in optoacoustic signals,'' \emph{Physics in Medicine and
  Biology}, vol.~56, no.~18, p. 6129, 2011. [Online]. Available:
  \url{http://stacks.iop.org/0031-9155/56/i=18/a=021}
\BIBentrySTDinterwordspacing

\bibitem{SzaboJASA1994}
\BIBentryALTinterwordspacing
T.~L. Szabo, ``Time domain wave equations for lossy media obeying a frequency
  power law,'' \emph{The Journal of the Acoustical Society of America},
  vol.~96, no.~1, pp. 491--500, 1994. [Online]. Available:
  \url{http://dx.doi.org/doi/10.1121/1.410434}
\BIBentrySTDinterwordspacing

\bibitem{SzaboBook2004}
------, ``Diagnostic ultrasound imaging,'' in \emph{Diagnostic Ultrasound
  Imaging: Inside Out}.\hskip 1em plus 0.5em minus 0.4em\relax Elsevier, 2004.

\bibitem{MorseBook1987}
P.~M. Morse, ``Theoretical acoustics,'' in \emph{Theoretical Acoustics}.\hskip
  1em plus 0.5em minus 0.4em\relax Princeton University Press, 1987.

\bibitem{HuangSPIE2010}
\BIBentryALTinterwordspacing
C.~Huang, A.~A. Oraevsky, and M.~A. Anastasio, ``Investigation of limited-view
  image reconstruction in optoacoustic tomography employing a priori structural
  information,'' in \emph{Image Reconstruction from Incomplete Data VI}, P.~J.
  Bones, M.~A. Fiddy, and R.~P. Millane, Eds., vol. 7800, no.~1.\hskip 1em plus
  0.5em minus 0.4em\relax SPIE, 2010, p. 780004. [Online]. Available:
  \url{http://link.aip.org/link/?PSI/7800/780004/1}
\BIBentrySTDinterwordspacing

\bibitem{AubryJASA2003}
\BIBentryALTinterwordspacing
J.-F. Aubry, M.~Tanter, M.~Pernot, J.-L. Thomas, and M.~Fink, ``Experimental
  demonstration of noninvasive transskull adaptive focusing based on prior
  computed tomography scans,'' \emph{The Journal of the Acoustical Society of
  America}, vol. 113, no.~1, pp. 84--93, 2003. [Online]. Available:
  \url{http://dx.doi.org/doi/10.1121/1.1529663}
\BIBentrySTDinterwordspacing

\bibitem{XingjinPMB}
\BIBentryALTinterwordspacing
X.~Jin and L.~V. Wang, ``Thermoacoustic tomography with correction for acoustic
  speed variations,'' \emph{Physics in Medicine and Biology}, vol.~51, no.~24,
  p. 6437, 2006. [Online]. Available:
  \url{http://stacks.iop.org/0031-9155/51/i=24/a=010}
\BIBentrySTDinterwordspacing

\bibitem{JoseMP2012}
\BIBentryALTinterwordspacing
J.~Jose, R.~G.~H. Willemink, W.~Steenbergen, C.~H. Slump, T.~G. van Leeuwen,
  and S.~Manohar, ``Speed-of-sound compensated photoacoustic tomography for
  accurate imaging,'' \emph{Medical Physics}, vol.~39, no.~12, pp. 7262--7271,
  2012. [Online]. Available: \url{http://link.aip.org/link/?MPH/39/7262/1}
\BIBentrySTDinterwordspacing

\bibitem{Fessler94}
J.~A. {Fessler}, ``Penalized weighted least-squares reconstruction for positron
  emission tomography,'' \emph{IEEE Transactions on Medical Imaging}, vol.~13,
  pp. 290--300, 1994.

\bibitem{Lee1980IJPP}
\BIBentryALTinterwordspacing
D.~T. Lee and B.~J. Schachter, ``Two algorithms for constructing a delaunay
  triangulation,'' \emph{International Journal of Parallel Programming},
  vol.~9, pp. 219--242, 1980, 10.1007/BF00977785. [Online]. Available:
  \url{http://dx.doi.org/10.1007/BF00977785}
\BIBentrySTDinterwordspacing

\bibitem{treeby2010k}
B.~Treeby and B.~Cox, ``k-wave: {{\rm MATLAB}} toolbox for the simulation and
  reconstruction of photoacoustic wave fields,'' \emph{Journal of Biomedical
  Optics}, vol.~15, p. 021314, 2010.

\bibitem{TabeiJASA2002}
\BIBentryALTinterwordspacing
M.~Tabei, T.~D. Mast, and R.~C. Waag, ``A k-space method for coupled
  first-order acoustic propagation equations,'' \emph{The Journal of the
  Acoustical Society of America}, vol. 111, no.~1, pp. 53--63, 2002. [Online].
  Available: \url{http://link.aip.org/link/?JAS/111/53/1}
\BIBentrySTDinterwordspacing

\bibitem{KatsibasIEEE2004}
T.~Katsibas and C.~Antonopoulos, ``A general form of perfectly matched layers
  for for three-dimensional problems of acoustic scattering in lossless and
  lossy fluid media,'' \emph{Ultrasonics, Ferroelectrics and Frequency Control,
  IEEE Transactions on}, vol.~51, no.~8, pp. 964 --972, aug. 2004.

\bibitem{BeckIEEE2009}
A.~Beck and M.~Teboulle, ``Fast gradient-based algorithms for constrained total
  variation image denoising and deblurring problems,'' \emph{Image Processing,
  IEEE Transactions on}, vol.~18, no.~11, pp. 2419 --2434, nov. 2009.

\bibitem{WangkunPMB2012}
\BIBentryALTinterwordspacing
K.~Wang, R.~Su, A.~A. Oraevsky, and M.~A. Anastasio, ``Investigation of
  iterative image reconstruction in three-dimensional optoacoustic
  tomography,'' \emph{Physics in Medicine and Biology}, vol.~57, no.~17, p.
  5399, 2012. [Online]. Available:
  \url{http://stacks.iop.org/0031-9155/57/i=17/a=5399}
\BIBentrySTDinterwordspacing

\bibitem{Jacket}
B.~Zhang, S.~Xu, F.~Zhang, Y.~Bi, and L.~Huang, ``Accelerating matlab code
  using gpu: A review of tools and strategies,'' in \emph{Artificial
  Intelligence, Management Science and Electronic Commerce (AIMSEC), 2011 2nd
  International Conference on}, aug. 2011, pp. 1875 --1878.

\bibitem{KaipioJCAM2007}
\BIBentryALTinterwordspacing
J.~Kaipio and E.~Somersalo, ``Statistical inverse problems: discretization,
  model reduction and inverse crimes,'' \emph{J. Comput. Appl. Math.}, vol.
  198, no.~2, pp. 493--504, Jan. 2007. [Online]. Available:
  \url{http://dx.doi.org/10.1016/j.cam.2005.09.027}
\BIBentrySTDinterwordspacing

\bibitem{CGAFaq}
CGAFaq, ``Evenly distributed points on sphere,''
  $\url{http://cgafaq.info/wiki/Evenly_distributed_points_on_sphere}$.

\bibitem{NielimingJBO2011}
\BIBentryALTinterwordspacing
L.~Nie, Z.~Guo, and L.~V. Wang, ``Photoacoustic tomography of monkey brain
  using virtual point ultrasonic transducers,'' vol.~16, no.~7, p. 076005,
  2011. [Online]. Available: \url{http://dx.doi.org/doi/10.1117/1.3595842}
\BIBentrySTDinterwordspacing

\bibitem{BobJBO2012}
R.~W. Schoonover, L.~V. Wang, and M.~A. Anastasio, ``Numerical investigation of
  the effects of shear waves in transcranial photoacoustic tomography with a
  planar geometry,'' \emph{Journal of Biomedical Optics}, vol.~17, no.~6, 2012.

\bibitem{iterTR}
T.~B. and C.~B., ``2d iterative image improvement using time reversal
  example,''
  $\url{http://www.k-wave.org/documentation/example_pr_2D_tr_iterative.php}$.

\bibitem{GHarris1981}
\BIBentryALTinterwordspacing
G.~R. Harris, ``Review of transient field theory for a baffled planar piston,''
  \emph{The Journal of the Acoustical Society of America}, vol.~70, no.~1, pp.
  10--20, 1981. [Online]. Available:
  \url{http://link.aip.org/link/?JAS/70/10/1}
\BIBentrySTDinterwordspacing

\bibitem{VAndreev2002}
\BIBentryALTinterwordspacing
V.~G. Andreev, A.~A. Karabutov, A.~E. Ponomaryov, and A.~A. Oraevsky,
  ``Detection of optoacoustic transients with a rectangular transducer of
  finite dimensions,'' in \emph{Biomedical Optoacoustics III}, A.~A. Oraevsky,
  Ed., vol. 4618, no.~1.\hskip 1em plus 0.5em minus 0.4em\relax SPIE, 2002, pp.
  153--162. [Online]. Available: \url{http://link.aip.org/link/?PSI/4618/153/1}
\BIBentrySTDinterwordspacing

\bibitem{KunTMI2011}
K.~Wang, S.~Ermilov, R.~Su, H.-P. Brecht, A.~Oraevsky, and M.~Anastasio, ``An
  imaging model incorporating ultrasonic transducer properties for
  three-dimensional optoacoustic tomography,'' \emph{Medical Imaging, IEEE
  Transactions on}, vol.~30, no.~2, pp. 203 --214, feb. 2011.

\end{thebibliography}

\newpage

\end{document}